\documentclass[fleqn,twoside]{article}
\usepackage{amsmath,amssymb,espcrc2,color,cite}
\usepackage[usenames,dvipsnames]{xcolor}

\usepackage{float}
\usepackage{graphicx,soul}
\usepackage[figuresright]{rotating}
\usepackage{ulem}
\usepackage{capt-of}
\newcommand{\url}[1]{{\tt #1}}
\newcommand{\lsim}
{\;\raisebox{-.3em}{$\stackrel{\displaystyle <}{\sim}$}\;}

\newcommand{\gmt}{\ensuremath{(g-2)_\mu}}
\newcommand{\br}{{\rm BR}}
\newcommand{\bsg}{BR($B_s \to X_s \gamma$)}

\newcommand{\bsmm}{\ensuremath{\br(B_s \to \mu^+\mu^-)}}
\newcommand{\bdmm}{\ensuremath{\br(B_d \to \mu^+\mu^-)}}
\newcommand{\bsdmm}{\ensuremath{\br(B_{s, d} \to \mu^+\mu^-)}}

\newcommand{\ssi}{\ensuremath{\sigma^{\rm SI}_p}}

\newcommand{\ssd}{\ensuremath{\sigma^{\rm SD}_p}}

\newcommand{\MW}{\ensuremath{M_W}}

\newcommand{\Mh}{\ensuremath{M_h}}

\newcommand{\MA}{\ensuremath{M_A}}

\newcommand{\msusy}{M_{\rm SUSY}}
\newcommand{\gl}{\ensuremath{{\tilde g}}}
\newcommand{\mgl}{\ensuremath{m_{\tilde g}}}
\newcommand{\msq}{\ensuremath{m_{\tilde q}}}

\newcommand{\sto}[1]{\ensuremath{\tilde t_{#1}}}

\newcommand{\sbot}[1]{\ensuremath{\tilde b_{#1}}}

\newcommand{\msqt}{\ensuremath{m_{\tilde q_3}}}
\newcommand{\msl}{\ensuremath{m_{\tilde l}}}

\newcommand{\ino}[1]{\tilde \chi_{#1}}

\newcommand{\cha}[1]{\tilde \chi^\pm_{#1}}
\newcommand{\chap}[1]{\tilde \chi^+_{#1}}
\newcommand{\cham}[1]{\tilde \chi^-_{#1}}

\newcommand{\mcha}[1]{\ensuremath{m_{\tilde \chi^\pm_{#1}}}}
\newcommand{\neu}[1]{\ensuremath{\tilde \chi^0_{#1}}}
\newcommand{\mneu}[1]{\ensuremath{m_{\tilde \chi^0_{#1}}}}

\newcommand{\sq}{\tilde q}
\newcommand{\sqr}{\tilde{q}_R}
\newcommand{\mst}[1]{m_{\tilde t_{#1}}}
\newcommand{\msb}[1]{m_{\tilde b_{#1}}}
\newcommand{\mstau}[1]{\ensuremath{m_{\tilde \tau_{#1}}}}
\newcommand{\msmu}[1]{\ensuremath{m_{\tilde \mu_{#1}}}}
\newcommand{\msel}[1]{\ensuremath{m_{\tilde e_{#1}}}}
\newcommand{\mslep}{\ensuremath{m_{\tilde \ell}}}

\newcommand{\staue}{\tilde \tau_1}
\newcommand{\mstaue}{m_{\staue}}

\newcommand{\tb}{\ensuremath{\tan\beta}}

\newcommand{\tev}{\ensuremath{\,\, \mathrm{TeV}}}
\newcommand{\gev}{\ensuremath{\,\, \mathrm{GeV}}}
\newcommand{\mev}{\ensuremath{\,\, \mathrm{MeV}}}

\def\reffi#1{\mbox{Fig.~\ref{#1}}}

\def\refse#1{\mbox{Sect.~\ref{#1}}}

\definecolor{orange}{rgb}{1,0.5,0}

\definecolor{Gray}{named}{Gray}

\newcommand{\ETslash}{\ensuremath{/ \hspace{-.7em} E_T}}

\title{\vspace{-4.25cm}
\bf \LARGE Likelihood Analysis of the pMSSM11 in Light of LHC 13-TeV Data} \vspace{0.5em}

\author{
{\bf E.~Bagnaschi}\address[DESY]
   {DESY, Notkestra{\ss}e 85, D--22607 Hamburg, Germany},
{\bf K.~Sakurai}\address[Warsaw]{Institute of Theoretical Physics, Faculty of Physics, University of Warsaw, ul.~Pasteura 5, PL--02--093 Warsaw, Poland},
{\bf M.~Borsato}\address[USdC]{Instituto Galego de F{\' i}sica de Altas Enerx{\' i}as, Universidade de Santiago de Compostela, Spain},
{\bf O.~Buchmueller}{\address[Imperial]
   {High\,Energy\,Physics\,Group,\,Blackett\,Laboratory,\,Imperial\,College,\,Prince\,Consort\,Road,\,London\,SW7\,2AZ,\,UK},
{\bf M.~Citron}\addressmark[Imperial],
{\bf J.~C.~Costa}\addressmark[Imperial],
{\bf A.~De~Roeck}\address[CERNEP]
   {Experimental Physics Department, CERN, CH--1211 Geneva 23, Switzerland; \\  Antwerp University, B--2610 Wilrijk, Belgium},
 {\bf M.J.~Dolan}\address[SLAC]
{ARC Centre of Excellence for Particle Physics at the Terascale, School of Physics, University of Melbourne, 3010, Australia},
{\bf J.R.~Ellis}\address[KCL]{Theoretical Particle Physics
  and Cosmology Group, Department of Physics, King's College London, London~WC2R~2LS, UK; \\
  National Institute of Chemical Physics and Biophysics, R{\" a}vala 10, 10143 Tallinn, Estonia; \\
  Theoretical Physics Department, CERN, CH--1211 Geneva 23, Switzerland},
{\bf H.~Fl\"acher}\address[Bristol]
   {H.H.~Wills Physics Laboratory, University of Bristol, Tyndall Avenue, Bristol BS8 1TL, UK},
{\bf S.~Heinemeyer}\address[Madrid]
{Campus of International Excellence UAM+CSIC, Cantoblanco, E--28049 Madrid, Spain;\\
  Instituto de F\'{\i}sica Te{\'o}rica UAM-CSIC, C/ Nicolas Cabrera 13-15, E--28049 Madrid, Spain; \\
   Instituto de F\'{\i}sica de Cantabria (CSIC-UC), Avda. de Los Castros s/n,
    E--39005 Santander, Spain},
{\bf M.~Lucio}\addressmark[USdC],
{\bf D.~Mart\'inez~Santos}\addressmark[USdC],
{\bf K.A.~Olive}\address[Minnesota]
{William I.\ Fine Theoretical Physics Institute, School of Physics and
 Astronomy, University of Minnesota, Minneapolis, Minnesota 55455, USA},
{\bf A.~Richards}\addressmark[Imperial],
{\bf V.C.~Spanos}\address[UA]{Section of Nuclear \& Particle Physics, Department of Physics,
National and Kapodistrian University of Athens, GRﾐ15784 Athens, Greece},
{\bf I.~Su{\'a}rez~Fern{\'a}ndez}\addressmark[USdC],
{\bf G.~Weiglein}\addressmark[DESY]
}}

\begin{document}
\begin{abstract}
\vspace{-0.15cm}

We {use {\tt MasterCode} to} perform a {frequentist} analysis of the constraints on a
phenomenological MSSM model with 11 parameters, the pMSSM11, including
constraints from {$\sim 36$/fb of LHC data at 13 TeV and PICO, XENON1T and PandaX-II
searches for dark matter scattering}, as well as previous
accelerator and astrophysical measurements, presenting fits {both} with and
without the \gmt\ constraint. The pMSSM11 is specified by the following
parameters: 3~gaugino masses $M_{1,2,3}$,  a common mass for the
first-and second-generation squarks $\msq$ and a distinct
third-generation squark mass $\msqt$, a common mass for the first-and
second-generation sleptons \mslep\ and a distinct third-generation
slepton mass $m_{\tilde \tau}$, a common trilinear mixing parameter
{$A$}, the Higgs mixing parameter $\mu$, the pseudoscalar Higgs mass
$\MA$ and $\tb$. In the fit including \gmt, a Bino-like $\neu1$ is preferred,
whereas a Higgsino-like $\neu1$ is mildly favoured when the \gmt\ constraint is
dropped. We identify the mechanisms that operate in different
regions of the pMSSM11 parameter space to bring the relic density of the
lightest neutralino, $\neu1$, into the range indicated by cosmological
data. {In the fit including \gmt, coannihilations with $\neu2$ and the Wino-like
$\cha1$ or with nearly-degenerate first- and second-generation sleptons are active,
whereas coannihilations with the $\neu2$ and the Higgsino-like $\cha1$
or with first- and second-generation squarks may be important when the \gmt\
constraint is dropped.} In the two cases, we present $\chi^2$ functions in
two-dimensional mass planes as well as their one-dimensional profile
projections and best-fit spectra. {Prospects remain for discovering strongly-interacting
sparticles at the LHC, in both the scenarios with and without the \gmt\ constraint,
as well as for discovering electroweakly-interacting sparticles at a future linear
$e^+ e^-$ collider such as the ILC or CLIC.
}
\vspace{0.25cm}
\begin{center}
{\tt KCL-PH-TH/2017-22, CERN-PH-TH/2017-087, DESY 17-059, IFT-UAM/CSIC-17-035\\
FTPI-MINN-17/17, UMN-TH-3701/17}
\end{center}
\end{abstract}
\thispagestyle{empty}
\newpage


\maketitle



\section{Introduction}
\label{sec:intro}
Supersymmetric (SUSY) models of TeV-scale physics are being subjected to
increasing pressure by the
strengthening constraints imposed by LHC experiments~\cite{CMSWiki,ATLASWiki} and
searches for Dark Matter (DM)~\cite{pandax,lux16,PICO,XENON1T}.
In particular, in the context of models with soft supersymmetry-breaking
parameters constrained to be universal at a high unification scale, the
LHC limits {on sparticle masses} have been in increasing tension with a supersymmetric
interpretation of the anomalous magnetic moment of the muon, $\gmt$,
which would require relatively light sleptons and electroweak
gauginos~\cite{mcold,mc9,mc10,mc11}.
This pressure has been ratcheted up by the advent of $\sim 36$/fb of data from Run 2 of
the LHC at a centre-of-mass energy of 13
TeV~\cite{cms_0lep-mt2,cms_1lep-MJ,sus-16-039}~\footnote{{We use here results from SUSY searches by the
CMS Collaboration:
the results from ATLAS~\cite{ATLASWiki} yield similar constraints.}},
which probe supersymmetric
models at significantly higher mass scales than was possible in Run~1 at
7~and 8~TeV in the centre of mass. In parallel, direct searches for DM
scattering have also been making significant progress towards the
neutrino `floor'~\cite{Snowmass}, in particular with the recent data releases from the LUX, PICO, XENON1T {and PandaX-II}
experiments~\cite{lux16,PICO,XENON1T,pandax}. Here we
analyze these constraints in the minimal supersymmetric extension of
the Standard Model (MSSM), which, because of $R$-parity, has a stable
cosmological relic particle that we assume to be the lightest
neutralino,~\neu1, \cite{EHNOS}.

The strengthening phenomenological, experimental and astrophysical
constraints on supersymmetry (SUSY) were initially explored mainly in the
contexts of models in which SUSY breaking was
assumed to be universal at the GUT scale, such as the constrained MSSM
(CMSSM)~\cite{mcold,mc9,Fittino}, non-universal Higgs models
(NUHM1,2)~\cite{mc9,mc10}~\footnote{{For a recent analysis of
these models in light of $\sim 13$/fb of LHC data at 13~TeV, see~\cite{GAMBIT}.
This analysis does not include the PICO, XENON1T and most recent PandaX-II results, and has other
differences that are noted later in this paper.}},
the minimal anomaly-mediated SUSY-breaking
model (mAMSB)~\cite{mc-amsb}, and models based on the SU(5)
group~\cite{mc-su5}. These models are
tractable by virtue of having a relatively limited number of parameters,
though the universality assumptions they employ are not necessarily well
{supported} in scenarios motivated by {fundamental principles, such as} string theory. Their limited
parameter spaces are amenable to analysis, {e.g.,} in {the frequentist
approach we follow}, in which one constructs a global likelihood function that
embodies all the information provided by the multiple constraints.

Alternatively, one may study phenomenological models in which the soft
SUSY-breaking parameters are not constrained by any universality
assumptions, though subject to milder constraints emanating, in
particular, from upper limits on SUSY contributions to
flavour-changing processes. These phenomenological MSSM
(pMSSM)~\cite{pMSSM} models contain
many more parameters, whose exploration is computationally demanding.
There have been cut-based global analyses of variants of the pMSSM with as many as
19 parameters~\cite{pMSSM19} and global fits focused on specific sectors or parameter ranges~\cite{pMSSMfocused},
however in the past we have restricted our frequentist attentions to a variant of the pMSSM with 10 parameters, the
pMSSM10~\cite{mc11,mc12}.
These were taken to be 3 independent gaugino masses, $M_{1,2,3}$,  a common
electroweak-scale mass for the first-and second-generation squarks, $\msq$,
a distinct mass for the third-generation squarks, $\msqt$, a common
electroweak-scale mass {\msl} for the sleptons,
a single trilinear mixing parameter {$A$} that is universal at the
electroweak scale, the Higgs mixing parameter $\mu$, the pseudoscalar
Higgs mass, $\MA$ and the ratio of Higgs vevs, $\tb$~\footnote{{For a recent analysis of a
7-dimensional version of the MSSM in light of $\sim 13$/fb of LHC data at 13~TeV, see
\cite{GAMBIT7}.}}.

It is desirable to extend this type of analysis to more general variants of the pMSSM,
for a couple of reasons. One is that the lower
bounds on sparticle masses will, in general, be weaker in models with
more parameters, so one should explore such models before making
statements about the magnitudes of these lower bounds and prospects for
discovering sparticles at the LHC or elsewhere. Another reason is that
reconciling the strengthening LHC constraints with the cosmological DM
density constraint requires, in general, specific relations between
sparticle masses that suppress the relic density via coannihilation
effects and/or rapid annihilations through direct-channel
resonances. Therefore one should study models capable of
accommodating these DM mechanisms~\cite{mc12}.

Examples of DM mechanisms that have been studied extensively in the past~\cite{mc12} include coannihilation
with the lighter stau slepton, $\staue$, the lighter chargino, $\cha1$, or the lighter stop
squark, $\sto1$, and rapid annihilations via the $Z$ boson, the 125-GeV Higgs boson, $h$,
or the heavier MSSM Higgs bosons, $H/A$. More recently, the possibility of
coannihilation with gluinos, $\gl$, has been explored in models with non-universal
gaugino masses~\cite{ELO,EELO}, and coannihilation with the right-handed up-type squarks of the first
two generations, ${\tilde u_R}/{\tilde c_R}$, emerged as a possibility in an SU(5)
model with non-universal scalar masses $m_5, m_{10}$ for sfermions in
$\mathbf{\bar 5}$ and $\mathbf{10}$ representations~\cite{mc-su5}.

All of these were possibilities in the pMSSM10, {but in that scenario the stau
and smuon masses were fixed to be equal, putting the {LHC constraints on} stau
coannihilation in
tension with the possibility of a SUSY
interpretation of $\gmt$, a tension that has increased with the advent of the
first LHC data at 13 TeV.}
{In this paper we study two possible resolutions of this {issue}.
We study an extension of the parameter space of the pMSSM10
to 11 parameters by relaxing the equality between the soft SUSY-breaking
contributions to the stau mass and to the (still common) masses of the smuon and
selectron, the pMSSM11.} {In order to assess the importance of the \gmt\ constraint,
we also consider a fit omitting the
SUSY interpretation of \gmt.} The principal results of this paper are
comparisons between the likelihoods of different spectra in the pMSSM11
with and without \gmt, and
comparisons between the likelihoods of different DM mechanisms including
$\staue${, ${\tilde \ell}$, ${\tilde q}$} and $\gl$ coannihilation, highlighting the
impacts of the LHC 13 TeV {and recent DM scattering} data.

The layout of this paper is as follows. In \refse{sec:framework}
we specify the framework of our analysis. Subsection~\ref{sec:pMSSM11}
specifies the pMSSM11, establishes our notation for its
parameters and describes our procedure for sampling the pMSSM11
parameter space. In Subsection~\ref{sec:MC} we review the
{\tt MasterCode} tool to construct a global $\chi^2$ likelihood function
combining constraints on model parameters, Subsection~\ref{sec:others}
describes our treatments of the {electroweak and flavour
  constraints, including some updates compared with our previous analyses.
In Subsection~\ref{sec:DM} we {give details} on our DM
analysis, {which includes constraints on both spin-independent and -dependent
DM scattering~\cite{lux16,PICO,XENON1T,pandax}. Our implementations of
the constraints from $\sim 36$/fb of LHC at 13~TeV~\cite{cms_0lep-mt2,cms_1lep-MJ,sus-16-039} are discussed
in} Subsection~\ref{sec:LHC}.
Then, in Section~\ref{sec:planes} we present results for the global
likelihood function in various parameter planes, highlighting the regions
where different DM mechanisms operate and comparing results with
and without the \gmt\ constraint being applied. Section~\ref{sec:1D}
displays the one-dimensional profile likelihood functions for various masses, mass
differences and other observables in these two cases, {and also shows predictions for
spin-independent and -dependent DM scattering. Section~\ref{sec:impacts}
highlights the impacts of the LHC 13-TeV data~\cite{cms_0lep-mt2,cms_1lep-MJ,sus-16-039}
and the recent direct searches for astrophysical
DM~\cite{lux16,PICO,XENON1T,pandax}}. Section~\ref{sec:spectra}
discusses the best-fit points, favoured and allowed spectra
{in these pMSSM scenarios}. Finally, Section~\ref{sec:conx}
summarizes our conclusions.


\section{Analysis Framework}
\label{sec:framework}

\subsection{Model Parameters}
\label{sec:pMSSM11}

As mentioned above, in this paper we consider a pMSSM scenario with eleven parameters, namely
\begin{align}
{\rm 3~gaugino~masses}&: \; M_{1,2,3} \, ,  \nonumber \\
{\rm 2~squark~masses}&: \; {\msq \, \equiv} \, m_{\tilde q_1}, m_{\tilde q_2} \nonumber \\
                       &\ne \, \msqt \, = \, m_{\tilde t} , m_{\tilde b}, \nonumber \\
{\rm 2~slepton~masses}&: \; {\mslep} \, \equiv \, m_{\tilde \ell_1} \, = m_{\tilde \ell_2} \, = \, \msel, \msmu
\nonumber \\
                       &\ne \, m_{\ell_3} \, = \, m_{\tilde \tau}, \nonumber \\
\label{mc11}
{\rm 1~trilinear~coupling}&: \; A \, ,  \\
{\rm Higgs~mixing~parameter}&: \; \mu \, ,  \nonumber \\
{\rm pseudoscalar~Higgs~mass}&: \; \MA \, ,  \nonumber \\
{\rm ratio~of~vevs}&: \; \tb \, ,  \nonumber
\end{align}
{where $q_{1,2} \equiv u, d, s, c$, we assume soft SUSY-breaking parameters for left- and right-handed sfermions,
and the sneutrinos have the same
soft SUSY-breaking parameter as the corresponding charged sfermions.} All of these parameters are specified at a
renormalisation scale $\msusy$ given by
the geometric mean of the masses of the scalar top eigenstates,
$\msusy \equiv \sqrt{\mst1 \mst2}$, which is also the scale at which electroweak symmetry breaking conditions
are imposed. {We allow the sign of the mixing parameter $\mu$ to be either positive or negative.}
{The important difference from the pMSSM10 scenario we studied previously~\cite{mc11} is
that the first- and second-generation slepton mass $\mslep$ and the stau mass $m_{\tilde \tau}$ are decoupled
in the pMSSM11}~\footnote{{In comparison, the pMSSM7 scenario studied in~\cite{GAMBIT7}
assumes gaugino and squark/slepton mass universality at
some input scale $Q$, and has two trilinear couplings $A_{t,b}$, independent Higgs masses $H_{u,d}$ and $\tb$
as free parameters.}}.

The ranges of these parameters sampled in our analysis are displayed in
Table~\ref{tab:ranges}. In each case, we indicate in the third column of Table~\ref{tab:ranges}
how the ranges of most of these parameters are divided into segments, much as we did previously
for our analysis of the pMSSM10~\cite{mc11}.


\begin{figure*}[thb!]
  \begin{center}
    \renewcommand{\arraystretch}{1.25}
  \begin{tabular}{|c|c|c|c|} \hline
Parameter   &  \; \, Range      & Number of  & Prior \\
            &             & segments   & Type \\
\hline
$M_1$       &  (-4 ,  4 )\tev  & 6 & soft \\
$M_2$       &  ( 0 ,  4 )\tev  & 2 & soft \\
$M_3$       &  (-4 ,  4 )\tev  & 4 & soft \\
\msq        &  ( 0 ,  4 )\tev  & 2 & soft \\
\msqt       &  ( 0 ,  4 )\tev  & 2 & soft \\
{\mslep}        &  ( 0 ,  2 )\tev  & 1 & soft \\
$m_{\tilde \tau}$       &  ( 0 ,  2 )\tev  & 1 & soft\\
\MA         &  ( 0 ,  4 )\tev  & 2 & soft \\
$A$         &  (-5  , 5 )\tev  & 1 & soft \\
$\mu$        &  (-5  , 5 )\tev  & 1 & soft \\
\tb         &  ( 1  , 60)      & 1 & soft \\
\hline
$M_t$~\cite{ATLAS:2014wva}   & $\mu = 173.34$ GeV, $\sigma = 0.76$ GeV & 1 & Gaussian \\
$M_Z$~\cite{PDG2015} & $\mu = 91.1876$ GeV, $\sigma = 0.0021$ GeV & 1 & Gaussian \\
$\Delta \alpha^{(5)}_{\mathrm{had}}(M_Z)$~\cite{PDG2015} & $\mu = 0.02771$, $\sigma = 0.00011$ & 1 & Gaussian \\
\hline \hline
Total number of boxes &   & 384   &  \\
\hline
  \end{tabular}
\end{center}
\captionof{table}{\it The ranges of the pMSSM11 parameters sampled, which are divided into the indicated numbers of
  segments, yielding the total number of sample boxes shown in the last row. In the last column, we indicate the kind of prior used,
  where ``soft'' means a flat prior  with Gaussian tails.}
\label{tab:ranges}
\end{figure*}

These segments define boxes in the eleven-dimensional parameter space, which we
sample using the {\tt MultiNest} package~\cite{multinest}. In order to ensure a smooth overlap between boxes
and eliminate features associated with their boundaries, we choose for each box a prior such that 80\% of the sample has a flat distribution
within the nominal box, and 20\% of the sample is in normally-distributed tails extending outside the box.
An initial scan over all mass parameters with absolute values
$ \le 4 \tev$ showed that non-trivial behaviour of the global likelihood function
was restricted to $|M_1|\lesssim 1 \tev$ and ${\mslep}\lesssim 1 \tev$.
In order to achieve high resolution efficiently, we restricted the range of ${\mslep}$ to $<2 \tev$
in the full scan~\footnote{{Since $\mslep > \mneu1$, this entails
also the restriction to $\mneu1 < 2 \tev$ visible in subsequent figures.}}.
To study properly the impact of the $(g-2)_{\mu}$, we performed separate sampling campaigns with and without it.
On the other hand, during the sampling phase the constraints coming from LHC13 results have not been included.
  Since their impact consists in providing lower bounds to the sparticle masses, this choice allows
  for a proper assessment of their impact on the full parameter space.
Moreover, we also performed dedicated scans for various DM annihilation mechanisms, in such a way to improve the quality
of the sample in the description of the fine-tuned spectrum configurations that characterize them.
The data sets from the various campaigns have been merged into a single set on which the likelihood
  is computed dynamically including or excluding the \gmt~and/or the LHC13 constraints according to our interest.
{The total number of points in our pMSSM11 parameter scan is  $\sim 2 \times 10^{9}$.
}


\subsection{MasterCode}
\label{sec:MC}

We perform a global likelihood analysis of the pMSSM11 including
constraints from
direct searches for SUSY particles at the LHC, measurements of the Higgs boson mass and signal strengths,
LHC searches for SUSY Higgs bosons, precision electroweak observables, flavour constraints from
$B$- and $K$-physics observables, the cosmological constraint on the
overall cold dark matter {(CDM)} density,
and upper limits on spin-independent and -dependent LSP-nuclear scattering. We treat \gmt\ as an optional constraint,
presenting results from global fits with and without it, {and we treat $m_t$, $\alpha_s$ and $M_Z$
as nuisance parameters.}

The observables contributing to the likelihood are calculated using the
{\tt MasterCode} tool~\cite{mcold,mc9,mc10,mc11,mc12,mc-su5,mc-amsb,mcweb},
which interfaces and combines consistently various public and private codes
using the SUSY Les Houches Accord (SLHA)~\cite{SLHA}. The following codes are used in this analysis:
{\tt SoftSusy~3.3.9}~\cite{Allanach:2001kg}
for the spectrum, {\tt FeynWZ}~\cite{Svenetal} for the electroweak precision observables~\footnote{We use here
an updated version of {\tt FeynWZ} (not yet publicly available) in which the $\MW$ evaluation is based on \cite{HHWZ} and is identical to that
implemented in {\tt FeynHiggs}, which gives more reliable results in parameter regions with larger SUSY masses or
small SUSY mass splittings. The other EWPO are treated in the same way as in~\cite{Svenetal}.},
{\tt FeynHiggs~2.11.3}~\cite{FeynHiggs}
for the Higgs sector~\footnote{{We note that {\tt FeynHiggs} incorporates resummation effects in Higgs mass calculations
that are not included in the {\tt MSSM} {\tt FlexibleSUSY} generator~\cite{FlexibleSUSY} used in~\cite{GAMBIT,GAMBIT7}, although these are available
through other {\tt FlexibleSUSY} generators, {\tt HSSUSY/SplitSUSY}\cite{FlexibleSUSY2} and {\tt FlexibleEFT}\cite{FlexibleEFT}.
It should also be noted that {\tt FeynHiggs} has recently been improved for higher SUSY mass scales~\cite{FeynHiggs2}.
}} and \gmt, {\tt SuFla}~\cite{SuFla} and
{\tt SuperIso}~\cite{SuperIso}
for the {flavour} physics observables, {\tt Micromegas-3.2}~\cite{MicroMegas} for the DM
relic density, {\tt SSARD}~\cite{SSARD} for the spin-independent {and -dependent elastic scattering cross-sections
\ssi\ and \ssd}~\footnote{The SSARD computation of the scattering cross-section follows the computations detailed in
\cite{EFOSS,EOSavage}. The uncertainties in the cross-sections are derived from a straightforward
propagation of errors in in the input quantities which determine the cross-section. The dominant
uncertainties are  discussed below in more detail.}, {\tt SDECAY~1.3b}~\cite{Sdecay} for calculating sparticle branching ratios, and {\tt HiggsSignals~1.4.0}~\cite{HiggsSignals}
and {\tt HiggsBounds~4.3.1}~\cite{HiggsBounds} for calculating constraints on the SUSY Higgs sector.


\subsection{Electroweak {and Flavour Constraints}}
\label{sec:others}

Our treatments of many of these constraints follow those we have used previously, which
were summarized most recently in Table~1
in~\cite{mc-su5}. Table~\ref{tab:updates} summarizes
the updates we make in this paper. As noted there,
the only change in the electroweak sector is in $\MW$~\footnote{{We emphasize that,
although they are not displayed in Table~\ref{tab:updates} because they have not changed since\cite{mc-su5}, we
use a complete set of electroweak constraints, not restricted to $\MW$ as used in~\cite{GAMBIT,GAMBIT7}. We also note that
{the {\tt FeynWZ} code we use to calculate $\MW$} incorporates 2-loop corrections that are not included
in the {\tt FlexibleSUSY} code~\cite{FlexibleSUSY} used in~\cite{GAMBIT,GAMBIT7}.}}. Here we
follow~\cite{HEPFit} in combining naively the recent ATLAS measurement
$\MW = 80.370 \pm 0.019 \gev$ with the previous world average value $\MW
= 80.385 \pm 0.015 \gev$, obtaining $\MW = 80.379 \pm 0.012
\gev$~%
\footnote{In so doing, we neglect correlations in the uncertainties due
  to PDFs, QED and boson $p_T$ modelling, but our results are relatively
  insensitive to the details of this combination.}%
.

Since one of our objectives in this paper is to emphasize the impact on the
pMSSM11 parameter space of the \gmt\ constraint, for reference we also
include in Table~\ref{tab:updates} the implementation of this constraint
that we use as an option}~\footnote{The \gmt\ evaluation in {\tt FeynHiggs} contains less
sophisticated two-loop corrections than {\tt GM2CALC}~\cite{GM2CALC}.
However, the difference is small compared with other uncertainties in our analysis.
}

\begin{table*}[htb!]
\renewcommand{\arraystretch}{1.25}
\begin{center}
\small{
\begin{tabular}{|c|c|c|} \hline
Observable & Source & Constraint \\
& Th./Ex.  & \\
\hline \hline
$\MW$ [GeV]
     &\cite{Svenetal}
     /\cite{lepewwg,gfitter2013} & {$80.379 \pm 0.012\pm0.010_{\rm {MSSM}}$ } \\
\hline
$ a_{\mu}^{\rm EXP} - a_{\mu}^{\rm SM}$
     &\cite{g-2}
     /\cite{newBNL}
     &$(30.2 \pm 8.8 \pm 2.0_{\rm {MSSM}})\times10^{-10}$ \\
\hline\hline
$  {R_{{\mu\mu}}}$
     & \cite{CMSLHCbBsmm,ATLASBsmm,1703.05747}   & { 2D likelihood}, { {MFV}} \\
\hline
$\tau (B_s \to \mu^+ \mu^-)$ & \cite{1703.05747} & $2.04 \pm 0.44 ({\rm stat.}) \pm 0.05 ({\rm syst.})$~ps \\
\hline
{$$ {BR}${_{{b \to s \gamma}}^{\rm EXP/SM}}$}
     &{\cite{Misiak}/\cite{HFAG}}
     &{${0.988 \pm 0.045_{\rm EXP} \pm 0.068_{\rm TH, SM} \pm 0.050_{\rm TH,SUSY}}$ } \\
\hline
{$$ {BR}${_{B \to \tau\nu}^{\rm EXP/SM}}$}
     & {\cite{HFAG,Kronenbitter:2015kls}}
     &  ${0.883 \pm 0.158_{\rm EXP} \pm 0.096_{\rm SM}}$ \\
\hline
{$ {BR}_{B \to X_s \ell \ell}^{\rm EXP/SM}$}
     & \cite{Xsllth}/\cite{HFAG}
     & ${0.966 \pm 0.278_{\rm EXP} \pm 0.037_{\rm SM}}$ \\
\hline
{$ {\Delta M}_{B_s}^{\rm EXP/SM}$}
     & {\cite{Buras:2000qz, SuFla} /\cite{HFAG}}
     & {${0.968 \pm  0.001_{\rm EXP} \pm 0.078_{\rm SM}}$ } \\
\hline
{$ {\frac{{\Delta M}_{B_s}^{\rm EXP/SM}}
           {{\Delta M}_{B_d}^{\rm EXP/SM}}}$}
     & \cite{Buras:2000qz,SuFla} /{\cite{HFAG}}
     & {${1.007 \pm 0.004_{\rm EXP} \pm 0.116_{\rm SM}}$ } \\
\hline
{$ {BR}_{K \to \mu \nu}^{\rm EXP/SM}$}
     & \cite{SuFla,Marciano:2004uf} /\cite{PDG2016}
     & ${1.0005  \pm 0.0017_{\rm EXP}  \pm 0.0093_{\rm TH}}$ \\
\hline
{$ {BR}_{K \to \pi \nu \bar{\nu}}^{\rm EXP/SM}$}
     & \cite{BurasKpinn15}/\cite{Artamonov:2008qb}
     & ${2.01 \pm 1.30_{\rm EXP} \pm 0.18_{\rm SM}}$ \\
\hline\hline
{ \ssi\ } & {\cite{lux16,XENON1T,pandax}} & {Combined likelihood in the $(\mneu1, \ssi)$ plane} \\
\hline
{\ssd\ }& {\cite{PICO}} & {Likelihood in the $(\mneu1, \ssd)$ plane} \\
\hline\hline
{${\tilde g} \to q \bar q \neu1, b \bar b \neu1, t \bar t \neu1$} & {\cite{cms_0lep-mt2, cms_1lep-MJ}} &
{Combined likelihood in the
${(\mgl, m_{\tilde\chi_1^0})}$ plane} \\ \hline
{$ {\sq \to q \neu1}$} & \cite{cms_0lep-mt2} &  {Likelihood in the
$(\msq, \mneu1)$  plane} \\ \hline
{${\tilde b} \to b \neu1$} & \cite{cms_0lep-mt2} & {Likelihood in the
${(m_{\tilde b}, m_{\tilde\chi_1^0})}$,  plane} \\ \hline
{${\tilde t_1} \to t \neu1, c \neu1, b \cha1$} & \cite{cms_0lep-mt2} & {Likelihood in the
${(m_{\tilde t_1}, m_{\tilde\chi_1^0})}$,  plane} \\ \hline
{$\cha1 \to \nu \ell^\pm \neu1, \nu \tau^\pm \neu1, W^\pm \neu1$} &
\cite{sus-16-039} & {Likelihood in the $(\mcha1, \mneu1)$ plane} \\
\hline
{$\neu2 \to \ell^+ \ell^- \neu1, \tau^+ \tau^- \neu1, Z \neu1$} &
\cite{sus-16-039} & {Likelihood in the $(\mneu2, \mneu1)$ plane} \\
\hline
$$ {{Heavy stable charged particles}} & {\cite{CMS:2016ybj}} & {Fast simulation based on} \cite{CMS:2016ybj, Khachatryan:2015lla} \\
\hline
${H/A \to \tau^+ \tau^-}$ & \cite{CMSHA,HBtautau,ATLAS:2016fpj,HA13} &
{Likelihood in the $(\MA, \tb)$ plane}  \\
\hline
\end{tabular}
\caption{\it Experimental constraints that we update in this work
  compared to Table~1 in~\protect\cite{mc-su5}.
We indicate separately the experimental and applicable theoretical errors in the SM and SUSY
 {(sometimes in combination, labelled ``MSSM'')}.
The contribution of the $\tau (B_s \to \mu^+ \mu^-)$ constraint to the
global $\chi^2$ likelihood function
is essentially constant across the relevant region of the pMSSM11
parameter space, and it is not included in the fit. {The new LHC constraints are all based on $\sim 36$/fb of data
at 13~TeV. 
}  }
\label{tab:updates}}
\end{center}
\end{table*}

As can be seen in Table~\ref{tab:updates}, we have also updated a number of flavour constraints.
In particular, we have updated the global analysis of \bsdmm\ {to include the latest Run~2 result from
LHCb~\cite{1703.05747} as well as the Run~1 results of CMS, LHCb~\cite{CMSLHCbBsmm} and ATLAS~\cite{ATLASBsmm}.
We assume minimal flavour violation (MFV) when combining the
\bdmm\ constraint with that from \bsmm\ into the quantity  $R_{\mu\mu}$~\cite{mc9},
{and take into account the correlation between the theoretical calculations of $f_{B_s}$ and $f_{B_d}$.}

The LHCb Collaboration has also published~\cite{1703.05747} a first determination of the effective $B_s$ lifetime
as measured in $B_s \to \mu^+ \mu^-$ decays, providing a constraint on the quantity $A_{\Delta\Gamma}$
via the relation
\begin{equation}
\frac{\tau (B_s \to \mu^+ \mu^-)}{\tau (B_s \to \mu^+ \mu^-)|_{\rm SM}} \; = \; \frac{1 + 2 A_{\Delta\Gamma} y_s + y_s^2}{(1 + y_s)(1+ A_{\Delta\Gamma} y_s)} \, ,
\label{ADG}
\end{equation}
where~\cite{HFAG}
\begin{eqnarray}
y_s & = & \tau_{B_s} \frac{ \Delta\Gamma_s}{2} \; = \; 0.0675 \pm 0.004 \, , \nonumber \\
A_{\Delta \Gamma} & \equiv & -2 \frac{{\cal R}e(\lambda)}{(1 + |\lambda|^2)} \, , \label{Bsequations} \\
\lambda & \equiv & \frac{q}{p} \frac{A(\bar{B_s} \to \mu^+ \mu^-)}{A({B_s} \to \mu^+ \mu^-)} \, , \nonumber \\ \nonumber
\end{eqnarray}
where {$\tau_{B_s}$ is the inclusive $B_s$ decay lifetime}, the complex numbers $p, q$ specify the relation between
the mass eigenstates of the $B_s^0 - \bar{B^0_s}$ system and the flavour eigenstates~\cite{HFAG},
and $A(B_s^0 \to \mu^+ \mu^-)$ and $A(\bar{B^0_s} \to \mu^+ \mu^-)$ are the $B_s^0$ and $\bar{B^0_s}$ decay amplitudes.}
In the Standard Model {(SM)},  $A_{\Delta\Gamma}= 1$ so that
$\tau(B_s \to \mu^+ \mu^-){|_{\rm SM}} = \tau_{B_s}/(1 - y_s) = 1.619 \pm 0.009$~ps.
On general grounds, $A_{\Delta\Gamma} \in [-1, 1]$. The LHCb measurement
$\tau(B_s \to \mu^+ \mu^-) = 2.04 \pm 0.44 ({\rm stat.}) \pm 0.05 ({\rm syst.})$~ps
corresponds formally to $A_{\Delta\Gamma} = 7.7 \pm 10.0$, implying that the
current LHCb result does not constrain significantly the pMSSM11 parameter space, and we do not include it in our fit.
However, in the later discussion of
our fit results we present for information
the $\chi^2$ profile likelihood functions we find for $A_{\Delta \Gamma}$ and $\tau (B_s \to \mu^+ \mu^-)$.

We have also updated our implementations of $b \to s \gamma$, $B \to \tau \nu$,
$B \to X_s \ell \ell$, ${\Delta M}_{B_s}$ and ${\Delta M}_{B_d}$ to take account of updated theoretical
calculations within the SM. For the same reason, in the kaon sector we have also updated
our implementations of $K \to \mu \nu$ and $K \to \pi \nu \bar{\nu}$~\footnote{{We refer to Table~1
of~\cite{mc-su5} for a complete set of the $K$-decay constraints we implement.}}.
{Since there are, in general, supersymmetric contributions to the observables commonly used
in global fits to CKM parameters, we remove these contributions and {make a global fit to the CKM parameters
without them}.}

In general, we treat the electroweak precision observables, \gmt\ and all $B$- and $K$-physics
observables (except for \bsdmm) as Gaussian constraints, combining in quadrature the experimental
and applicable SM and SUSY theory errors.


\subsection{Dark Matter Constraints and\\ Mechanisms}
\label{sec:DM}

\noindent
{\it Cosmological density} \\
Since we work in the framework of the MSSM, $R$-parity is conserved, so
that the lightest SUSY particle (LSP) is a candidate
to provide the CDM. We assume that the LSP is the lightest
neutralino $\neu1$~\cite{EHNOS}, and that it is the dominant component of the CDM.
As in our recent papers~\cite{mc-su5,mc-amsb}, we use the Planck 2015 constraint on the total
CDM density: $\Omega_{\rm CDM} h^2 = 0.1186 \pm 0.0020_{\rm EXP} \pm
0.0024_{\rm TH}$~\cite{Planck15}. \\

\noindent
{\it Density mechanisms} \\
As one of the primary objectives in our analysis is to investigate the relevances of various
mechanisms for bringing the relic $\neu1$ density into the range allowed by astrophysics and cosmology,
we introduce a set of measures related to particle masses that were found in our previous analyses~\cite{mc12}
to indicate when specific mechanisms were dominant~\footnote{{We have checked specifically the validity of these
measures using {\tt Micromegas}, finding good consistency in most cases. However, in certain hybrid
regions where more than one mechanism satisfied the criteria we found that just one mechanism dominates.
Moreover, it might also happen that some regions of the parameter space are not classified by a given measure
  even if the corresponding mechanism is active.}}.
These may be grouped as follows.\\

\noindent
$\bullet${\it Coannihilation with an Ino}

This may be important if the $\neu1$ is not much lighter than {the lighter chargino, $\cha1$, {and the second neutralino, $\neu2$},}
or the gluino, $\gl$.
For these cases we introduce the coannihilation measures
\begin{equation}
{\rm Ino} {\rm ~coann.~:} \qquad \left(\frac{M_{\rm Ino}}{\mneu{1}} - 1 \right)  \,<\,  0.25 \, .
\label{Inoco}
\end{equation}
{We find that chargino {and $\neu2$} coannihilation is important in our analysis,
and in our 2-dimensional plots we shade green the regions where (\ref{Inoco}) is satisfied
when the Ino is the lighter chargino, $\cha1$ {(which is almost degenerate with the $\neu2$)}. On the other hand, we find that gluino coannihilation is not
important in the pMSSM11 {when the \gmt\ constraint is imposed}.
This is due to the fact that \gmt\ forces the neutralino mass to values for which a gluino of equivalent mass would be excluded
  by current LHC results.
}\\

\noindent
$\bullet${\it Coannihilation with sleptons}

{In the version of the pMSSM that we study here, the two stau mass eigenvalues are similar,
since the soft SUSY-breaking parameters are specified at the TeV scale {and the left-right mixing $\propto m_\tau$
is relatively small, but the stau masses} are not
degenerate with the selectron and smuon masses, in general. We find that smuon and selectron
coannihilation are {in general more important than stau coannihilation, thanks to the
greater multiplicity of near-degenerate states}.
We introduce the following coannihilation measure:
\begin{equation}
{{\tilde \ell}} {\rm ~coann.~:} \qquad \left(\frac{m_{\tilde \ell}}{\mneu{1}} - 1 \right)  \,<\,  0.15 \, ,
\label{Ellco}
\end{equation}
and shade in yellow (pink) the regions of our two-dimensional plots where (\ref{Ellco}) is satisfied
for $\ell = \mu, e$ ($\tau$), respectively.}\\

\noindent
$\bullet${\it Coannihilation with squarks}

Similarly, this may be important for squarks $\tilde q$ that are not much heavier than the $\neu1$.
The case considered most often has been ${\tilde q} = {\tilde t}_1$, {but here we consider all
possibilities, including coannihilations with first- and second-generation squarks,
which we find to be important when {the LHC 13-TeV constraint or} \gmt\ is dropped.} We introduce the coannihilation measure
\begin{equation}
{\tilde q} {\rm ~coann.~:} \qquad \left(\frac{m_{\tilde q}}{\mneu{1}} - 1 \right)  \,<\,  0.15 \, ,
\label{Sqco}
\end{equation}
and we use the following colours in our plots for the regions where (\ref{Sqco}) is satisfied:
${\tilde q} = {\tilde d}/{\tilde s}/{\tilde u}/{\tilde c}_{L,R}$ cyan, ${\tilde t}_{1}$ grey, ${\tilde b}_1$ purple. \\

\noindent
$\bullet${\it Annihilation via a direct-channel boson pole}

When there is a massive boson $B$ with mass $M_B \sim 2 \mneu1$, $\neu1 \neu1$ annihilation is
enhanced along a `funnel' in parameter space. We have found that such a mechanism is likely to
dominate if the following condition is satisfied:
\begin{equation}
B {\rm ~funnel~:} \qquad
\left|\frac{M_B}{\mneu{1}} - 2 \right| \,<\,  0.1 \, .
\label{Bfunnel}
\end{equation}
{We have considered the cases $B = h, Z$ and $H/A$, and use blue shading for the regions of our
subsequent plots where (\ref{Bfunnel}) is satisfied when $B = H/A$. We comment later on a small region where
rapid annihilation via the $h$ and $Z$ poles is important}. \\

\noindent
$\bullet${\it Enhanced Higgsino component}

{We have also considered a somewhat different possibility, namely that the $\neu1$ has an enhanced
Higgsino component because the following condition is satisfied,
{which is similar to the situation in the focus-point region of the CMSSM}}:
\begin{equation}
{\rm Higgsino~:} \qquad \left| \left( \frac{\mu}{\mneu1} \right) - 1 \right| \,<\, 0.3 \, .
\label{Higgsino}
\end{equation}
Regions where
the condition (\ref{Higgsino}) is satisfied {generally satisfy the chargino coannihilation
condition with a Higgsino-like LSP, and are also shaded green}.\\

\noindent
$\bullet${\it Hybrid regions}

In addition to the `primary' regions where only one of the conditions (\ref{Inoco}, \ref{Ellco}, \ref{Sqco}, \ref{Bfunnel}, \ref{Higgsino})
is satisfied, there are also `hybrid' regions where more than one condition is satisfied. These are indicated in the following
{by mixtures of the corresponding primary colours}. \\

\noindent
{\it Direct DM searches} \\
{We implement experimental constraints from direct searches for supersymmetric
DM via both spin-independent and -dependent scattering on nuclei. We use the
LUX~\cite{lux16}, XENON1T~\cite{XENON1T} and PandaX-II~\cite{pandax} constraints on the spin-independent DM scattering
cross section \ssi, which we implement via a combined two-dimensional likelihood function in the $(\mneu1, \ssi)$ plane.

Our treatment of the spin-independent nuclear scattering matrix element follows that in our previous work~\cite{mc10}
{and is based on {\tt SSARD}~\cite{SSARD}.}
As reviewed, for example, in~\cite{EOSavage} the largest uncertainties in the matrix element are those associated
with the pion-nucleon $\sigma$-term, $\Sigma_{\pi N}$, and the SU(3) octet symmetry-breaking contribution to
the nucleon mass, $\sigma_0$. These may be expressed as follows in terms of ${\bar q} q$ matrix elements in the nucleon:
\begin{eqnarray}
\Sigma_{\pi N} & = & \frac{m_u + m_d}{2} \langle N | {\bar u} u + {\bar d} d | N \rangle \, , \nonumber \\
\sigma_0 & = & \frac{m_u + m_d}{2} \langle N | {\bar u} u + {\bar d} d - 2 {\bar s} s | N \rangle \, ,
\label{sigmas}
\end{eqnarray}
from which we see that the ${\bar s} s$ matrix element
\begin{equation}
y \; \equiv \; \frac{ 2 \langle N | {\bar s} s | N \rangle}{ \langle N | {\bar u} u + {\bar d} d | N \rangle} \; = \; 1 - \frac{\sigma_0}{\Sigma_{\pi N}} \, .
\end{equation}
It is well known that \ssi\ is sensitive to the value of $y$, and hence to the values of $\sigma_0$ and $\Sigma_{\pi N}$.
{We follow~\cite{Meissner1} in interpreting the measured octet baryon mass differences
as yielding $\sigma_0 = 36 \pm 7 \mev$}~\footnote{{However, we note that this estimate has been challenged~\cite{larger},
and flag this as an issue requiring resolution.}}, and we follow our
previous work in assuming here that $\Sigma_{\pi N} = 50 \pm 7 \mev$~\footnote{ {For a
recent estimate with a very similar central value of $\Sigma_{\pi N}$ made using covariant baryon chiral perturbation theory, see~\cite{NewSigmaTermEstimate}.}}, corresponding to a central value of $y = 0.28$.
For comparison, two recent determinations of $\Sigma_{\pi N}$ give somewhat larger values
that are, however, compatible with the value we assume, within the quoted uncertainties:
$\Sigma_{\pi N} = 59.1 \pm 3.5 \mev$ (from pionic atoms)~\cite{Meissner2} and
$58 \pm 5 \mev$ (from $\pi$-nucleon scattering)~\cite{Meissner3} (see also~\cite{Alarcon},
which found the value $\Sigma_{\pi N} = 59 \pm 7 \mev$). On the other hand, lattice calculations~\cite{Lattice} yield systematically
smaller values of $\Sigma_{\pi N}$ that are in tension with
these data-driven estimates, as discussed in~\cite{Meissner3}. Our value of
$\Sigma_{\pi N}$ is intermediate and relatively conservative in that it implies a smaller value of $y$ than
the data-driven estimates of $\Sigma_{\pi N}$~\footnote{{For comparison, a similar value of
$\Sigma_{\pi N} = 59 \pm 9 \mev$
is assumed in~\cite{GAMBIT}, but with $\sigma_s \equiv m_s \langle N | {\bar s} s | N \rangle = 43 \pm 8 \mev$ inferred from
lattice calculations. This corresponds to $\Sigma_{\pi N} - \sigma_0 = (m_u + m_d) \sigma_s/m_s \sim 3.5 \mev$,
implying a value of $\sigma_0$ different from the value we use, which is based {on} octet baryon masses.}}}.

We also implement in this paper the PICO~\cite{PICO} constraint
on the spin-dependent DM scattering cross section \ssd, also using the {\tt SSARD} code~\cite{SSARD}.
As discussed in~\cite{EFR}, the spin-dependent
$\neu1 p$ scattering matrix element is determined by the light quark contributions to the proton spin,
which we take to be~\cite{EOSavage}
\begin{eqnarray}
\qquad \Delta u & = & + 0.84 \pm 0.03 \, , \nonumber \\
\qquad \Delta d & = & - 0.43 \pm 0.03 \, , \nonumber \\
\qquad  \Delta s & = & - 0.09 \pm 0.03 \, ,
\label{Deltas}
\end{eqnarray}
{where the uncertainties are dominated by those in measurements of polarized deep-inelastic scattering,
and hence are {correlated: the uncertainty in the combination
$\Delta u - \Delta d$ (from $g_A$) is very small, and that in $\Delta u + \Delta d - 2 \Delta s$ (from semileptonic octet
baryon decays) is also somewhat} smaller~\footnote{{{The} values (\ref{Deltas}) of the $\Delta q$ that we use
are similar to those used in~\cite{GAMBIT}.}}.}\\

\noindent
{\it Indirect astrophysical searches for DM} \\
{These include searches for $\gamma$-rays from DM annihilations near the
Galactic centre and in dwarf galaxies, and for energetic neutrinos produced by the
annihilations of DM particles trapped inside the Sun. There are large astrophysical uncertainties in
estimates of the possible $\gamma$-ray flux from the Galactic centre, and other studies have
indicated that the available limits on the fluxes from dwarf galaxies do not yet impose
competitive constraints on supersymmetric models - see, for example,~\cite{EOSgamma}
and~\cite{GAMBIT}. The strongest constraints on energetic
solar neutrinos are those provided by the IceCube Collaboration~\cite{IceCube}. Their impact depends on
the annihilation final states, being strongest for annihilations into $\tau^+ \tau^-$, somewhat
weaker for $W^+ W^-$, and much weaker for ${\bar b} b$ final states.

The capture of dark
matter particles in the Sun is often assumed to be dominated by energy loss due to
spin-dependent scattering on protons, in which case an upper limit on the neutrino flux
may be used to constrain the spin-dependent cross-section \ssd, as done by the IceCube
Collaboration~\cite{IceCube}. However, the
interpretation of this constraint~\cite{IceCube} depends on the importance of spin-independent
scattering on $^4$He and heavier nuclei inside the Sun, and whether the DM density
inside the Sun is in equilibrium between capture and annihilation~\cite{EOSSun}. {As discussed in Section~\ref{sec:IDM}, we} have found in an
exploratory study that the IceCube constraint has little impact once the
more recent PICO constraint~\cite{PICO} on \ssd\ is taken into account.
{{In view of the fact that it has fewer uncertainties, we use the
PICO result in our global fit, setting aside the IceCube result~\cite{IceCube}~\footnote{{In contrast,
\cite{GAMBIT} uses the IceCube result, but not the PICO result.}}.}}


\subsection{13 TeV LHC Constraints}
\label{sec:LHC}

The LHC constraints we consider are those from searches for {coloured sparticles in events with}
missing transverse energy, $\ETslash$, {accompanied by jets and possibly leptons}, {searches for
electroweak inos in events with multiple leptons},
searches for long-lived charged particles, measurements of the 125 GeV Higgs boson $h$, and searches
for the heavier SUSY Higgs bosons $H, A, H^\pm$. {Our principal focus in this paper is on the implications
of Run-2 LHC searches with $\sim 36$/fb of data at 13 TeV, though we
also make comparisons with the situation before these constraints were released.}
Our implementations of the constraints from LHC Run~1 at energies of 7 and 8 TeV used in
our previous analysis of the pMSSM10 model were described in~\cite{mc11}, {and our}
implementations of $\ETslash$ searches with $\sim 13$/fb of data at 13 TeV in the gluino and squark
production channels were described in~\cite{mc-su5}, as were our implementations of searches for
long-lived charged particles and {for} $H, A, H^\pm$ with similar data sets.
We refer the reader to these
publications for details of those implementations, focusing here on our implementations of the
Run~2 searches with $\sim 36$/fb of data.\\

\noindent
{\it Searches for gluinos and squarks}

We consider the constraints from CMS simplified model searches
using events with $\ETslash$ and jets but no leptons released in~\cite{cms_0lep-mt2}
and events with $\ETslash$ and jets and a single lepton released in~\cite{cms_1lep-MJ}.

In the approach taken, e.g., by {\tt CheckMATE} \cite{Drees:2013wra},
{\tt ColliderBit} \cite{Balazs:2017moi} and {\tt MadAnalysis 5} \cite{Conte:2012fm},
Monte Carlo simulations are used to estimate the signal yield from a model point after the event selection and to test it by
comparing it with the upper bound given by an experimental collaboration.
However, such a method is time-consuming and computationally prohibitive for our purpose.
To circumvent this issue, we take the {\tt Fastlim} \cite{Papucci:2014rja}
approach~\footnote{{The {\tt SmodelS} code~\cite{Kraml:2013mwa, Ambrogi:2017neo}
takes a similar approach, as described in~\cite{Papucci:2014rja}.}}
and consider the implications
of~\cite{cms_0lep-mt2} for the following supersymmetric topologies: $\gl \gl \to [ q {\bar q} \neu1 ]^2$ and
$[ b {\bar b} \neu1 ]^2$, and ${\tilde q} {\tilde {\bar q}} \to [ q \neu1 ] [{\bar q} \neu1 ]$, and the implications
of~\cite{cms_1lep-MJ} for the topology ${\tilde g} {\tilde g} \rightarrow [ t {\bar t} \neu1 ]^2$.
The kinematics of
each of these topologies depends on a reduced subset of sparticle masses,
e.g., $(\mgl, \mneu1)$ in the case of the $\gl \gl \to [ q {\bar q} \neu1 ]^2$ topology,
and the CMS publications~\cite{cms_0lep-mt2,cms_1lep-MJ} provide
in {\tt Root} files 95\% CL upper limits $\sigma_{\rm UL}$ on the cross sections in the corresponding parameter
planes.
For each point
in the main pMSSM11 sample, we calculate for the $\gl \gl$ initial state and various final states
contributions to the global $\chi^2$ likelihood function of the form
\begin{equation}
\chi^2_{\gl \to {\rm SM} \neu1} = 5.99 \cdot \Big[
\frac{\sigma_{\gl \gl}\;{\rm BR}^2_{\gl \to {\rm SM} \neu1}}{{\sigma_{\rm UL}^{\gl \to {\rm SM} \neu1}}(\mgl, \mneu1)}
\Big]^2 \, ,
\label{eq:chi2modelg}
\end{equation}
where SM denotes the Standard Model particles considered in each topology, ${\rm SM} \equiv q {\bar q}, b {\bar b}$
and $t {\bar t}$, and analogously for the ${\tilde q} {\tilde {\bar q}} \to [ q \neu1 ] [{\bar q} \neu1 ]$ topology,
where ${\rm SM} \equiv q$ and ${\bar q}$.
We use {\tt NLL-fast} \cite{NLL-fast} to compute the cross sections for coloured sparticle pair-production
up to NLO+NLL level.

If gluino and squarks have comparable masses, associated gluino-squark production
may be sizeable.
In the $m_{\gl} \gtrsim m_{\tilde q}$ region, a fraction of the $gq \to \gl \tilde q$ process
where the gluino decays into $\bar q + \tilde q$
may be regarded as the production of a squark-antisquark pair with a soft quark jet.
Ignoring this soft jet, we can constrain this process by considering
the $q {\bar q} \to \tilde q {\tilde {\bar q}}$ simplified model limit.
In the analyses we consider, jets are treated inclusively and
this extra quark jet tends to slightly increase the acceptance.
Ignoring the soft jet therefore results in underestimation of the signal acceptance,
leading to a conservative limit.
In order to constrain the $g q \to \gl \tilde q \to \tilde q {\tilde {\bar q}} q$ process in the same way as
$q {\bar q} \to \tilde q {\tilde {\bar q}}$, we rescale the squark cross-section as
$\sigma_{\tilde q \tilde q} \to \sigma_{\tilde q \tilde q} + \sigma_{\tilde g \tilde q} \cdot {\rm BR}_{\tilde g \to q \tilde q}$ before applying squark simplified model limit.

Similarly, in the $m_{\tilde q} \gtrsim m_{\gl}$ region we rescale the gluino cross-section as
$\sigma_{\tilde g \tilde g} \to \sigma_{\tilde g \tilde g} + \sigma_{\tilde g \tilde q} \cdot {\rm BR}_{\tilde q \to q \tilde g}$ to constrain
the $g q \to \gl \tilde q \to \tilde g \tilde g q$ process using
the gluino simplified model limit.

\begin{table*}[htb!]
\renewcommand{\arraystretch}{1.25}
\begin{center}
\small{
\begin{tabular}{|l|c|c|} \hline
Topology & Analysis & Ref. \\
\hline
\hline
$\gl \gl \to [\, q {\bar q} \neu1 \,]^2, \, [\, b {\bar b} \neu1 \,]^2$ & 0 leptons + jets with $\ETslash$ & \cite{cms_0lep-mt2} \\
\hline
$\gl \gl \to [\, t {\bar t} \neu1 \,]^2$ & 1 lepton + jets with $\ETslash$ & \cite{cms_1lep-MJ} \\
\hline
${\tilde q} {\tilde {\bar q}} \to [\, q \neu1 \,] [\, {\bar q} \neu1 \,]$ & 0 leptons + jets with $\ETslash$ & \cite{cms_0lep-mt2} \\
\hline
${\tilde b} {\tilde {\bar b}} \to [\, b \neu1 \,] [\, {\bar b} \neu1 \,]$ & 0 leptons + jets with $\ETslash$ & \cite{cms_0lep-mt2} \\
\hline
${\tilde t_1} {\tilde {\bar t}}_1 \to [\, t \neu1 \,] [\, {\bar t} \neu1 \,]$, $[\, c \neu1 \,] [\, {\bar c} \neu1 \,]$ & 0 leptons + jets with $\ETslash$ & \cite{cms_0lep-mt2} \\
\hline
${\tilde t_1} {\tilde {\bar t}}_1 \to [\, \bar b {\chap1} \,] [\, \bar b {\cham1} \,] \to [\, \bar b W^+ \neu1 \,] [\, \bar b W^- \neu1 \,] $
 & 0 leptons + jets with $\ETslash$ & \cite{cms_0lep-mt2} \\
\hline
$\cha1 \neu2 \to [\, \nu \ell^\pm \neu1 \,] [\, \ell^+ \ell^- \neu1 \,] ~({\rm via}~ \tilde \ell^\pm)$
 & multileptons with $\ETslash$ & \cite{sus-16-039} \\
\hline
$\cha1 \neu2 \to [\, \nu \tau^\pm \neu1 \,] [\, \tau^+ \tau^- \neu1 \,] ~({\rm via}~ \tilde \tau^\pm)$
 & multileptons with $\ETslash$ & \cite{sus-16-039} \\
\hline
$\cha1 \neu2 \to [\, W^\pm \neu1 \,] [\, Z \neu1 \,]$
 & multileptons with $\ETslash$ & \cite{sus-16-039} \\
\hline
\end{tabular}
\caption{\it Summary of the simplified model limits from $\sim 36$/fb of CMS data at 13 TeV used in our study.}
\label{tab:smslim}}
\end{center}
\end{table*}

~\\
\noindent{\it Stop and sbottom searches}

{Our treatment of LHC 13 TeV limits on stops and sbottoms is similar in principle to our implementation of the
gluino and squark constraints described above. It is based on CMS simplified model searches
in the jets + 0 \cite{cms_0lep-mt2} and 1 \cite{cms_1lep-MJ} lepton final states,
where the results are interpreted as limits on the following topologies:
${\tilde t_1} {\tilde {\bar t}}_1 \to [ t \neu1 ] [{\bar t} \neu1 ]$, $[ c \neu1 ] [{\bar c} \neu1 ]$
in the compressed-spectrum region, $[ b W^{+} \neu1 ] [{\bar b} W^{-} \neu1 ]$ via $\cha1$
intermediate states and
${\tilde b_1} {\tilde {\bar b}}_1 \to [ b \neu1 ] [{\bar b} \neu1 ]$.
We also use {\tt Fastlim} to implement the CMS constraints in all these channels, following the
same procedure as described above for gluinos and squarks, and estimating the corresponding
contributions to the global $\chi^2$ likelihood function as
\begin{equation}
\chi^2_{\tilde q_3 \to {\rm SM} \neu1} = 5.99 \cdot \Big[
\frac{\sigma_{\tilde q_3 {\tilde {\bar q}}_3}\;{\rm BR}^2_{\tilde q_3 \to {\rm SM} \neu1}}{{\sigma_{\rm UL}^{\tilde q_3 \to {\rm SM} \neu1}}(\mst1, \mneu1)}
\Big]^2 \, ,
\label{eq:chi2modelt}
\end{equation}
where ${\rm SM} = t, c$ and $b W^{+}$ for $\tilde q_3 = \tilde t_1$ and
${\rm SM} = b$ for $\tilde q_3 = \tilde b_1$, respectively.

In a significant part of the pMSSM11 parameter space,
the neutralino relic abundance is brought into the observed range
by Wino or Higgsino coannihilation mechanisms.
In these regions, $\cha1$ and $\neu1$ are highly mass degenerate,
with a mass difference that is typically smaller than $5$ GeV.
Since the decay products of the $\cha1 \to \neu1$ transition
are too soft to affect the signal acceptance,
we can replace $\cha1$ by $\neu1$
in the simplified topology.
This approximation allows us to constrain the $\tilde t_1 \to b \chap1$ ($\tilde b_1 \to t \cham1$) topology
using the $\tilde b_1 \to b \neu1$ ($\tilde t_1 \to t \neu1$) simplified model limit.
Thus, in the Wino and Higgsino coannihilation regions,
we replace, e.g., the numerator in (\ref{eq:chi2modelt})
by $\sigma_{\tilde t_1 {\tilde {\bar t}}_1} {\rm BR}^2_{\tilde t_1 \to t \neu1}
\to \sigma_{\tilde t_1 {\tilde {\bar t}}_1} {\rm BR}^2_{\tilde t_1 \to t \neu1} +
\sigma_{\tilde b_1 {\tilde {\bar b}}_1} {\rm BR}^2_{\tilde b_1 \to t \cham1}$,
enhancing the sensitivity.\\
}

\noindent{\it Searches for electroweak inos}

{
The CMS Collaboration has also released results from searches for electroweak ino production at the LHC
in multilepton final states with $\sim 36$/fb of data at 13~TeV~\cite{sus-16-039}.
The signatures we have implemented are $\cha1 \neu2 \to [W \neu1] [Z \neu1], 3 \ell^\pm + 2 \neu1$
via ${\tilde \ell}^\pm/\tilde \nu$ intermediate states, and $3 \tau^\pm + 2 \neu1$
via ${\tilde \tau}^\pm$ intermediate states. As in the cases of searches for strongly-interacting
sparticles described above, we use {\tt Fastlim}
to compare the cross-section times branching ratio
with the 95\% CL upper limit released by CMS~\cite{sus-16-039}.
We obtain the corresponding contributions to the global $\chi^2$ likelihood function as
\begin{eqnarray}
& & \chi^2_{\cha1 \to {\rm SM} \neu1, \neu2 \to {\rm SM} \neu1} \simeq \nonumber \\
& & 5.99 \cdot \Big[ \frac{\sigma_{\cha1 \neu2} {\rm BR}_{\cha1 \to {\rm SM} \neu1} {\rm BR}_{\neu2 \to {\rm SM} \neu1}}{\sigma_{\rm UL}^{(\cha1 \to {\rm SM} \neu1)(\neu2 \to {\rm SM} \neu1) }}  \Big]^2 \, ,
\label{eq:chi2modelEWK}
\end{eqnarray}
where ${\rm SM} \equiv W$ or $Z$, one or two $\ell^\pm$ and one or two $\tau^\pm$, respectively.
One complication compared to the previous coloured sparticle cases is that $\sigma_{\cha1 \neu2}$
depends on many MSSM parameters:
\begin{eqnarray}
& & \sigma(pp \to \cha1 \neu2) \; = \; \nonumber \\
& &  F \left( M_1, M_2, \mu, \tan \beta, m_{\tilde q_L}, m_{\tilde u_R}, m_{\tilde d_R} \right) \, ,
\label{notabulation}
\end{eqnarray}
and it is not feasible to tabulate the cross section directly
in a multi-dimensional look-up table.
We have therefore used the code
{\tt EWK-fast}~\cite{EWKfast}, which is based on the observation that $\sigma(pp \to \cha1 \neu2)$ factorizes
mathematically {(where $\ino{i}$ and $\ino{j}$ represent any chargino and/or
  neutralino)}:
\begin{equation}
\sigma(pp \to {\tilde \chi_i} {\tilde \chi_j}) \; = \; \sum_a T_a ({\cal U}) F_a \left( m_{\tilde \chi_i}, m_{\tilde \chi_j}, m_a \right) \, ,
\label{EWKfast}
\end{equation}
where $T_a ({\cal U})$ is a function of the
mixing matrices ${\cal U} = \{U, V, N\}$ that can be calculated analytically.
The factor $F_a(m_{\tilde \chi_i}, m_{\tilde \chi_j}, m_a)$
captures the kinematics and the effect of the parton distribution function
and is tabulated in 3-dimensional look-up tables as a function of
$m_{\tilde \chi_i}, m_{\tilde \chi_j}$ and $m_a$,
where $m_a = m_{\tilde q_L}, m_{\tilde u_R}$
or $m_{\tilde d_R}$.

{The electroweak ino analyses described above can be extended to constrain models in
which electroweak inos can be produced in the decays of coloured sparticles.
This is because these searches do not impose conditions on the number of jets
and the final states in such events resemble those arising from the direct production of electroweak inos
associated with initial-state QCD radiation. In order to constrain this class of events
we include an extra contribution to the electroweak ino cross-section, much as
we discussed above in the case of the $\tilde q \tilde g$ constraint.
For example, in order to constrain
$\tilde q {\tilde {\bar q}} \to \tilde \chi_i \tilde \chi_j + {\rm jets}$,
we rescale the cross-section:
$\sigma_{{\tilde \chi_i} {\tilde \chi_j}}
\to \sigma_{{\tilde \chi_i} {\tilde \chi_j}} +
\sigma_{\tilde q {\tilde {\bar q}}} \, {\rm BR}_{\tilde q \to j \tilde{\chi}_i} \, {\rm
  BR}_{{\tilde {\bar q}} \to j \tilde{\chi}_j}$
before applying the electroweak ino simplified limit~\footnote{{We note here for completeness that
the LHC searches for sleptons~\cite{CMSWiki,ATLASWiki} do not constrain the pMSSM11 parameter space significantly.}}.}
}

\subsection{Combination of contributions to global \boldmath{$\chi^2$} function from LHC sparticle searches}

The total contribution of LHC Run-2 sparticle searches is obtained by adding the contributions from
the coloured sparticle (\ref{eq:chi2modelg}, \ref{eq:chi2modelt}) and electroweak
ino searches (\ref{eq:chi2modelEWK}):
\begin{equation}
\chi^2_{\rm LHC~Run~2} \; = \; \sum_i^{\rm Topologies} \chi^2_i \, ,
\label{totalLHCchi2}
\end{equation}
where the sum is over all the distinct SM final states mentioned above.
The simple sum is justified because event samples with different final states
are statistically independent, {so that their correlations are not important for our analysis.}
We summarise the simplified model limits we use in our scan in Table~\ref{tab:smslim}.

{\subsection{Measurements of the \boldmath{$h(125)$} Boson}}

{These are incorporated via the {\tt HiggsSignals} code~\cite{HiggsSignals}, which implements the information
from ATLAS and CMS measurements from LHC Run~1, as summarized in the joint ATLAS
and CMS publication~\cite{ATLAS+CMSH}.}

{\subsection{Searches for Heavy MSSM Higgs Bosons}}

{These are incorporated via the {\tt HiggsBounds} code~\cite{HiggsBounds}, which implements the information
from ATLAS and CMS measurements from LHC Run~1, supplemented by the constraint from
$\sim 36$/fb of data from the LHC at 13 TeV provided by ATLAS~\cite{HA13}.}

{\subsection{Searches for long-lived or stable charged particles}}

{The CMS Collaboration has published a search for charged particles with
  lifetimes $\gtrsim 3$ ~ns~\cite{Khachatryan:2015lla},
and a search for massive charged particles that leave the detector without decaying~\cite{LLsearches2}. We do not
include the results of these searches in our global likelihood analysis, but comment later on their
potential impacts. The only constraint that we impose on long-lived charged sparticles {\it a priori}  is to
require the lifetime to be {smaller than} $10^3$~s so as to avoid modifying the successful predictions of cosmological nucleosynthesis
calculations~\cite{BBN}.}\\


\section{Global Fit Results}

{The input parameter values for our best-fit points with and without \gmt\ are
shown in the second and fourth columns of Table~\ref{tab:points}, and
the spectra and dominant decays shown in Fig.~\ref{fig:spectra}. The third and fifth
columns show input values for other points of interest that we discuss below.
Lower rows of Table~\ref{tab:points} show the total $\chi^2$ per degree of freedom (d.o.f.) for each point,
dropping the contributions from {\tt HiggsSignals} that are shown in the last line. We also show the
corresponding p-values, as calculated using the prescription described in~\cite{mc-su5} to estimate the number
of degrees of freedom~\footnote{In previous studies (see, e.g., the first paper in~\cite{mcold})
we have validated our naive p-value approximation with
toy experiments, and found that it provides a reasonably accurate and conservative estimate of the underlying
p-value of the likelihood distribution. This was confirmed by a study in the last paper in~\cite{Fittino}, which
compared for different scenarios the naive p-value calculation with that obtained from toys.}.
We ignored the contribution to the likelihood coming from the nuisance parameters,
and we removed the contribution to the likelihood from {\tt HiggsSignals}, so as to avoid biasing our results by
giving too much importance to the Higgs signal rates. Since all the other constraints contribute significantly to $\chi^2$
function somewhere in the pMSSM11, we include them all in the d.o.f. count.
However, we merged into a single constraint the LHC direct searches for sparticle production at 8 and 13 TeV,
and also combined the 8- and 13-TeV limits on heavy Higgs bosons from $A/H \to \tau^+ \tau^-$
searches. This results in totals of 31 and 30 constraints for the cases with and without \gmt, respectively.
Since the number of free parameters is 11, this yields 20 and 19 for the numbers of d.o.f. in the two cases,
as stated in Table~\ref{tab:points}. We note that the p-values are all comfortably high, whether \gmt\ is
included, or not.}

\begin{table*}[htb!]
  \begin{center}
    \def\arraystretch{1.3}%
\begin{tabular}{|c||c|c||c|c|} \hline
Parameter   &  \multicolumn{2}{|c||}{With LHC 13 TeV and \gmt} & \multicolumn{2}{|c|}{With LHC 13 TeV, not \gmt} \\
            &      Best fit       & `Nose' region  &      Best fit       & `Nose' region \\
\hline
$M_1$           &  ~~~~~0.25 \tev~~~~  & -~0.39 \tev &~~~~-~1.3 \tev~~~~ & -~1.5 \tev \\
$M_2$           &  ~0.25 \tev          & 1.2 \tev  & 2.3 \tev          & 2.0 \tev \\
$M_3$           &  -~3.86 \tev         & -~1.7 \tev & ~1.9 \tev        & 1.0 \tev \\
\msq            &  4.0  \tev           & 2.00 \tev & 0.9 \tev          & 0.9 \tev \\
\msqt           &  1.7  \tev           & 4.1 \tev & 2.0 \tev           & 1.9 \tev \\
{\mslep}        &  0.35 \tev           & 0.36 \tev & 1.9 \tev          & 1.4 \tev \\
$m_{\tilde \tau}$ &  0.46 \tev           & 1.4 \tev &  1.3 \tev         & 1.4 \tev \\
\MA             &   4.0 \tev           & 4.2 \tev & 3.0 \tev           & 3.3 \tev \\
$A$             &  2.8  \tev           & 5.4 \tev & -~3.4 \tev         & -~3.4 \tev \\
$\mu$           &  1.33 \tev           & -~5.7 \tev & -~0.95 \tev      & -~0.93 \tev \\
\tb             &  36                  & 19 & 33                       & 33 \\
\hline \hline
$\chi^2$/d.o.f.       & 22.1/20 & 24.46/20 & 20.88/19 & 22.57/19 \\
p-value  & 0.33 & 0.22 & 0.34 & 0.25 \\
\hline
$\chi^2(HS)$   & 68.01 & 67.97 & 68.06 & 68.05 \\
\hline
\end{tabular}
\caption{\it {Values of the pMSSM11 input parameters and values of the global $\chi^2$ function at the best-fit points
including the LHC 13-TeV constraints, with and without the \gmt\ constraint, as well as at representative points
in the `nose' regions in the top left and right panels of Fig.~\protect\ref{fig:2dsquarkgluinoplanes}.}
{Lower rows show the total $\chi^2/$d.o.f. and the corresponding p-values for each point. As discussed in the text, we
calculate these omitting the contributions from {\tt HiggsSignals}, which are shown separately in the last line.}
The SLHA files for these points are available on our website, at the following URL \url{https://mastercode.web.cern.ch/mastercode/downloads.php}.}
\label{tab:points}
\end{center}
\end{table*}

\begin{figure*}
\begin{center}
\vspace{-2cm}
        \includegraphics[width=0.8\textwidth]{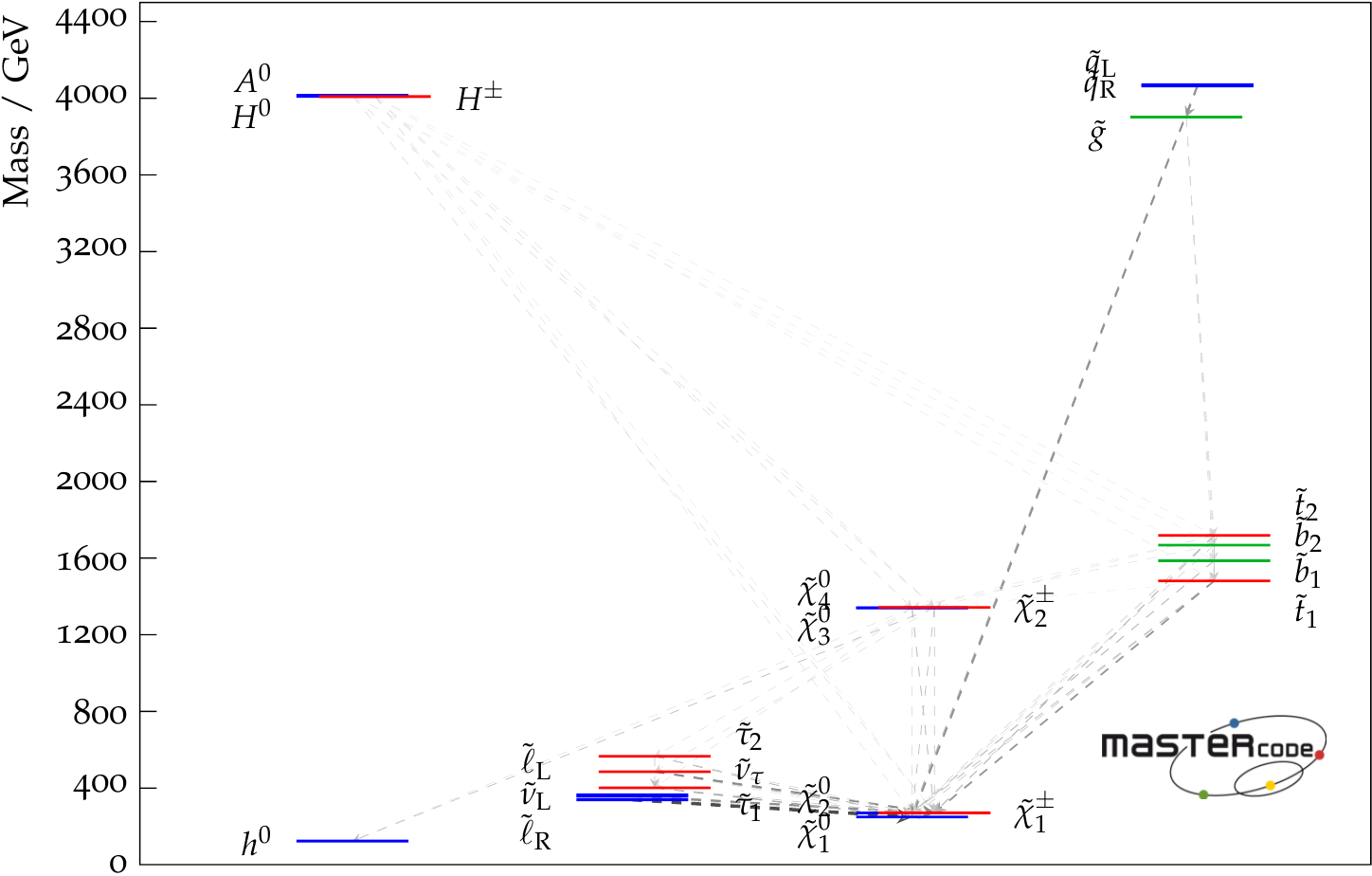} \\
        \vspace{0.5cm}
\includegraphics[width=0.8\textwidth]{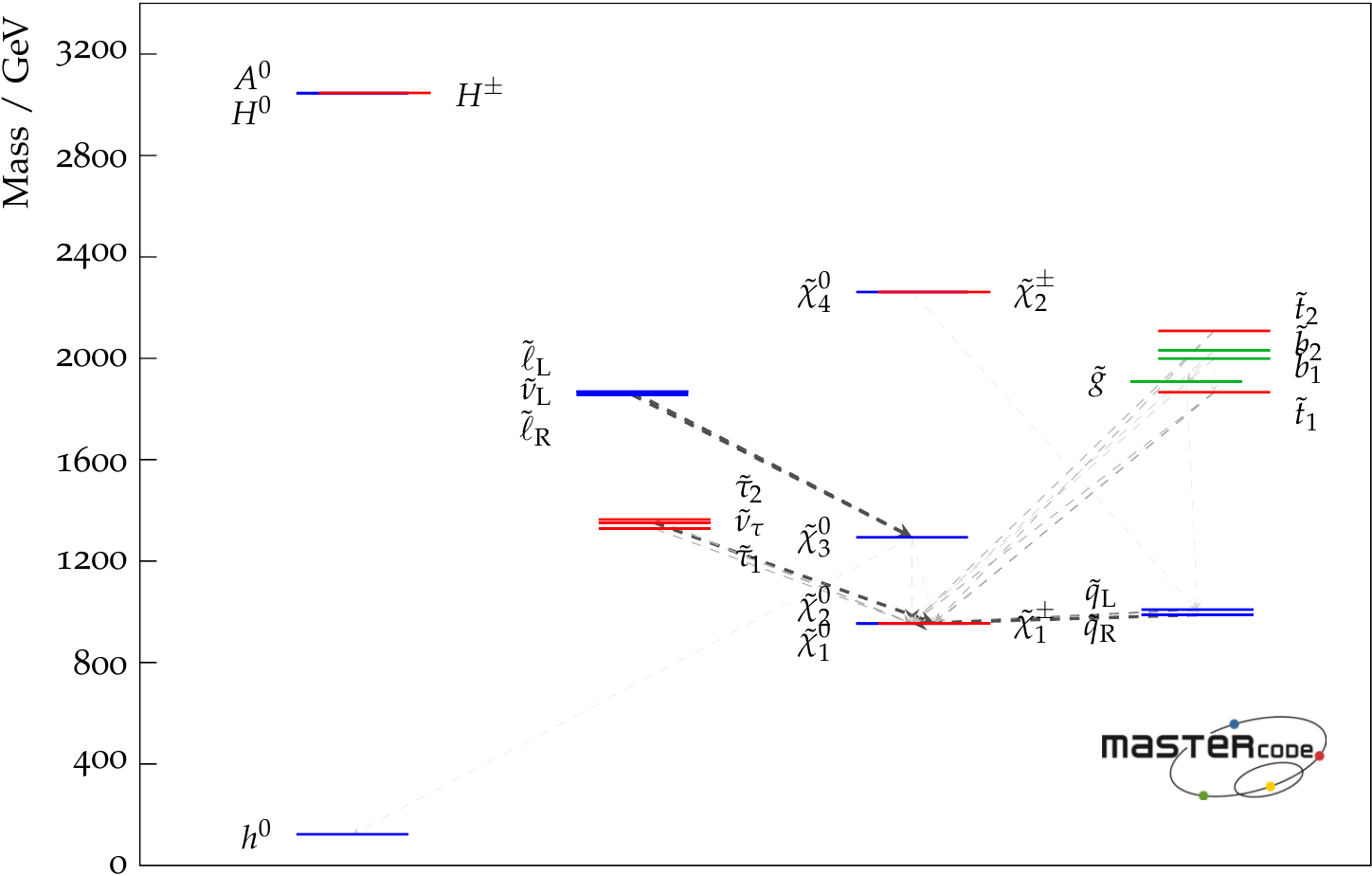} \\
\end{center}
\vspace{-0.5cm}
  \caption{\it Higgs and sparticle spectra for the best-fit points for the pMSSM11 with (top) and without the \gmt\
  constraint (bottom), showing also decay paths with branching ratios  $> 5\%$,
  the widths of the lines being proportional to the branching
  ratios. {These plots were prepared using the code presented in~\protect\cite{PySLHA}.}
}
  \label{fig:spectra}
\end{figure*}

\subsection{Parameter Planes}
\label{sec:planes}

{We now display results from our global fits with and without \gmt\ in pairs of 2-dimensional
pMSSM11 parameter planes. We indicate the locations of the best-fit points in
these two-dimensional projections by green stars,
We also show in these planes the $\Delta \chi^2 = 2.30, 5.99$ and $11.3$
contours, corresponding approximately to the boundaries of the regions preferred/allowed/possible at the
1-/2-/3-$\sigma$ levels (68\%, 95\% and 99.7\% CL), as red, blue and green solid lines, respectively.
Within the 2-$\sigma$ contours, we use colour coding to indicate the dominant DM mechanisms,
as discussed in \refse{sec:DM}, for the parameter sets that minimize $\chi^2$ at each
point in the plane.}\\

\noindent
{\it Squarks and gluinos}\\
The top row of plots in \reffi{fig:2dsquarkgluinoplanes} show $(\msq, \mgl)$ planes,
where $\msq$ {is an average over the masses of the left- and right-handed
first- and second-generation squarks, which are very similar in the pMSSM11~\footnote{{This and later
figures were prepared using {\tt Matplotlib}~\cite{Matplotlib}, except where otherwise noted.}}. In the top left panel,
where \gmt\ is included, we see 95\% CL lower bounds $\msq \gtrsim 2000 \gev$ and $\mgl \gtrsim 1400 \gev$,
with regions favoured at the 68\% CL appearing at slightly larger masses. We note that the best-fit point, denoted
by the green star, is at large $\msq > 4000 \gev$ and $\mgl \sim 3900 \gev$. {The full set of pMSSM parameter
values at this point, as well as the value of the global $\chi^2$ function, are listed in
the second column of Table~\ref{tab:points}. Important sparticle production cross-sections and decay modes at this
best-fit point are shown in the top panel of Table~\ref{tab:chains}.}

Within the 2-$\sigma$ contour, the dominant DM mechanism is
slepton coannihilation, {with stau coannihilation also playing a role for $\msq \sim 2.5 \tev$,
and} $\cha1$ coannihilation playing a role {at $\mgl \sim 1500 \gev$ and when $\mgl \gtrsim 2500 \gev$ and
$\msq \gtrsim 2800 \gev$.}
{Finally, we observe that
at the 3-$\sigma$ level much smaller values of $\msq$ are allowed, and that there is also {a peninsula at
  small $\mgl$ and larger $\msq$} that appears at the same level.}
These regions avoid the LHC exclusion searches in virtue of the same mechanisms which allow lower masses when the \gmt\ constraint is not applied
  and which will be described more in detail below. However, they are not able to satisfy the $(g-2)_{\mu}$ and this is why they take a $\Delta \chi^2 \simeq 11$
  penalty which makes them allowed only at 3-$\sigma$.

\begin{figure*}[htbp!]
\centering
\includegraphics[width=0.45\textwidth]{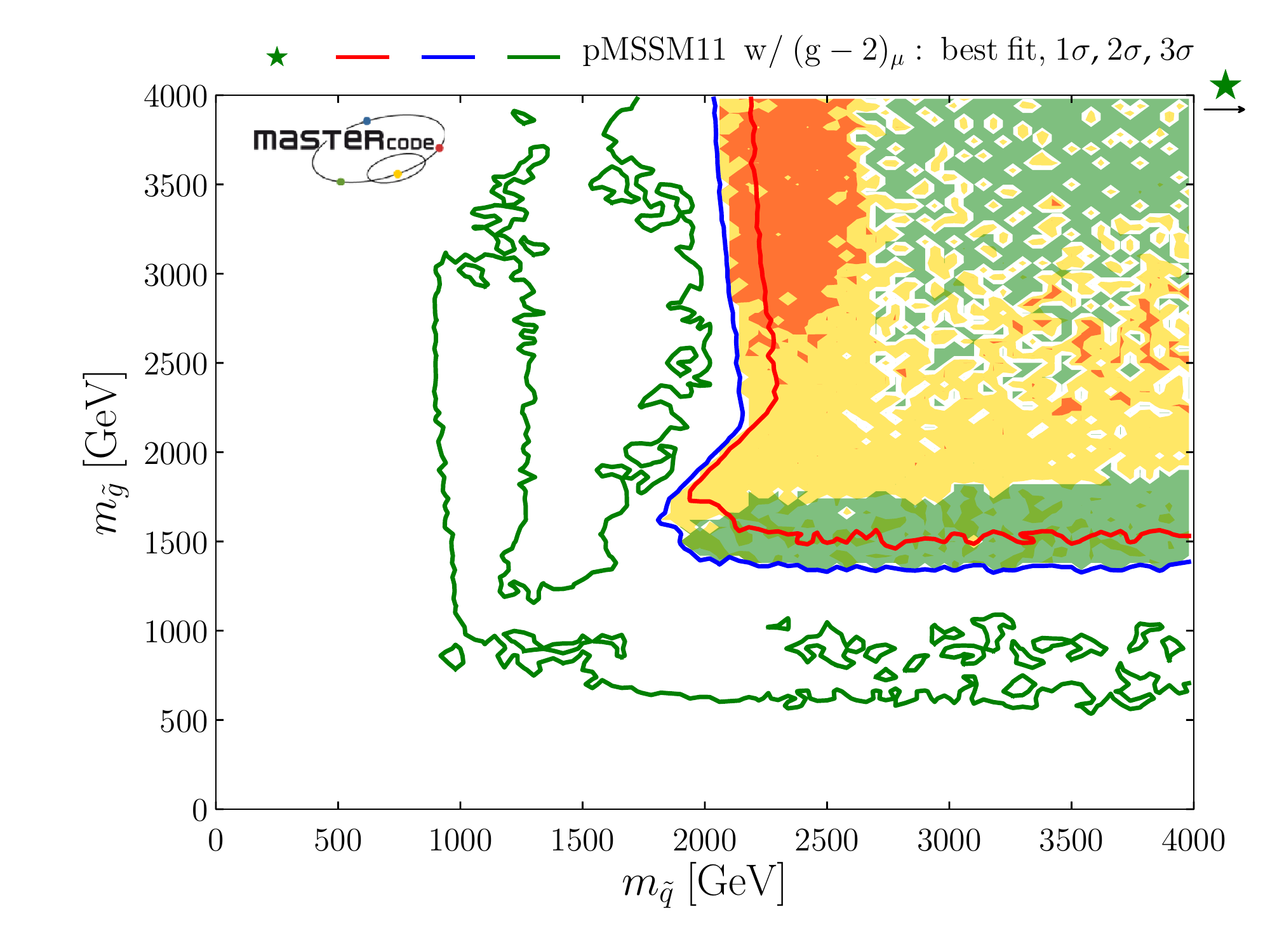}
\includegraphics[width=0.45\textwidth]{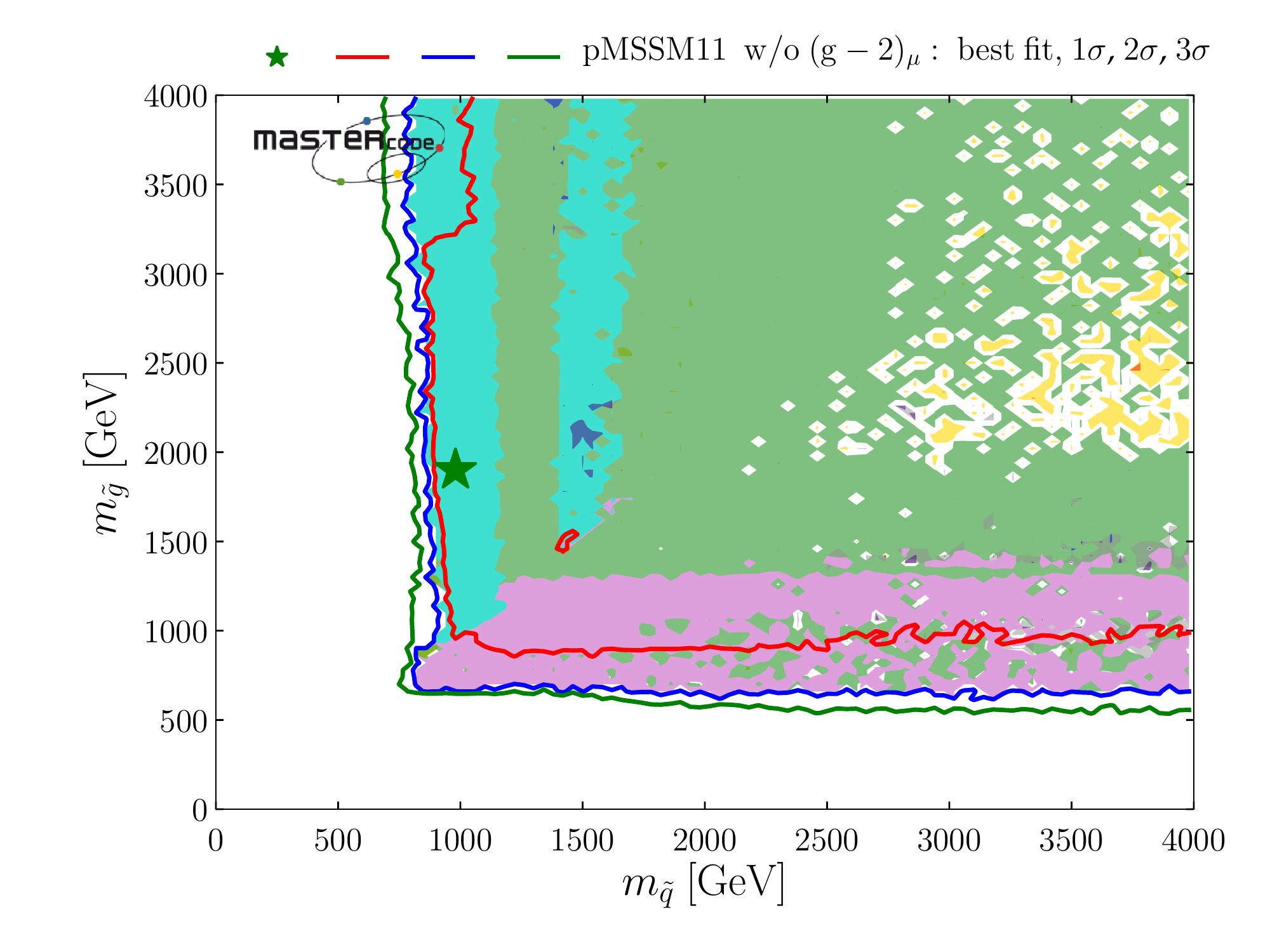}
\\[1em]
\centering
\includegraphics[width=0.45\textwidth]{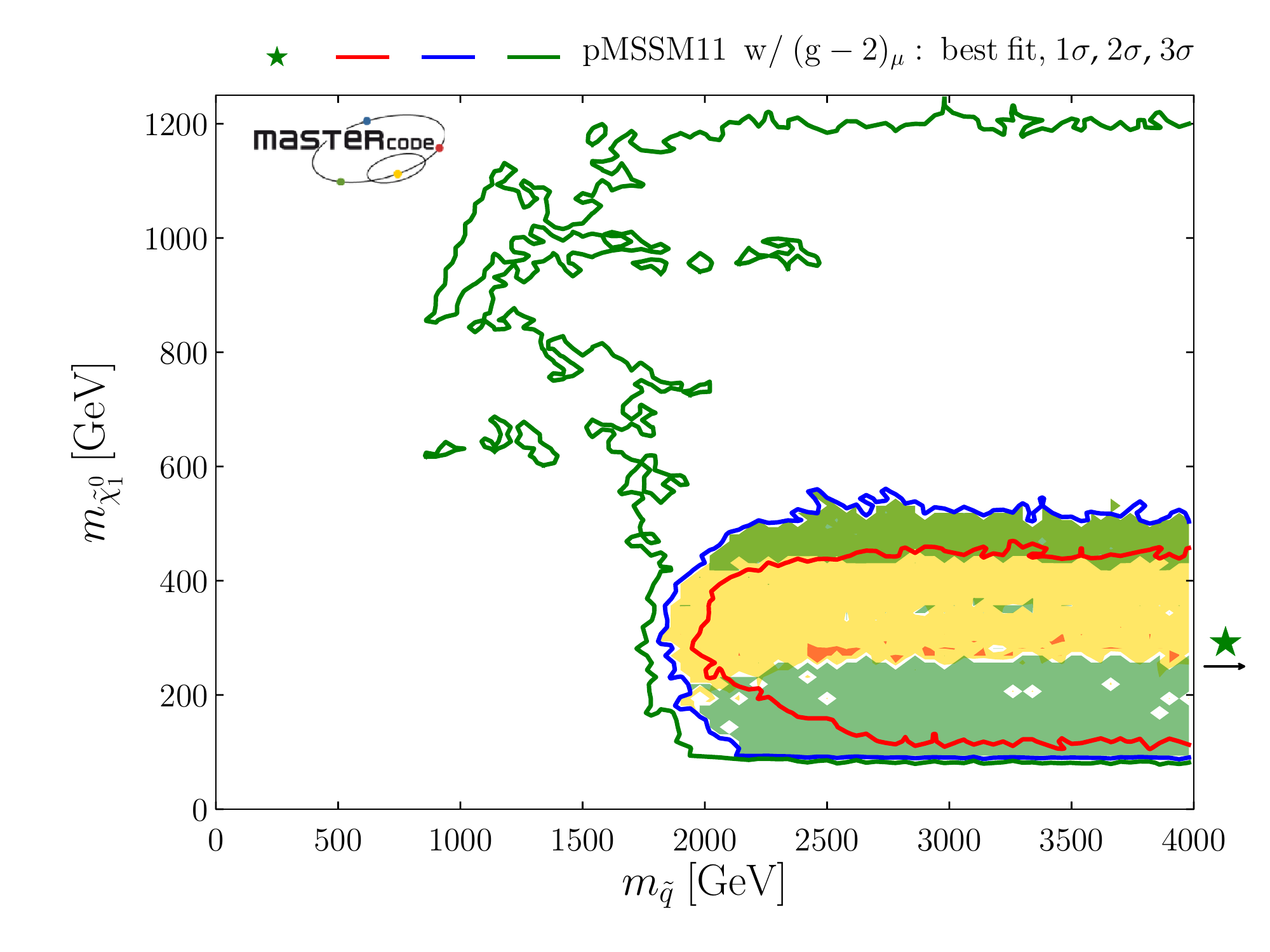}
\includegraphics[width=0.45\textwidth]{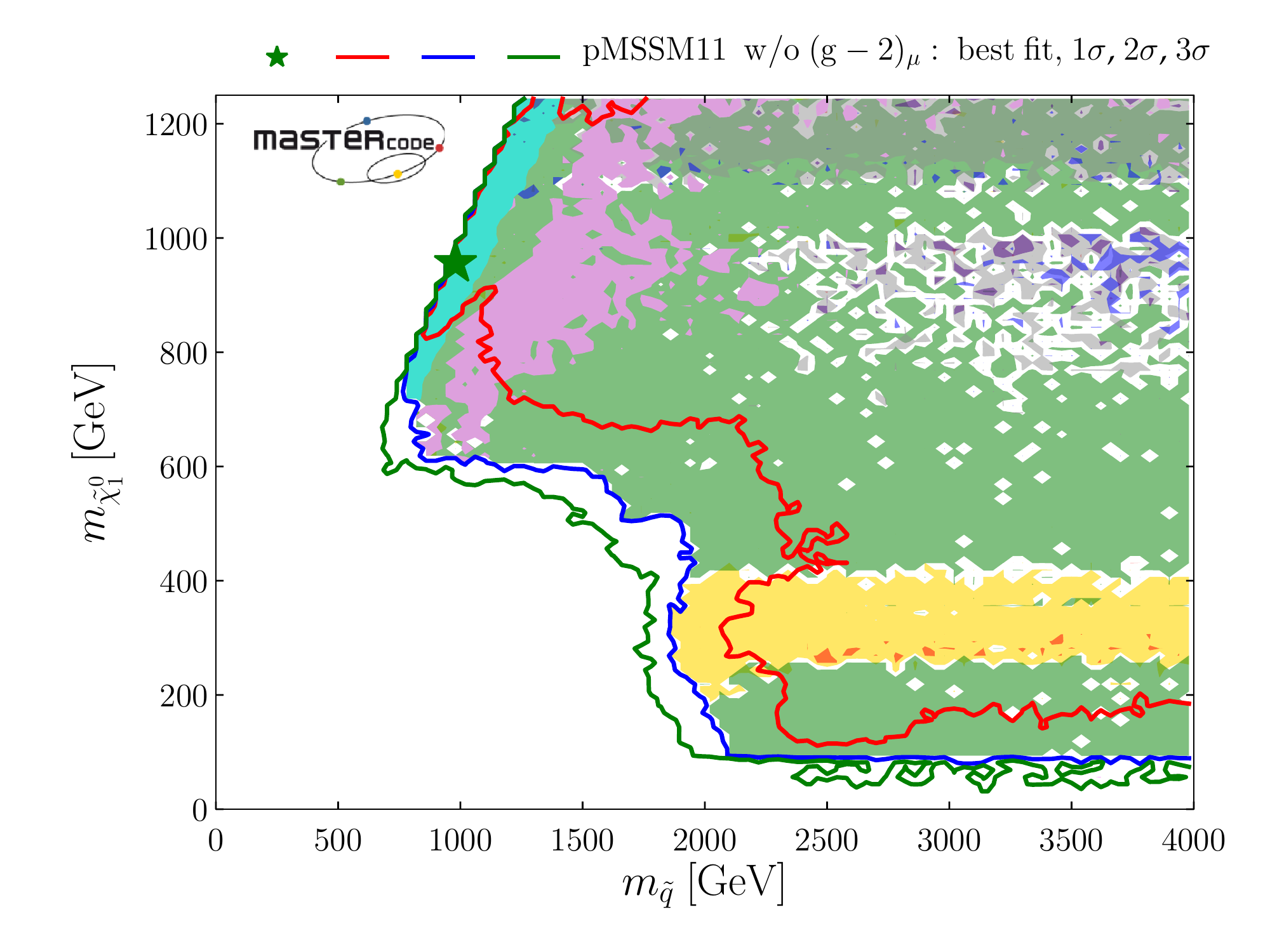} \\[1em]
\centering
\includegraphics[width=0.45\textwidth]{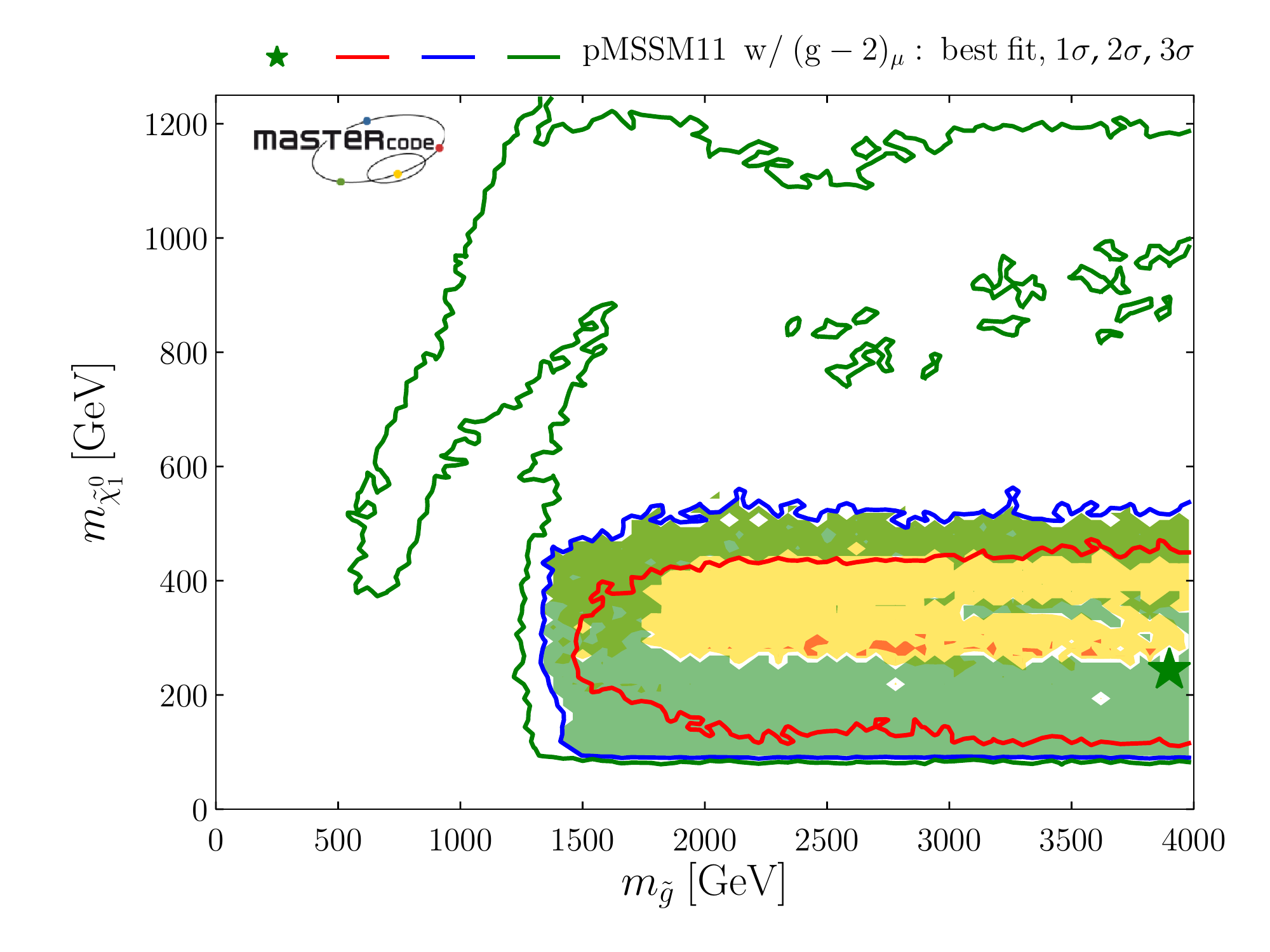}
\includegraphics[width=0.45\textwidth]{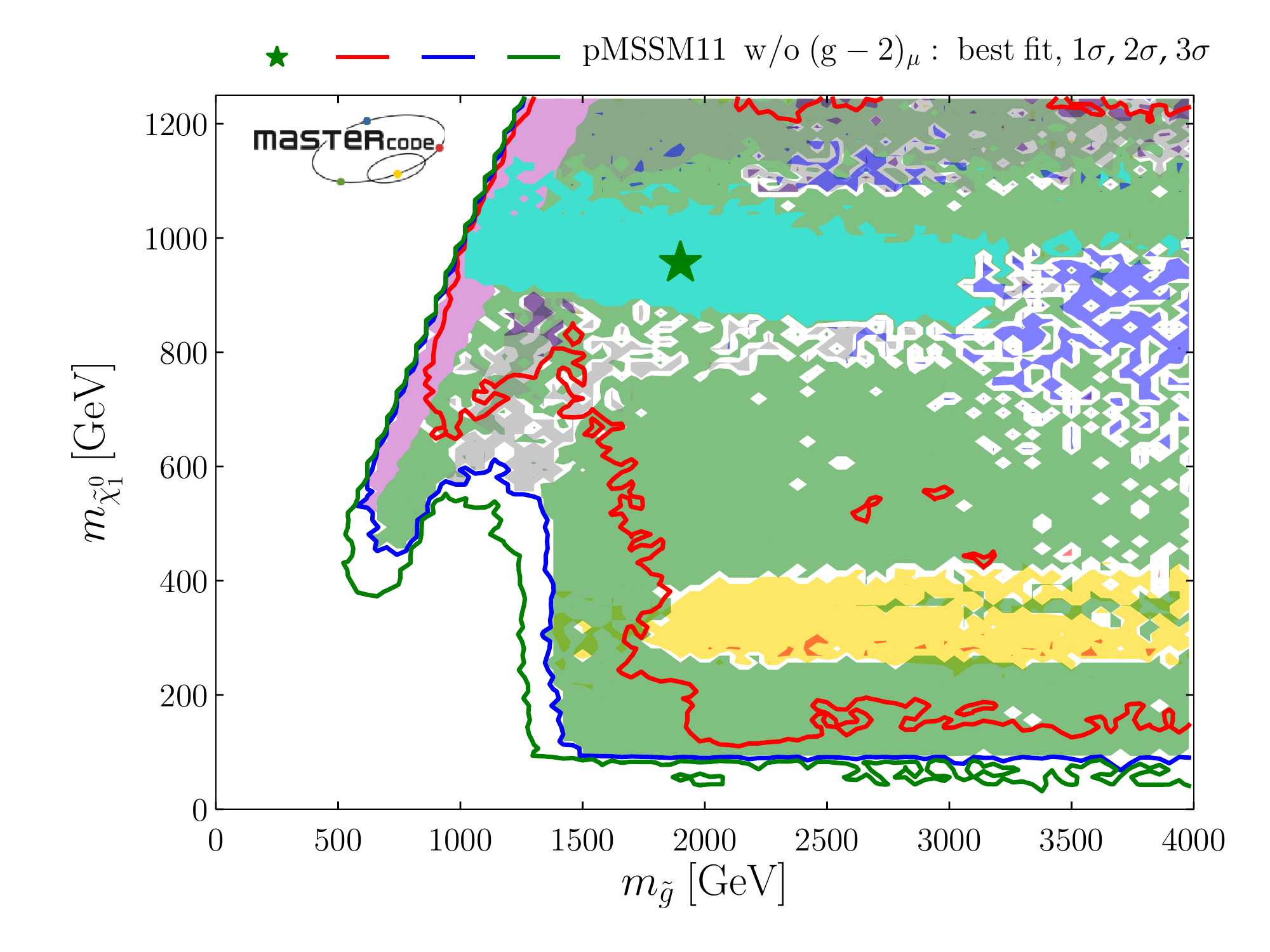} \\
\centering
\includegraphics[width=0.8\textwidth]{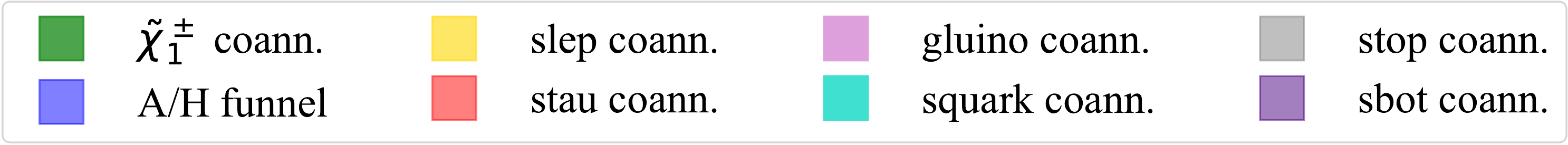}
\caption{\it Two-dimensional projections of the global likelihood function for the pMSSM11 in the $(\msq, \mgl)$ planes
(top panels), the $(\msq, \mneu1)$ planes (middle panels) and the $(\mst1, \mneu1)$ planes
(bottom panels), including the \gmt\ constraint (left panels) and dropping it (right panels).
}
\label{fig:2dsquarkgluinoplanes}
\end{figure*}

{We also note a `nose' feature corresponding to a reduction in the lower bounds when
$\msq \sim 2.2 \tev$ and $0 < \msq - \mgl \lesssim 200 \gev$. {We have verified that}
this is due to a loss of search sensitivity when {$\sqr \to \gl + q$}, the ${q}$ jet is soft, and $\gl \to q {\bar q} + {\tilde \chi}^*$,
where ${\tilde \chi}^*$ denotes any electroweak ino other than the LSP, {compared
to a high sensitivity for $\sqr \to q \tilde{\chi}^0_1$  in the $\mgl > \msq$ case.} {The input pMSSM11 parameter
values at a representative point in this `nose' region are listed in the third column of Table~\ref{tab:points}.
The upper panel of Fig.~\ref{fig:decaychains}
displays relevant sparticle masses and the most important sparticle decay chains at this point, and
numerical values are given in the second panel of Table~\ref{tab:chains}. We see
that the {right-handed squarks decay into a variety of final states involving heavier neutralinos and charginos
via intermediate gluinos due to $\mgl < \msq$,
reducing the effectiveness of {$\ETslash$-based} searches in this `nose' region, compared to simple
$\sq \to q + \neu1$ decays}.}}

\begin{figure*}[hp!]
\vspace{-1.2cm}
\begin{center}
\includegraphics[width=0.8\textwidth]{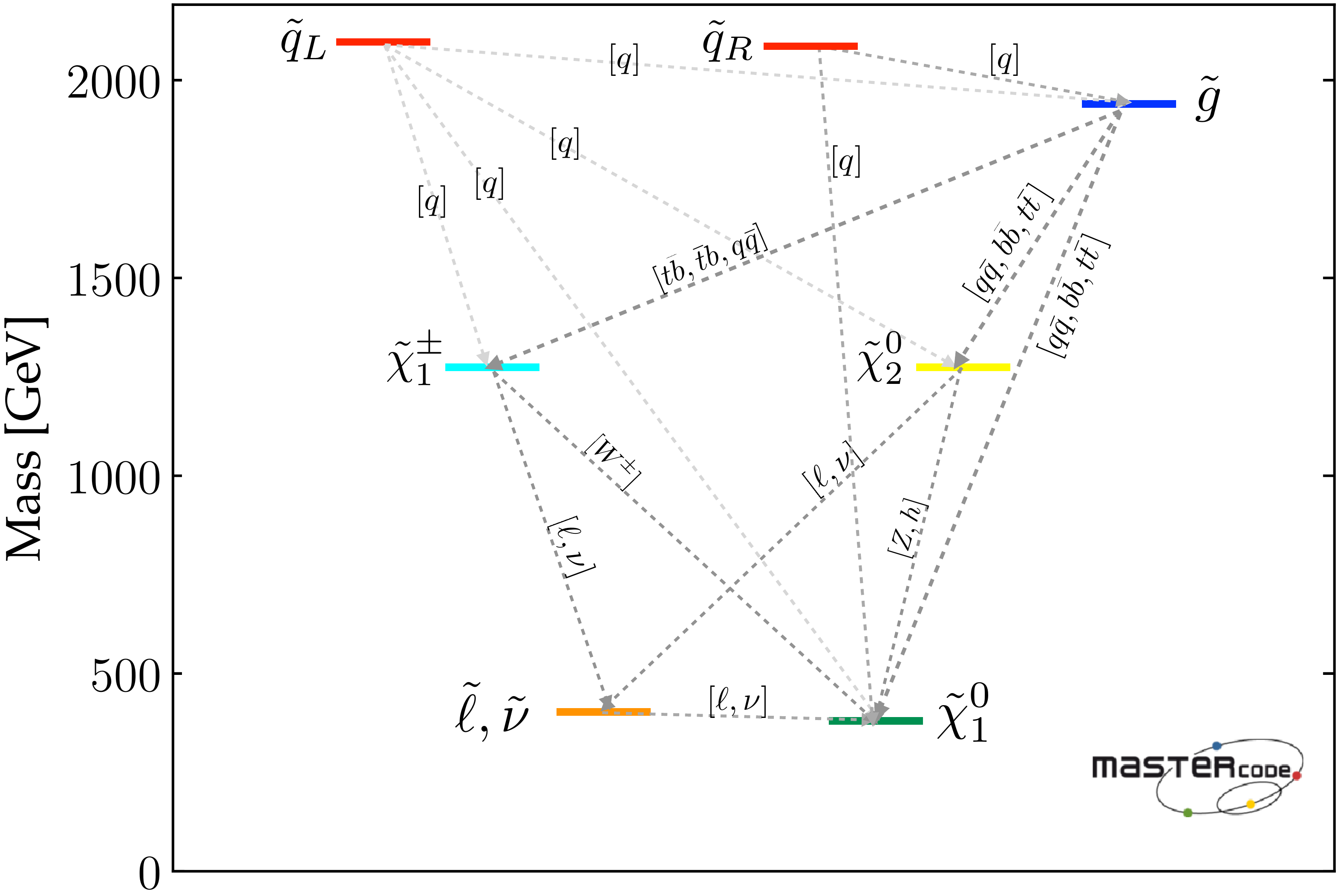} \\
\includegraphics[width=0.8\textwidth]{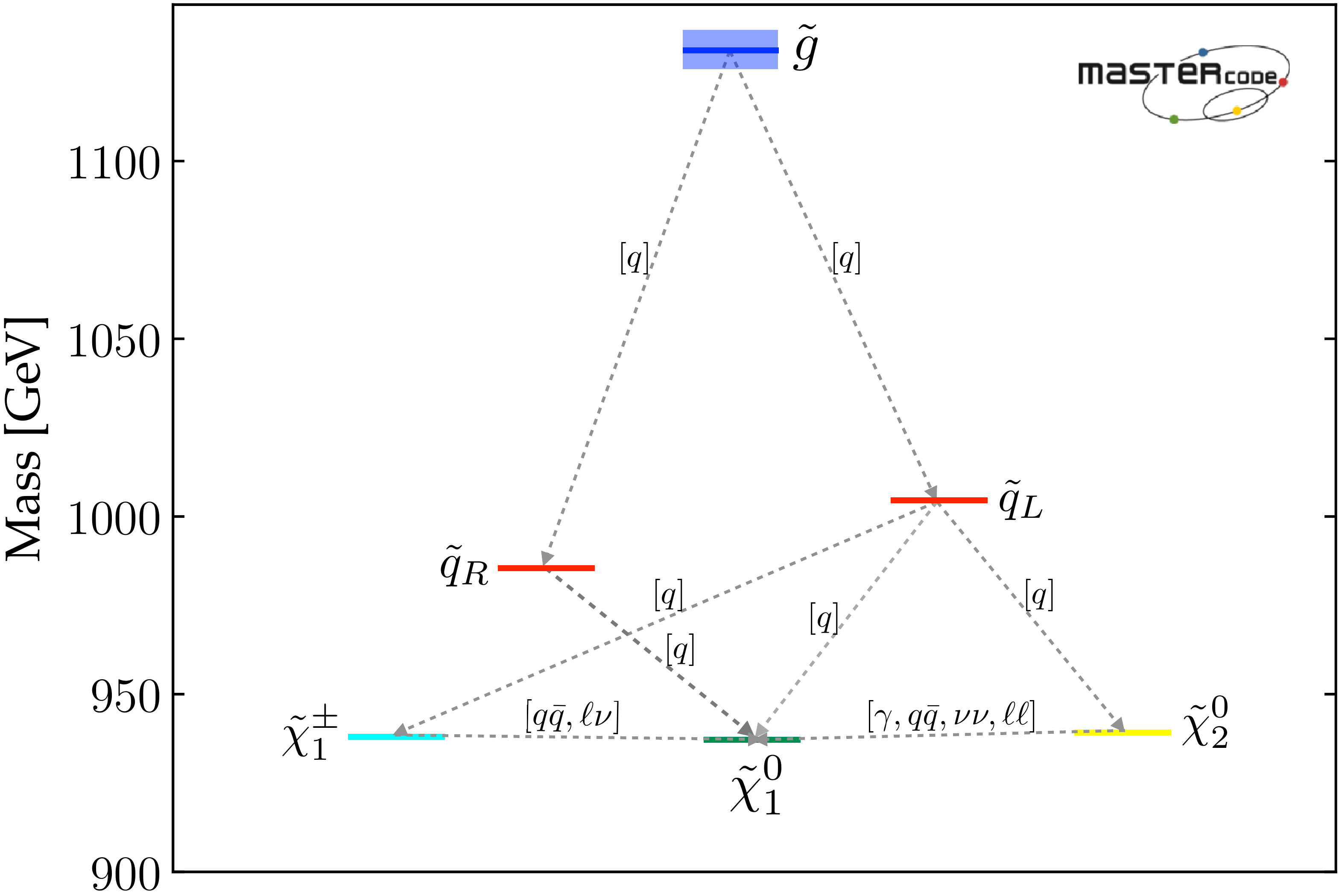} \\
\end{center}
\caption{\it {Upper panel: The dominant sparticle decay chains at the representative point in the `nose'
region in the top left panel of Fig.~\protect\ref{fig:2dsquarkgluinoplanes} (with \gmt) whose parameters are listed in the
second column of Table~\protect\ref{tab:points}. Lower panel: The dominant sparticle decay chains
at the representative point in the `nose' region in the top right panel of Fig.~\protect\ref{fig:2dsquarkgluinoplanes}
(without \gmt) whose parameters are listed in the fourth column of Table~\protect\ref{tab:points} - note that the
vertical scale has a suppressed zero. {In both plots the widths of the sparticles are represented as semi-transparent bands around the bar representing the nominal
mass value and of the same color.}}
}
\label{fig:decaychains}
\end{figure*}

\begin{table*}
    \def\arraystretch{1.12}%
\begin{center}
\vspace{-2cm}
{\it Dominant sparticle production and decay modes at best-fit point with \gmt}\\
\vspace{0.15cm}
\begin{tabular}{ |l|r| }
 \hline
 Production & $\sigma$ [fb] \\ \hline
$pp \to \tilde t_1 \tilde t_1$ + X & 0.25  \\
 $pp \to \tilde b_1 \tilde b_1$ + X & 0.13  \\
 \hline
Decays (mass [GeV])  & BR [\%] \\ \hline
~$\tilde t_1 (1481) \,\to\, b \tilde \chi_1^\pm (270)  \,/\, t \tilde \chi_2^0 (270) \,/\, t \tilde \chi_1^0 (249)$ & $56 \,/\, 25 \,/\, 19$  \\
~$\tilde b_1 (1586) \,\to\, t \tilde \chi_1^\pm (270)  \,/\, b \tilde \chi_2^0 (270) \,/\, b \tilde \chi_{3/4}^0 (270) \,/\, b \tilde \chi_1^0 (249) $ & $60 \,/\, 29 \,/\, 5 \,/\, 4$  \\
~\,$\tilde \chi_1^\pm (270) \,\to\, \ell^\pm \nu_\ell \tilde \chi_1^0 (249) \, / \, q q^\prime \tilde \chi_1^0(249)
\,/\, \tau^\pm \nu_\tau \tilde \chi_1^0 (249) $ & $52 \,/\, 38 \,/\, 1$
\\
~\,\,$\tilde \chi_2^0 (270) \,\to\, \nu \bar \nu \tilde \chi_1^0 (249)
\, / \,
\ell^\pm \ell^\mp \tilde \chi_1^0 (249)
\, / \,
\tau^\pm \tau^\mp \tilde \chi_1^0 (249)
$ & $53 \, / \, 37 \, / \, 1$
\\
\hline
\end{tabular}\\
\vspace{0.3cm}
{\it Dominant sparticle production and decay modes at `nose' point in fit with \gmt}\\
\vspace{0.15cm}
\begin{tabular}{ |l|r| }
 \hline
 Production & $\sigma$ [fb] \\ \hline
$pp \to \tilde q \tilde q$ + X & 3.4  \\
 $pp \to \tilde g \tilde q$ + X & 3.4  \\
 $pp \to \tilde g \tilde g$ + X & 0.5  \\
 \hline
Decays (mass [GeV])  & BR [\%] \\ \hline
~~\,$\tilde g (1942) \,\to\, q q \tilde \chi_1^0 (380)  \,/\, q q^\prime \tilde \chi_1^\pm (1273) \,/\, q q \tilde \chi_2^0 (1273)$ & $45 \,/\, 37 \,/\, 18$  \\
\,\,$\tilde q_L (2099) \,\to\, q \tilde \chi_1^\pm (1273)  \,/\, q \tilde g (1942)  \,/\, q \tilde \chi_2^0 (1273) \,/\, q \tilde \chi_1^0 (380)$ & $48 \,/\, 26 \,/\, 24 \,/\, 2$  \\
\,\,$\tilde q_R (2086) \,\to\, q \tilde g (1942)  \,/\, q \tilde \chi_1^0 (380)$ & $57 \,/\, 43$  \\
\,$\tilde \chi_1^\pm (1273) \,\to\, [\ell^\pm \tilde \nu_\ell (400) \to \ell^\pm \nu_\ell \tilde \chi_1^0(380)]  \,/\, [\nu_\ell \tilde \ell^\pm (404) \to \nu_\ell \ell^\pm \tilde \chi_1^0(380)]$ & $50 \,/\, 50$  \\
~$\tilde \chi_2^0 (1273) \,\to\, [\ell^\pm \tilde \ell^\mp (404) \to \ell^+ \ell^- \tilde \chi_1^0(380)]  \,/\, [\nu \tilde \nu_\ell (400) \to \nu_\ell \nu_\ell \tilde \chi_1^0(380)]$ & $50 \,/\, 50$  \\
\hline
\end{tabular}\\
\vspace{0.3cm}
{\it Dominant sparticle production and decay modes at best-fit point without \gmt}\\
\vspace{0.15cm}
\begin{tabular}{ |l|r| }
 \hline
 Production & $\sigma$ [fb] \\ \hline
$pp \to \tilde q \tilde q$ + X & 386  \\
 $pp \to \tilde g \tilde q$ + X & 51  \\
 $pp \to \tilde g \tilde g$ + X &  1  \\
 \hline
Decays (mass [GeV])  & BR [\%] \\ \hline
~\,\,$\tilde g (1908) \,\to\, q \tilde q_R (988)  \,/\, q \tilde q_L (1008)$ & $51 \,/\, 49$  \\
\,$\tilde q_L (1008) \,\to\, q \tilde \chi_1^\pm (955)  \,/\, q \tilde \chi_1^0 (954) \,/\, q \tilde \chi_2^0 (954) $ & $55 \,/\, 39 \,/\, 6$  \\
~~$\tilde q_R (988) \,\to\, q \tilde \chi_2^0 (954)  \,/\, q \tilde \chi_1^0 (954)$ & $98 \,/\, 2$  \\
\hline
\end{tabular}\\
\vspace{0.3cm}
{\it Dominant sparticle production and decay modes at `nose' point in fit without \gmt}\\
\vspace{0.15cm}
\begin{tabular}{ |l|r| }
 \hline
 Production & $\sigma$ [fb] \\ \hline
$pp \to \tilde q \tilde q$ + X & 619  \\
 $pp \to \tilde g \tilde q$ + X & 586  \\
 $pp \to \tilde g \tilde g$ + X & 87  \\
 \hline
Decays (mass [GeV])  & BR [\%] \\ \hline
~~$\tilde g (1131) \,\to\, q \tilde q_R (984)  \,/\, q \tilde q_L (1003)$ & $44 \,/\, 56$  \\
\,$\tilde q_L (1003) \,\to\, q \tilde \chi_1^\pm (939)  \,/\, q \tilde \chi_1^0 (937)  \,/\, q \tilde \chi_2^0 (938)$ & $58 \,/\, 38 \,/\, 4$  \\
~~$\tilde q_R (984) \,\to\, q \tilde \chi_2^0 (938)  \,/\, q \tilde \chi_1^0 (937)$ & $96 \,/\, 4$  \\
\hline
\end{tabular}
\end{center}
\vspace{-0.4cm}
\caption{\it {Dominant particle production and decay modes for various pMSSM11
parameter sets. Top panel: best-fit point with \gmt. Second panel: representative point in the `nose' region
in fit with \gmt. Third panel: best-fit point without \gmt. Bottom panel: representative point in the `nose' region
in fit without \gmt.}}
\label{tab:chains}
\end{table*}

We see significant differences in the top right panel where \gmt\ is dropped.
{The best-fit in this case is close to the 68\% CL boundary at
$(\msq, \mgl) \sim (1000, 1600) \gev$, with the parameters and $\chi^2$ value shown
in the fourth column of Table~\ref{tab:points}. As we discuss later, \bsdmm\ and the DM density
constraint play important roles in preferring a relatively low value of $\msq$.
The dominant particle production and decay modes for this
best-fit point are shown in the third panel of Table~\ref{tab:chains}. It is notable that the 95\% CL
lower limits on $\msq$ and $\mgl$ are reduced to $\sim 1000 \gev$, and {a less-pronounced} `nose' feature now appears
when $\msq \sim 1 \tev$ and $0 < \mgl - \msq \lesssim 200 \gev$. {Again, we have verified that this
reflects} a loss of search sensitivity when $\gl \to \sq + {\bar q}$,
the ${\bar q}$ jet is soft, and {$\sq \to q + \ino{}^* (\tilde{\chi}^0_1)$,
  where ${\tilde \chi}^0_1$ is much heavier than in the fit with \gmt\
  (for which a large SUSY contribution requires $\mneu1$ to be small),
  since the direct decay  $\tilde{g} \to q \bar{q} \ino{}^* (\tilde{\chi}^0_1)$
in the $\msq > \mgl$ case is more sensitive than the above cascade decay in the compressed spectrum.
}
{The lower panel of Fig.~\ref{fig:decaychains}
shows the most important sparticle decay chains at the representative point in this region whose
parameters are listed in the fourth column of Table~\ref{tab:points}, and the numerical values
of branching ratios are given in the bottom panel of Table~\ref{tab:chains}.} }

The differences {between the fits with and without the \gmt\ constraint are driven primarily by the fact that
{the fit with \gmt\ prefers small $\mneu1$, in which case the LHC 13-TeV searches require large $\msq$ and $\mgl$,
whereas the fit without \gmt\ favours} a
region with larger $\mneu1$. In this case, the loss of search efficiency due to a compressed spectrum allows
$\msq$ and $\mgl$ to be smaller than in the fit with \gmt. As we see later, in this compressed region the
LSP is mainly a neutral Higgsino, and coannihilations with a nearby charged Higgsino and the $\neu2$ are
important in determining the relic neutralino density.} {Coannihilations with first- and second-generation
squarks are also relevant here and in a band with $\msq \sim 1 \tev \lesssim \mgl$ (coloured cyan), whereas coannihilations
with gluinos are important along a band with $(1 \tev, 2 \tev) \ni \mgl \lesssim \msq$ (coloured magenta).}
{In this plane the 1-, 2- and 3-$\sigma$ contours lie relatively close to each other.}

{In the middle row of Fig~\ref{fig:2dsquarkgluinoplanes} we display the corresponding $(\msq, \mneu1)$ planes.
We see a preference for $\mneu1 \lesssim 550 \gev$ in the left panel, where the \gmt\ constraint is included,
whereas much larger values of $\mneu1$ are {allowed at the 3-$\sigma$ level.
These larger values of $\mneu1$ appear within the 1- and 2-$\sigma$ contours} in the middle right panel where the \gmt\ constraint
is dropped. We also see again that larger values of $\msq$ are favoured when \gmt\ is included, whereas a small
$\msq - \mneu1$ mass difference is preferred when the \gmt\ constraint is dropped.
{In both the middle panels the dominant DM mechanisms are slepton and $\cha1$ coannihilation,
with the rapid annihilation via the heavy $H/A$ Higgs bosons becoming important at large masses when \gmt\ is dropped.}
Similar features are seen in the $(\mgl, \mneu1)$ planes displayed in the bottom row of Fig~\ref{fig:2dsquarkgluinoplanes}.\\

\noindent
{\it Third-generation squarks}\\
Fig.~\ref{fig:2dstopsbottomplanes} displays the $(\mst1, \mneu1)$ planes in the upper panels
and the $(m_{\tilde b_1}, \mneu1)$ planes in the lower panels, again including the \gmt\ constraint
in the left panels and dropping it in the right panels. We see that both the third-generation squark masses
may be considerably smaller than those in the first two generations. Specifically, an isolated,
low stop-mass region where $(\mst1, \mneu1) \sim (500, 300) \gev$ is allowed at the 95\% CL~\footnote{For
relatively low stop masses, large values of $X_t/M_S \simeq A_t/M_S \simeq \sqrt{6}$ are required to avoid tension with the
Higgs mass measured at the LHC. Constraints from Charged and Color Breaking (CCB) minima can be relevant~\cite{CCB} in such a case,
but we have not taken these in account~\cite{vevacious} in our analysis.
This is because our best-fit point region is characterized by relatively small values of $A_t/M_S$, as it can be seen from Table~\ref{tab:points}, for which this issue is not relevant.} in
both the cases with and without \gmt, {which is connected in the latter case to the rest of the 95\% CL region
  at the 3-$\sigma$ level.
The low stop-mass island is allowed and defined by different physics mechanisms.
First, the third-generation-squark spectra are sufficiently compressed to allow the points to
bypass the LHC13 constraints. Moreover, it is characterized by
compressed-slepton spectra as well, which explains the fact that the region is shaded in
yellow in the plots. We also note that it can not be extended to lower stop masses
because otherwise it would be disallowed by sbottom searches, since in
our scenario the masses of the stop and sbottom squarks are defined by a
single soft SUSY-breaking mass term and the sbottoms would not be
sufficiently compressed to be allowed by LHC searches.
LHC constraints also limit its extensions in the direction of lower neutralino (too light third-generation squarks)
or higher stop masses (due to the loss of compression).
Finally, at heavier neutralino masses slepton coannihilation is insufficient to reduce the relic density into the allowed range.4
When \gmt\ is dropped, extended 95\% CL regions with
$\mneu1 \gtrsim 500 \gev$ appear when} $\mst1 \gtrsim 1100 \gev$ and
$m_{\tilde b_1} \gtrsim 1250 \gev$.} {When \gmt\ is included, there are extended regions with $\mneu1 \gtrsim 500 \gev$
that appear at the 3-$\sigma$ level. Within the 1- and 2-$\sigma$ contours, the dominant DM mechanisms
are slepton and $\cha1$ coannihilation, with rapid annihilation via the heavy $H/A$ Higgs bosons
again becoming important at large $\mneu1$ when \gmt\ is dropped. The same mechanism is also active
  inside the white regions between $ 800 \gev~(1 \tev) \lesssim \mst1~(\msb1)  \lesssim 1.1~(1.2) \tev$ and $ 400 \gev \lesssim \mneu1  \lesssim 600 \gev$,
  the blue shading being absent due to the proxy-measure being not sufficiently descriptive in this parameter space region.}
Stop and sbottom coannihilation are also important for small $\mst1 - \mneu1$ and $\msb1 - \mneu1$.}\\

\begin{figure*}[htb!]
\centering
\includegraphics[width=0.45\textwidth]{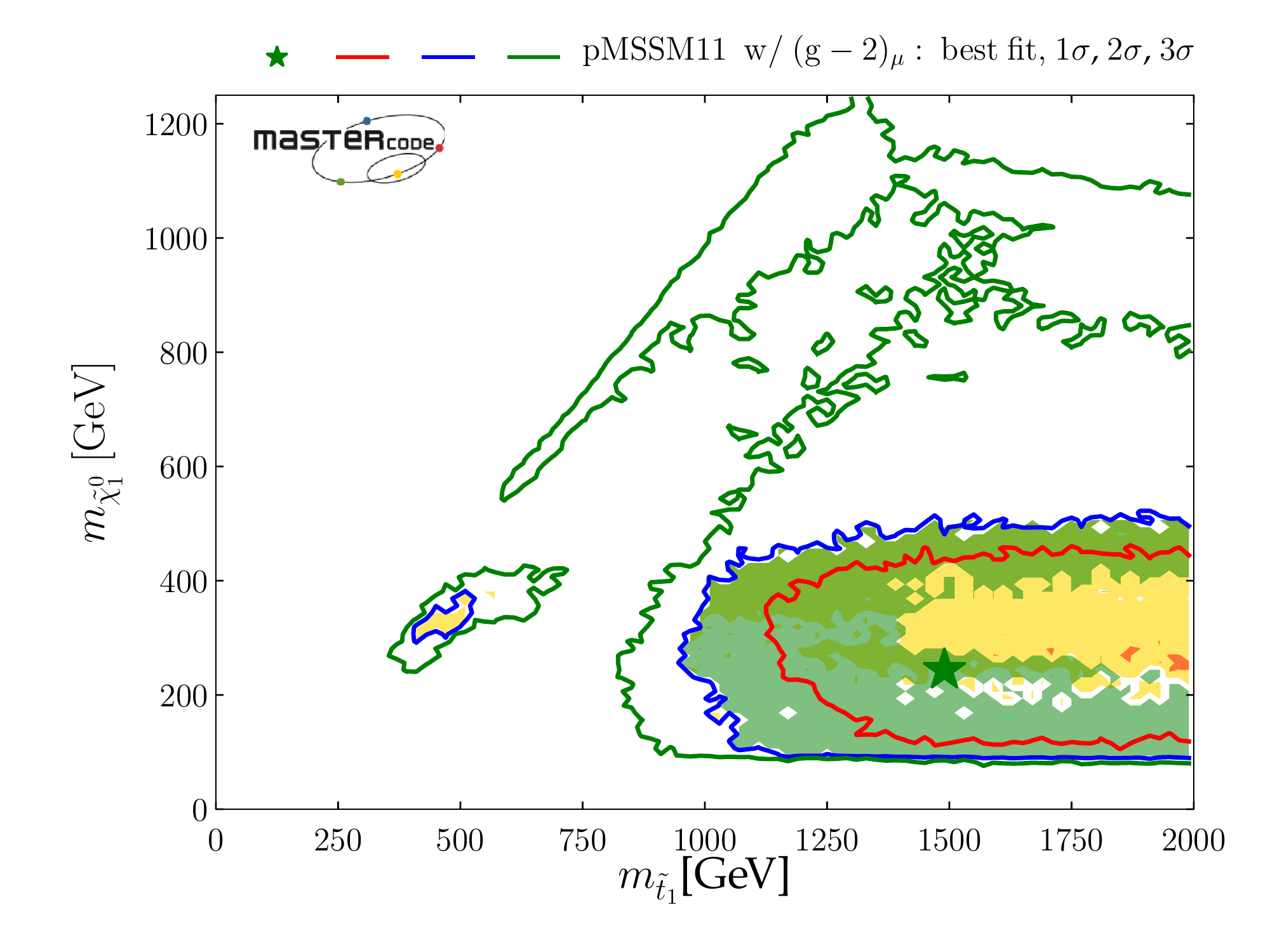}
\includegraphics[width=0.45\textwidth]{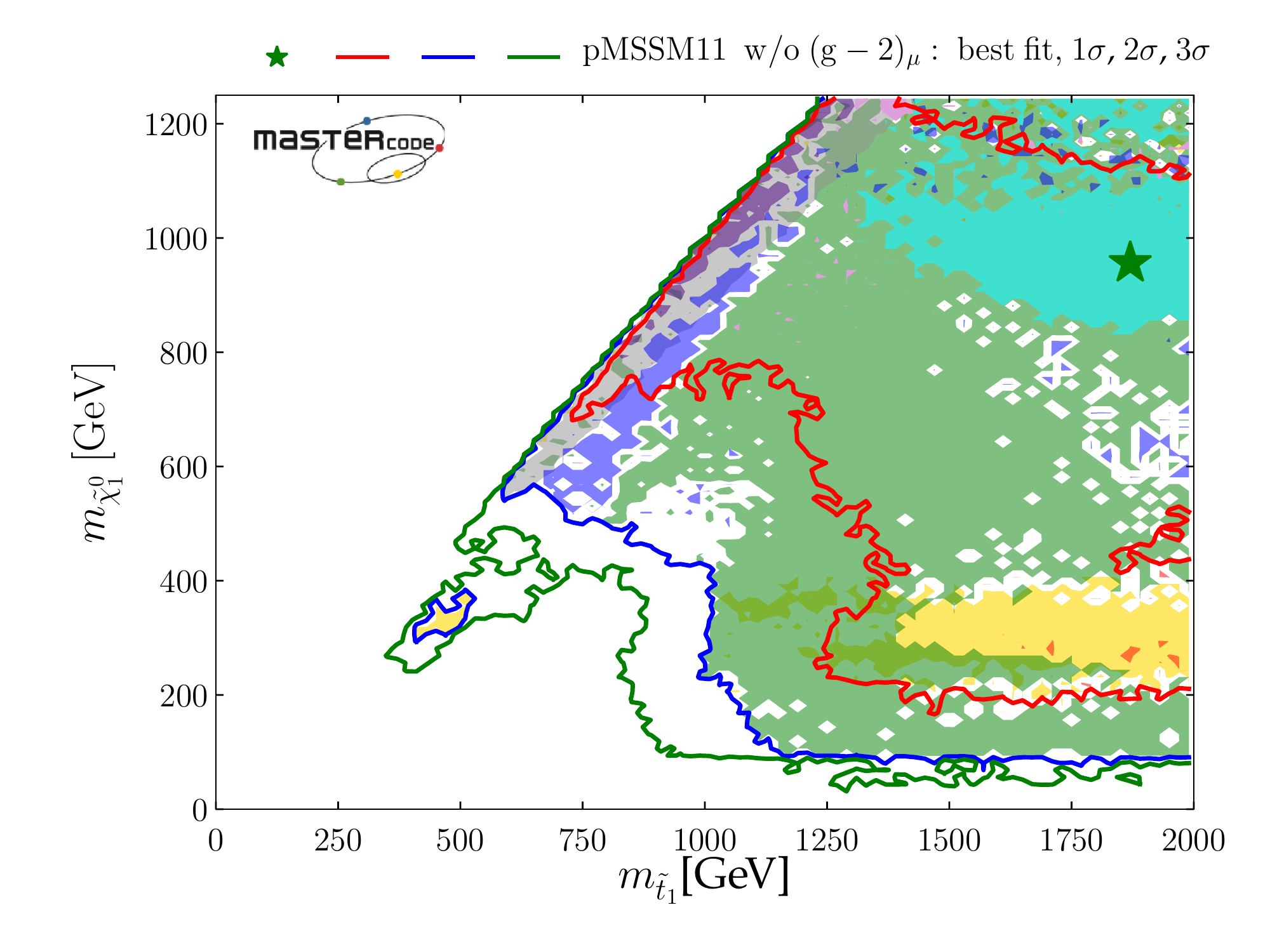} \\
\centering
\includegraphics[width=0.45\textwidth]{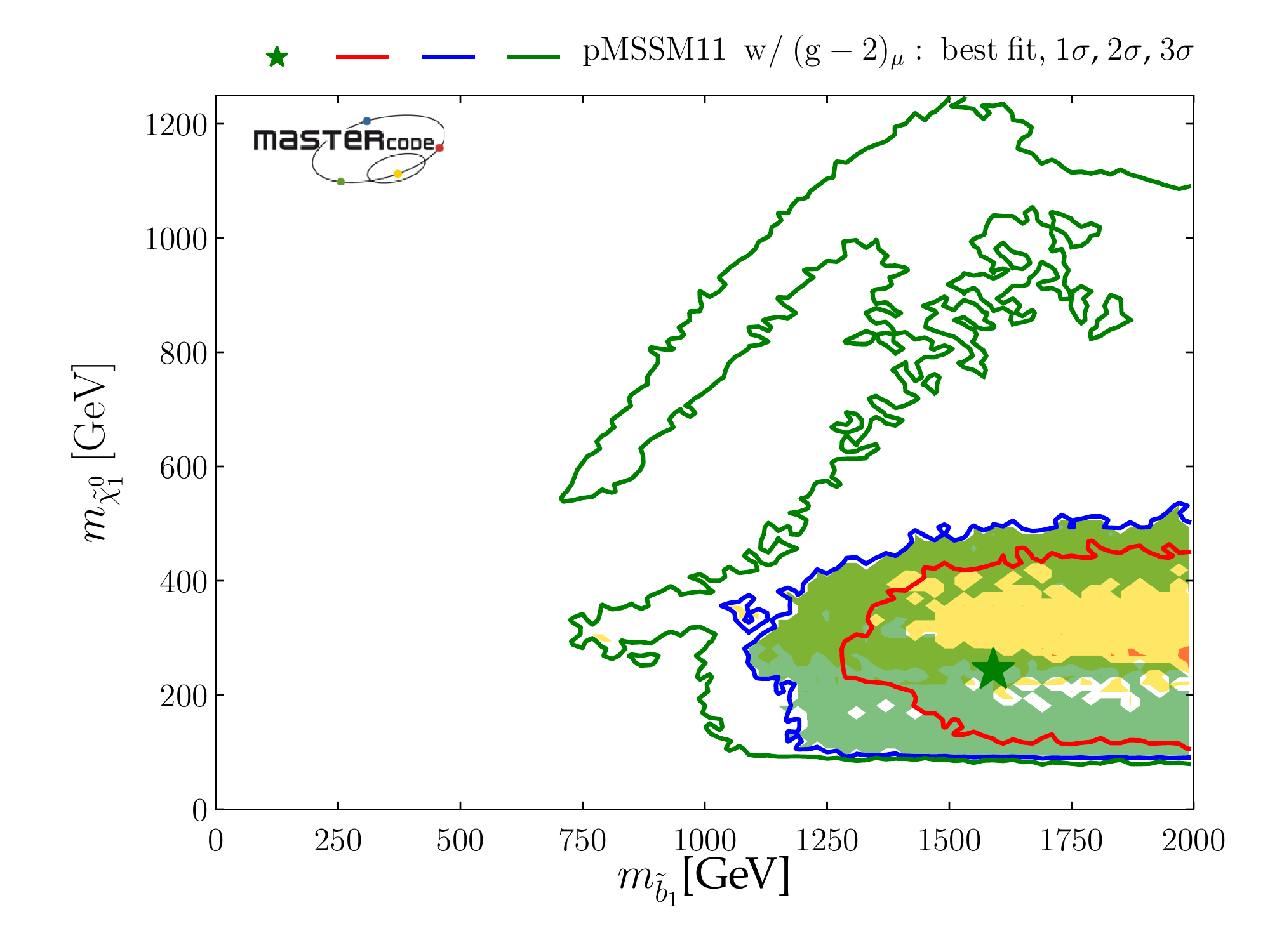}
\includegraphics[width=0.45\textwidth]{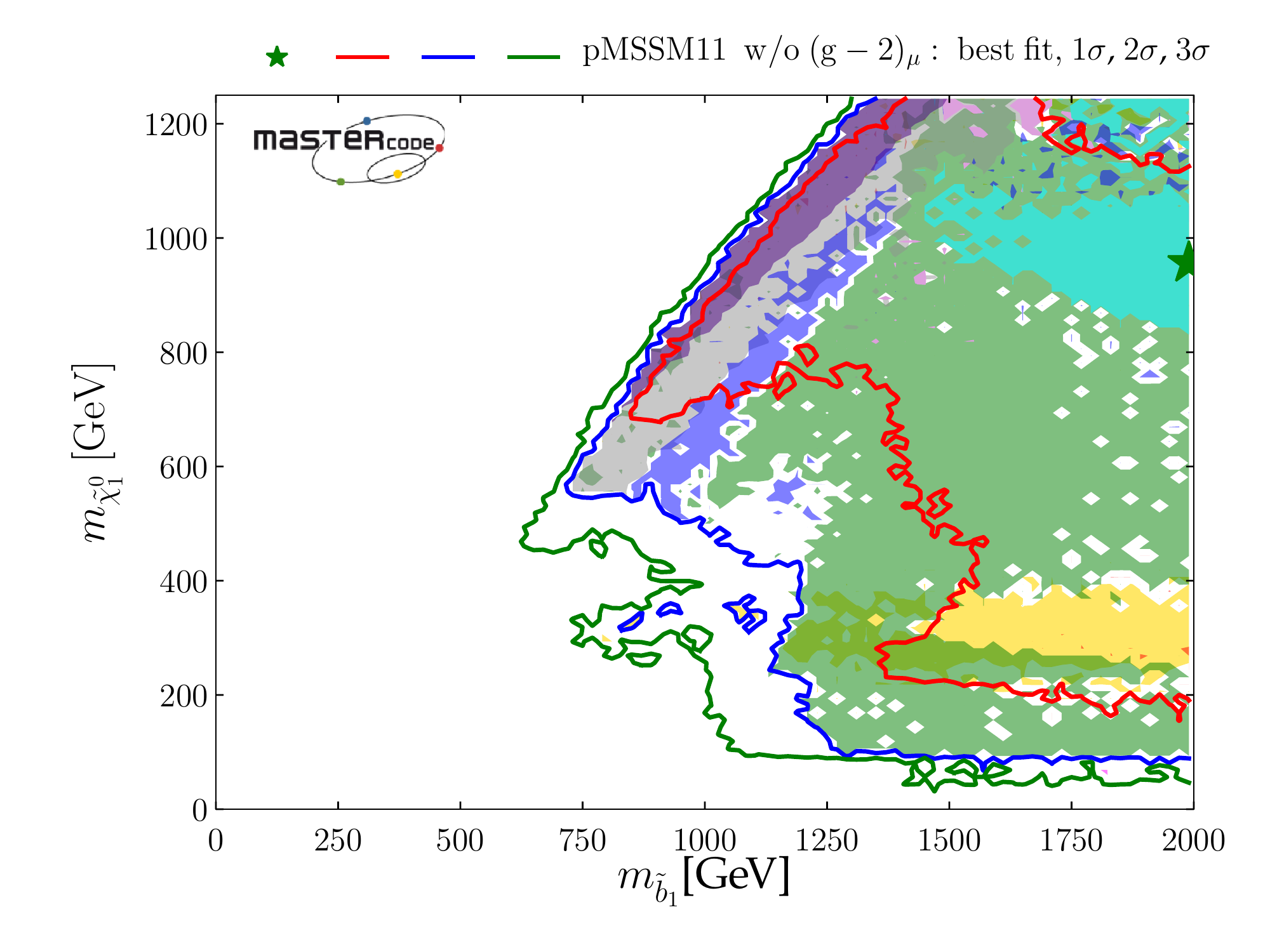} \\
\centering
\includegraphics[width=0.8\textwidth]{figs/pMSSM11_dm_legend.pdf}
\caption{\it Two-dimensional projections of the global likelihood function for the pMSSM11 in the $(\mst1, \mneu1)$ planes
(upper panels) and the $(m_{\tilde b_1}, \mneu1)$ planes (lower panels),
including the \gmt\ constraint (left panels) and dropping it (right panels).
}
\label{fig:2dstopsbottomplanes}
\end{figure*}

\noindent
{\it Sleptons}\\
{As was to be expected, there are large differences between the $(m_{\tilde \mu_R}, \mneu1)$ planes
with and without the \gmt\ constraint, shown in the upper panels in \reffi{fig:2dsmustauplanes}.
We see in the upper left plane a preference for $m_{\tilde \mu_R} \lesssim 550 (750) \gev$ and $\mneu1 \lesssim 500 (550) \gev$
at the 68 (95)\% CL, enforced by the \gmt\ constraint, {with larger masses allowed at the 3-$\sigma$ level}.
There is also a 68\% CL region with similar ranges of
$m_{\tilde \mu_R}$ and $\mneu1$ in the case without \gmt\ (upper right panel), but the 95\% CL region extends to
much larger values of $m_{\tilde \mu_R}$ and $\mneu1$, and there is also a
second, extended 68\% CL region that is separated by a band of points with only slightly higher $\chi^2$.
{In both these plots, we see a very narrow strip where slepton-$\neu1$ coannihilation is important, whereas
$\cha1$ coannihilation dominates in most of the regions allowed at the 95\% CL, supplemented by annihilation
via the $H/A$ bosons at large $\mneu1$ when \gmt\ is dropped}.
We do not display the corresponding $(m_{\tilde \mu_L}, \mneu1)$ and $(m_{\tilde e_{L,R}}, \mneu1)$ planes,
which are very similar because we impose universality on the soft SUSY-breaking masses of the first
two slepton generations.

However, in the pMSSM11 the soft SUSY-breaking stau masses are allowed
to be different, with the result seen in the lower panels of \reffi{fig:2dsmustauplanes} that large values of
$\mstaue$ are allowed at the 68 and 95\% CL even when \gmt\ is imposed. The main differences between
the cases with and without \gmt\ are that larger values of $\mneu1$ are allowed in the latter case - indeed,
the best-fit point has  $\mstaue \sim \mneu1 \sim 1 \tev$. {We see, once again, the importance of
the slepton and $\cha1$ coannihilation mechanisms, supplemented by annihilation via $H/A$ at large $\mneu1$
in the case without \gmt. The small `nose' at $(\mstaue, \mneu1) \sim (100, 50) \gev$ is a remnant of rapid annihilations via
direct-channel $Z$ and $h(125)$ poles.}} \\

\begin{figure*}[htb!]
\centering
\includegraphics[width=0.45\textwidth]{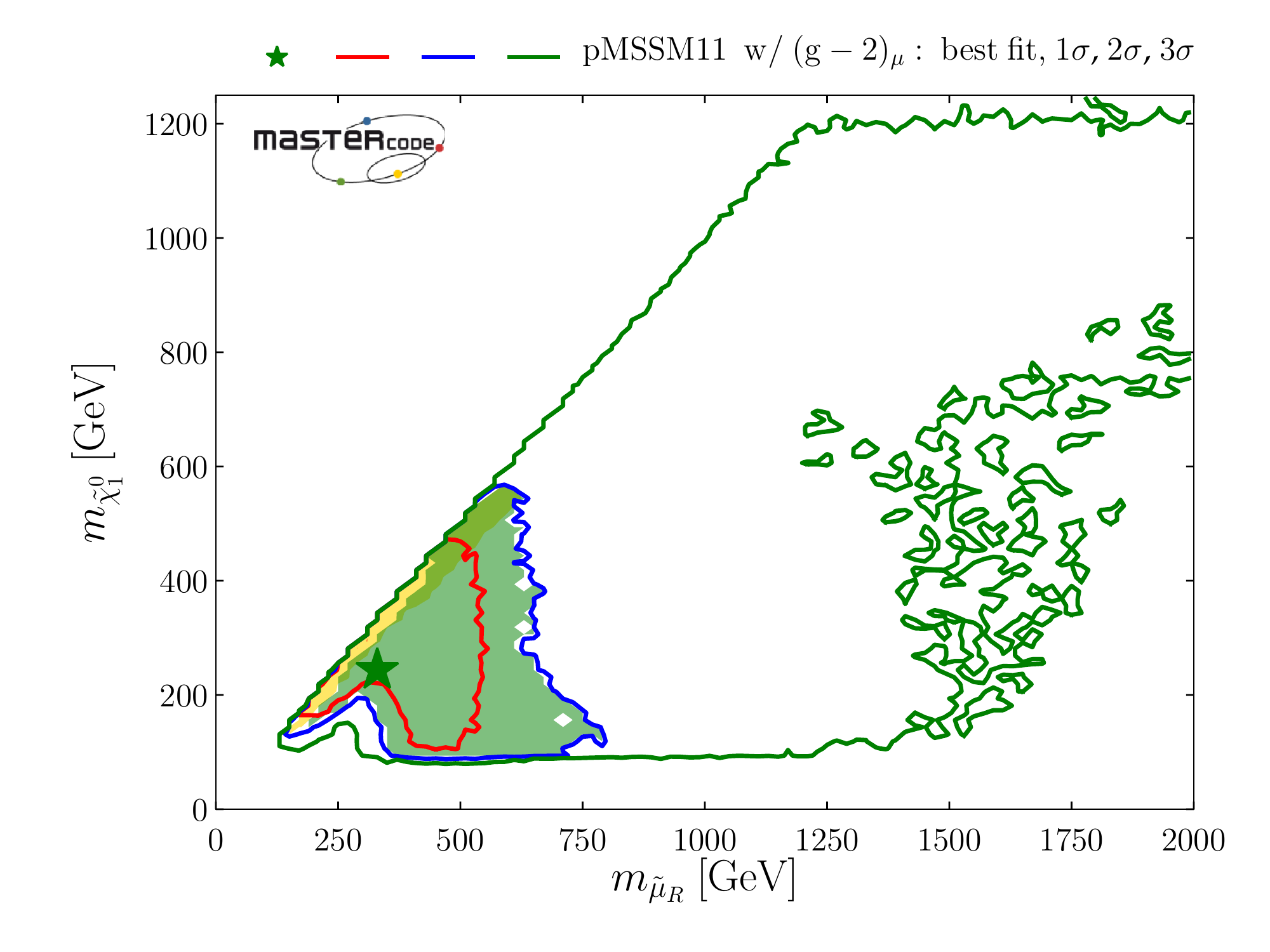}
\includegraphics[width=0.45\textwidth]{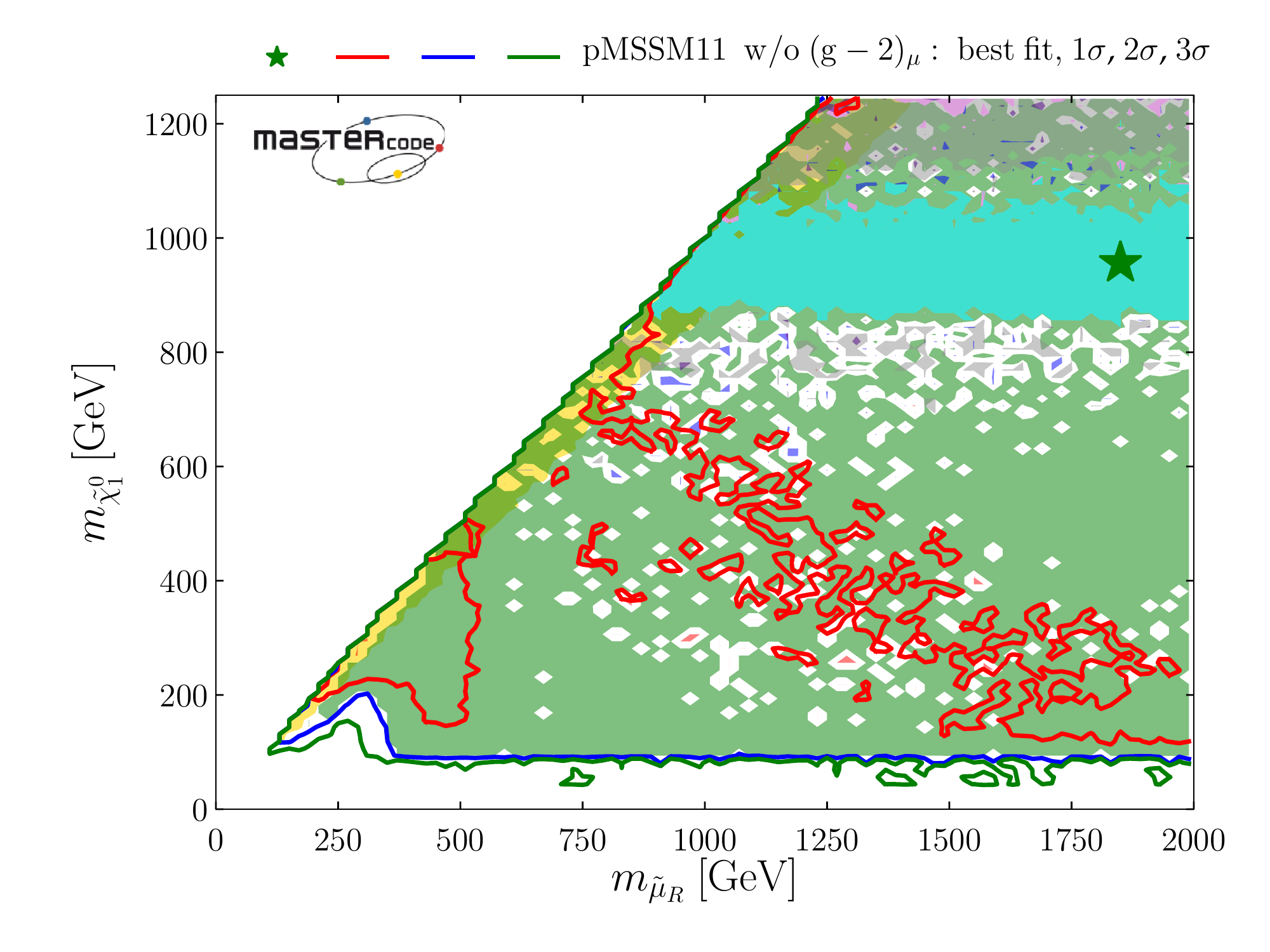} \\
\centering
\includegraphics[width=0.45\textwidth]{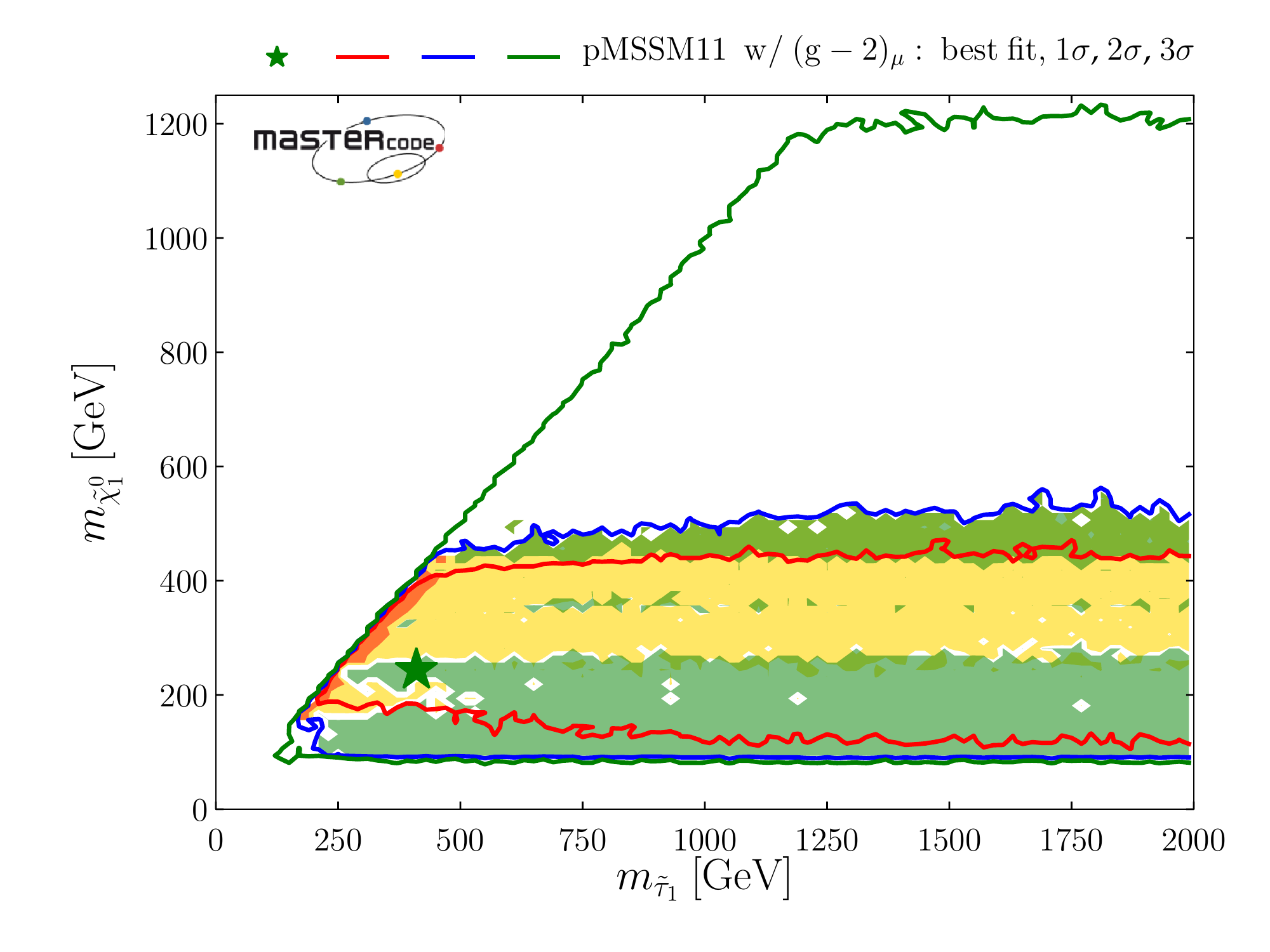}
\includegraphics[width=0.45\textwidth]{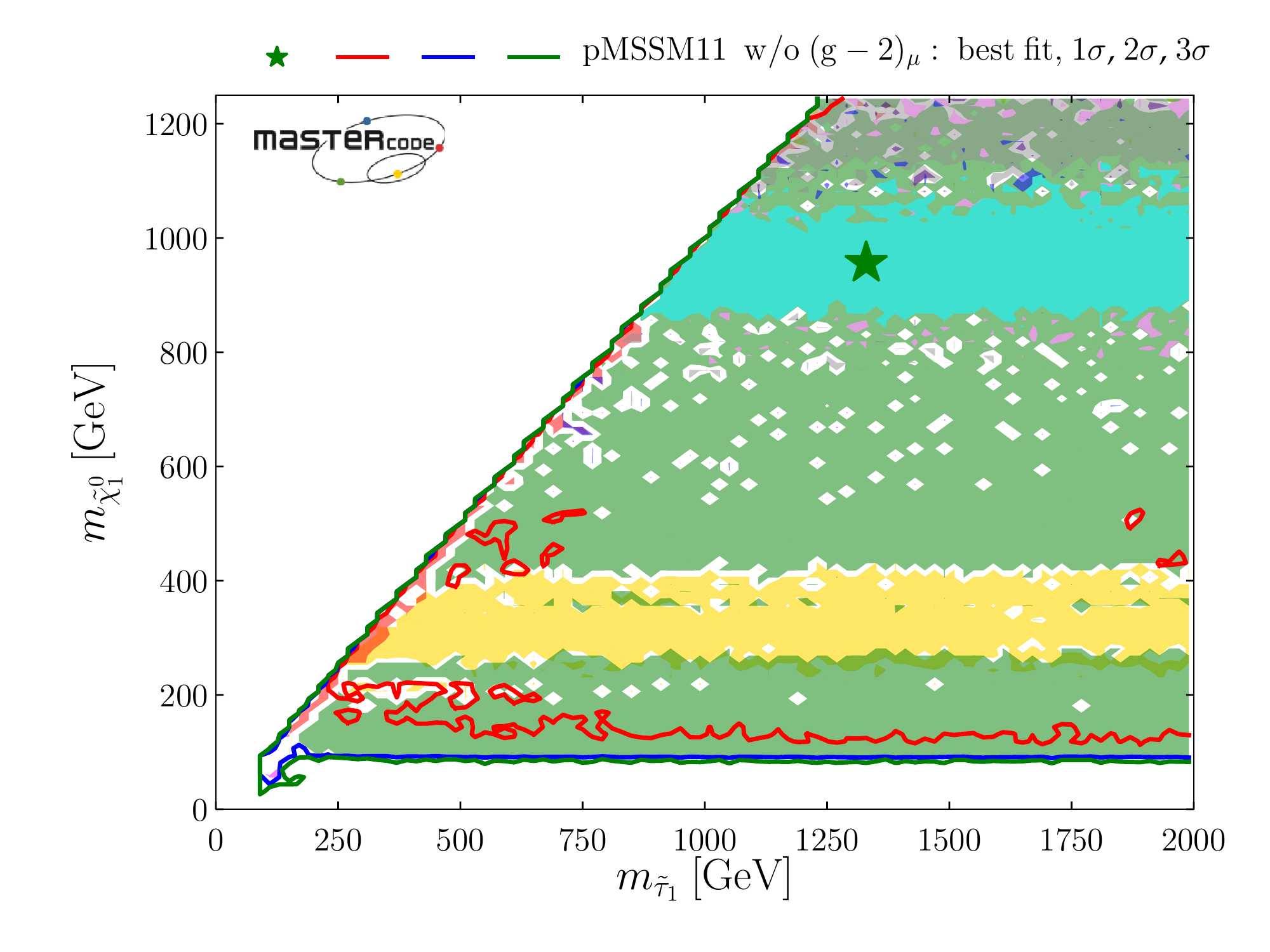} \\
\centering
\includegraphics[width=0.8\textwidth]{figs/pMSSM11_dm_legend.pdf}
\caption{\it Two-dimensional projections of the global likelihood function for the pMSSM11 in the $(m_{\mu_R}, \mneu1)$ planes
(upper panels) and the $(\mstaue, \mneu1)$ planes (lower panels), including the \gmt\ constraint (left panels) and dropping it (right panels).
}
\label{fig:2dsmustauplanes}
\end{figure*}

\noindent
{\it Electroweak inos}\\
{In the upper panels of \reffi{fig:2dEWplanes} we show the $(\mcha1, \mneu1)$ planes with (left panel)
and without (right panel) the \gmt\ constraint. In both panels we see a $\cha1$ coannihilation
strip starting at $(\mcha1, \mneu1) \sim (100, 100) \gev$, and extending to larger $\mcha1$
in the latter case.  This $\cha1$ coannihilation strip is isolated in the \gmt\ case, but connected to an extended 95\% CL
region at large $\mcha1$ in the case without \gmt. In both panels there is a broad band with
$\mneu1 \sim 150$ to $400 \gev$ where slepton coannihilation dominates.
A major difference between the plots is the extensive region at large $\mneu1$
in the case without \gmt\ where annihilation via $H/A$ is important.
The best-fit points are at $\mcha1 \sim \mneu1 \sim 250 \gev$
in the \gmt\ case and $\sim 1000 \gev$ in the case without it.}\\

\begin{figure*}[htb!]
\centering
\includegraphics[width=0.45\textwidth]{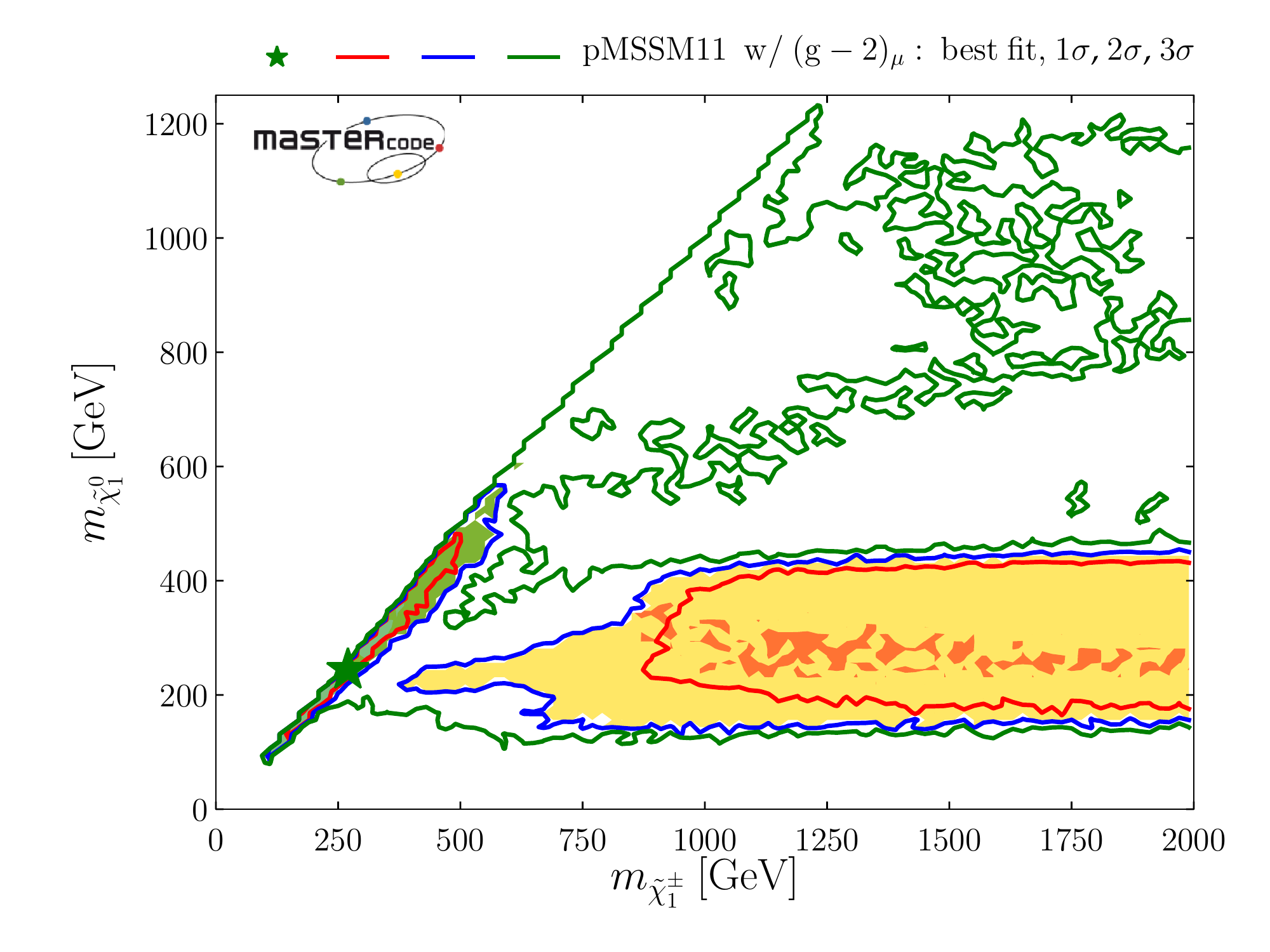}
\includegraphics[width=0.45\textwidth]{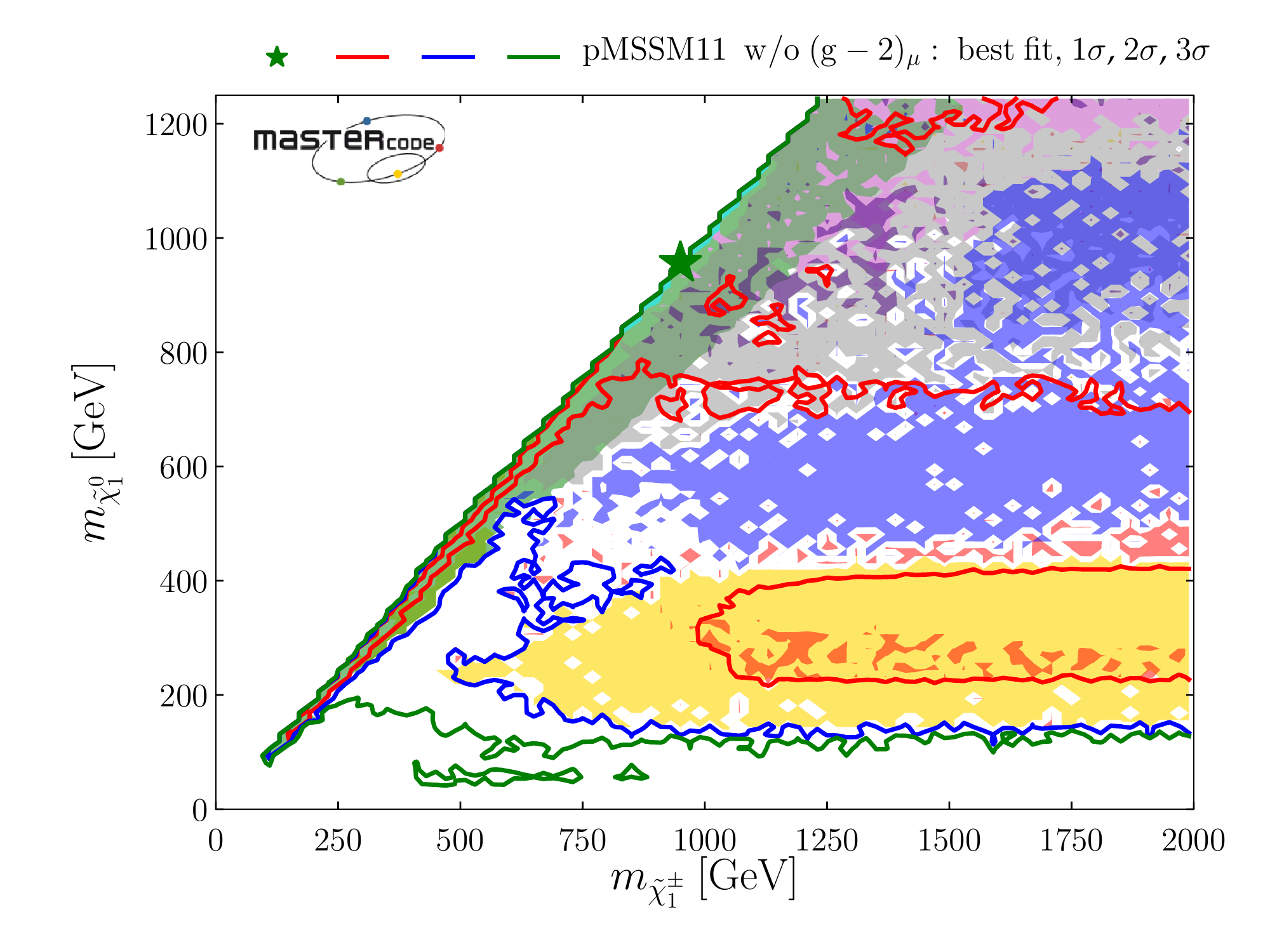} \\
\centering
\includegraphics[width=0.45\textwidth]{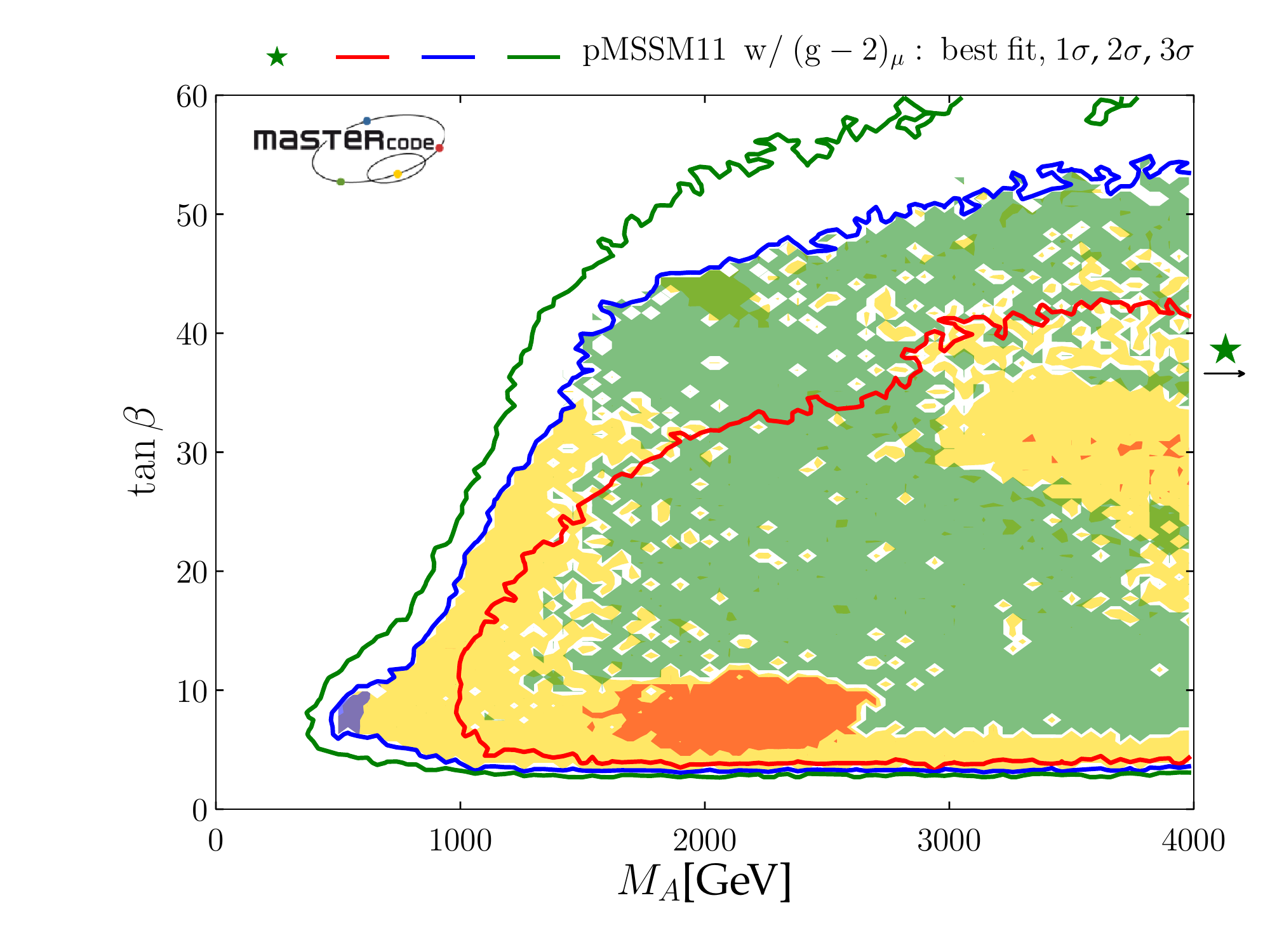}
\includegraphics[width=0.45\textwidth]{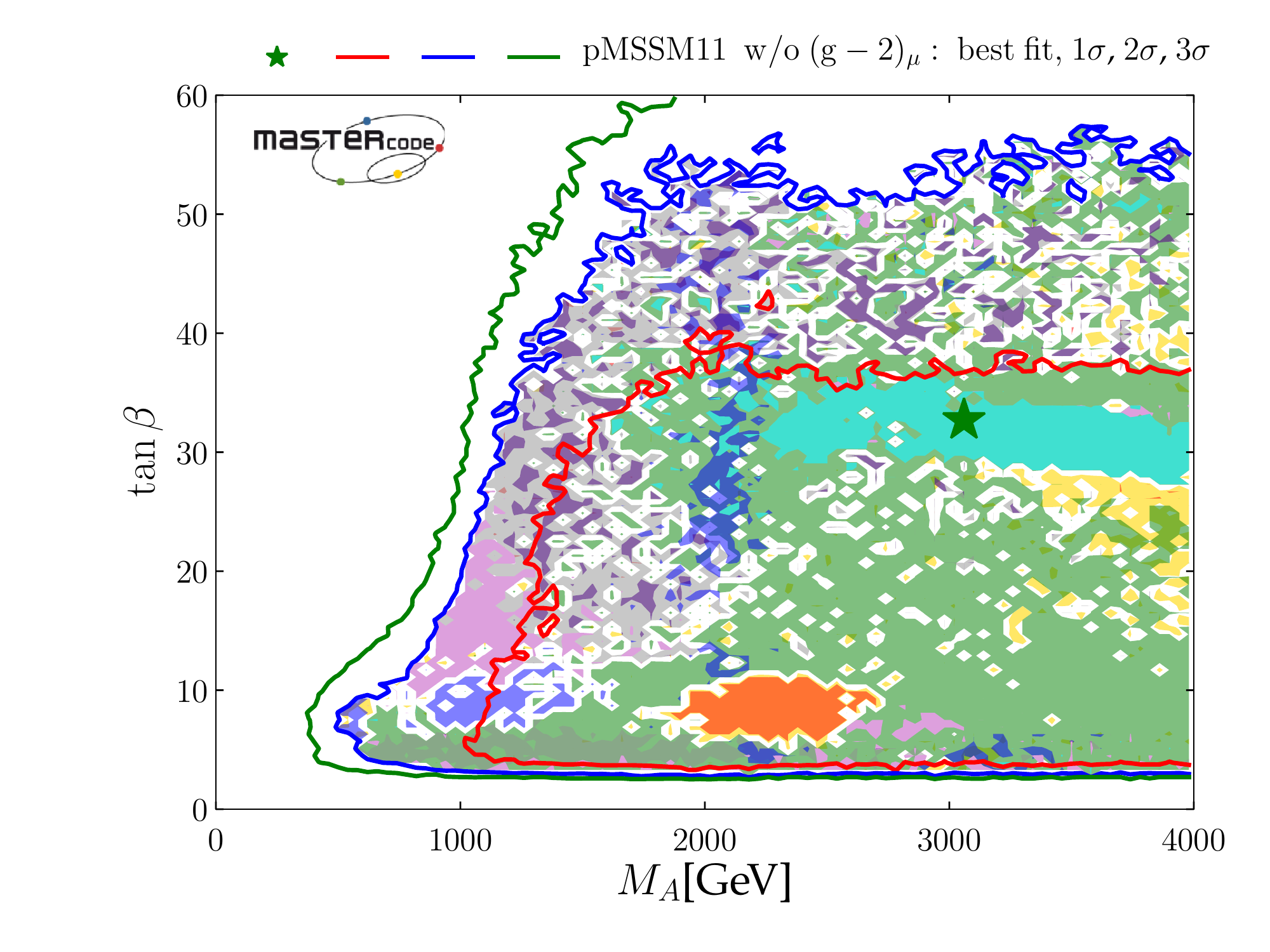} \\
\centering
\includegraphics[width=0.8\textwidth]{figs/pMSSM11_dm_legend.pdf}
\caption{\it Two-dimensional projections of the global likelihood function for the pMSSM11 in the $(\mcha1, \mneu1)$ planes
(upper panels) and the $(\MA, \tb)$ planes (lower panels),
including the \gmt\ constraint (left panels) and dropping it (right panels).
}
\label{fig:2dEWplanes}
\end{figure*}

\noindent
{\it Heavy Higgs bosons}\\
{The 68 and 95\% CL regions in the pair of $(\MA, \tb)$ planes shown
in the lower panels of Fig.~\ref{fig:2dEWplanes} {display the importance of the latest
ATLAS constraint on $A/H \to \tau^+ \tau^-$ decays with $\sim 36$/fb of data at 13 TeV~\cite{HA13},
which disfavours regions with $\MA \lesssim 1 \tev$ at larger \tb.
We also note that} the dominant DM mechanisms display significant differences.
Chargino coannihilation is important in both planes, but slepton coannihilation appears only in the
case where \gmt\ is included. In this case annihilation via the $H/A$ poles appears only when
$\MA \lesssim 1 \tev$, but it appears also at larger $\MA$ when \gmt\ is dropped. We see
in both cases a limited region with $\MA \sim 2 \tev$ and $\tb \lesssim 10$ where stau
coannihilation dominates. In our previous pMSSM10 analysis~\cite{mc11}
the interplay of the LHC electroweak searches, \gmt\ and the DM constraints, heavily relying on the fact that only one
  independent slepton mass parameter was allowed, led to a
region with $25 \lsim \tb \lsim 45$ being preferred at the 68\% CL. However, in the pMSSM11, dropping
the restriction $\mstau\ = {\mslep}$ now allows values of $\tb < 5$
for a wide range of $\MA$ values. Also, despite the updated (stronger) constraints on
$H/A \to \tau \tau$, values down to $\MA \sim 500 \gev$ are still allowed
at the 95\% CL.}


\section{One-Dimensional Likelihood Functions}
\label{sec:1D}

{In this Section we present the profile $\chi^2$ likelihood functions corresponding to
various one-dimensional projections of the results from our global fits, again comparing
those with and without the \gmt\ constraint. {In the following series of plots, results
including the LHC 13-TeV constraints are shown as solid lines, and those using only
8-TeV results are shown as dashed lines. Results obtained including \gmt\ are shown in blue
and those obtained without \gmt\ are shown in green.}

\subsection{\gmt}

As a preliminary,
\reffi{fig:1dg-2} shows the one-dimensional profile likelihood functions
for \gmt\ with ({blue}) and ({green}) without applying the \gmt\ constraint {\it a priori}. {Comparing
the solid and dashed lines, we see very little difference between the results using and
discarding the LHC 13-TeV data. The results including \gmt\
(blue lines) largely reflect} our
implementation of the \gmt\ constraint shown in Table~\ref{tab:updates}. Interestingly, when this constraint is not
applied {\it a priori} (green lines), whilst a very small SUSY contribution to \gmt\ is preferred,
a wide range of values of \gmt\ are found to be allowed at the $\Delta
\chi^2 {\sim 2}$ level {and the experimental value can be accommodated at the 1.5-$\sigma$ level}.
Although the other data certainly do not favour a large SUSY contribution to \gmt, neither do
they exclude it.
}

\begin{figure*}[h]
\centering
\includegraphics[width=0.6\textwidth]{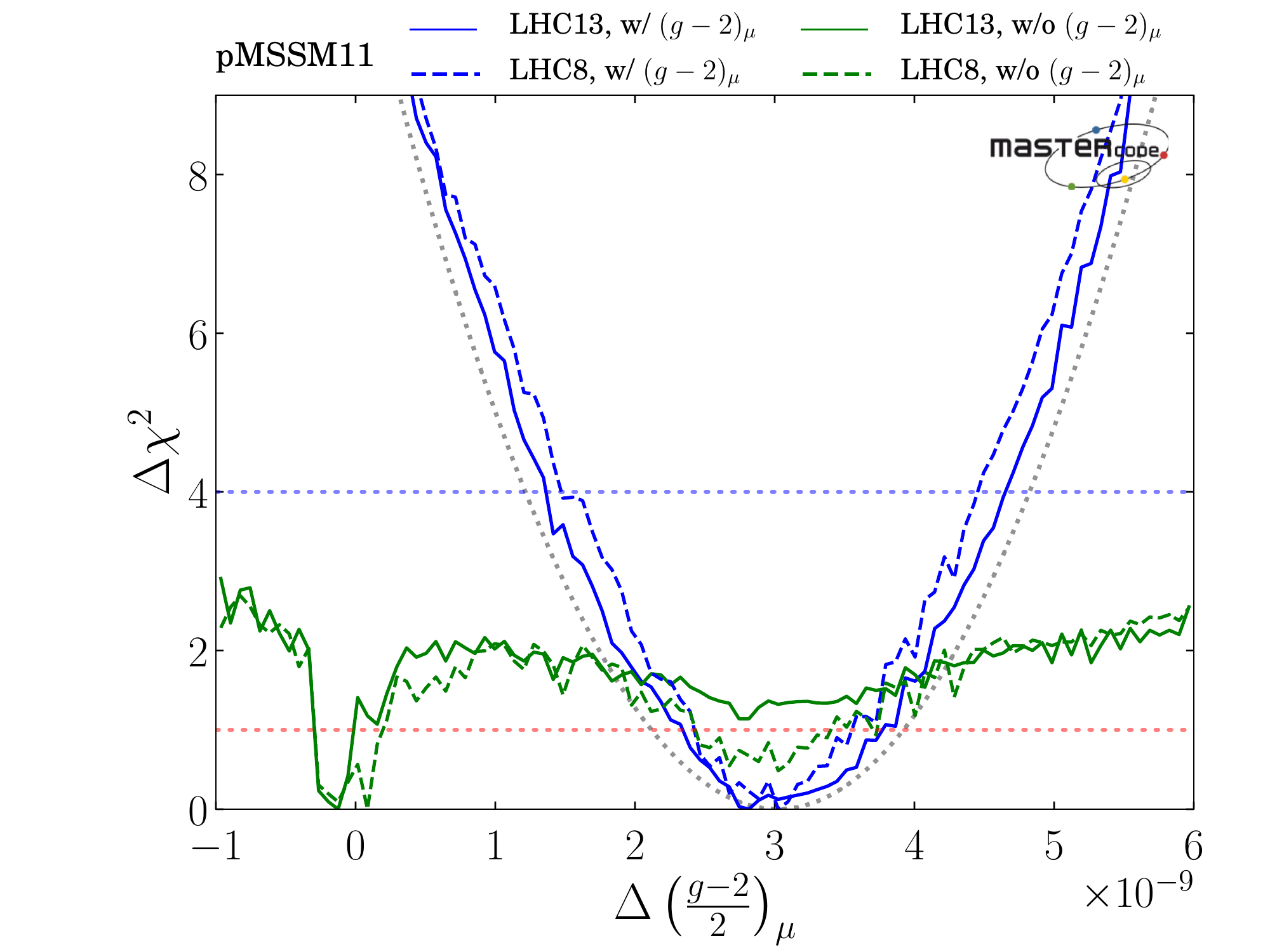}
\caption{\it One-dimensional profile likelihood functions for \gmt\ in the pMSSM11,
  with (blue) and without (green) applying the \gmt\ constraint {\it a priori} and with (solid) and without (dashed)
applying the constraints coming from the LHC run at 13 TeV. {Also shown
as a dotted line is the experimental constraint~\protect\cite{newBNL},
taking into account the theoretical uncertainty~\protect\cite{g-2} within the Standard Model.}}
\label{fig:1dg-2}
\end{figure*}


\subsection{Sparticle Masses}
\label{sec:masses}

~\\
\noindent
{\it Squarks and gluinos}\\
{The profile likelihood functions for squarks and gluinos are shown in
\reffi{fig:1dglsq}. The left panel is for $\msq$, where we see that when the
13-TeV LHC data and \gmt\ constraint are included (solid blue line),
there is a monotonic decrease in $\chi^2$ as ${\msq}$ increases,
with {$\msq \gtrsim 1.9 \tev$} at the 95\% CL
(horizontal dotted line). This constraint is much stronger than that obtained with 8-TeV data alone
(dashed blue and green lines): $\msq \gtrsim 1.0 \tev$ {at the 95\% CL}. In particular, the 13-TeV data exclude a
squark coannihilation strip that had been allowed by the 8-TeV data.
When \gmt\ is dropped but the 13-TeV data retained (solid green line),
the $\chi^2$ function exhibits a global minimum at $\msq \sim 1 \tev$, with a plateau at $\Delta \chi^2 \simeq 1.5$
at larger $\msq$. {Important roles in the location of this global minimum are played by
the \bsdmm\ constraint as discussed in Subsection~\ref{sec:flavour}, whose contribution to
the global $\chi^2$ function at this point is $\sim 1.1$ lower than at large $\msq$, and by
the relic DM density constraint, which is satisfied thanks to multiple coannihilation
processes as discussed in Subsection~\ref{sec:measures}.}}

{In the right panel of \reffi{fig:1dglsq} for $\mgl$, we see that with both the LHC 13-TeV data and \gmt\ included
$\mgl \gtrsim 1.8 \tev$ (solid blue line), whereas without \gmt\ we find $\mgl \gtrsim 1.0 \tev$ (solid green line).
On the other hand, in the absence of the LHC 13-TeV data (dashed lines), $\mgl \gtrsim 500 \gev$
would have been allowed at the 95\% CL, whether \gmt\ is included, or not. The LHC 13-TeV run
has excluded a region of gluino coannihilation that was allowed by the 8-TeV data.}\\

\begin{figure*}[h]
\centering
\includegraphics[width=0.49\textwidth]{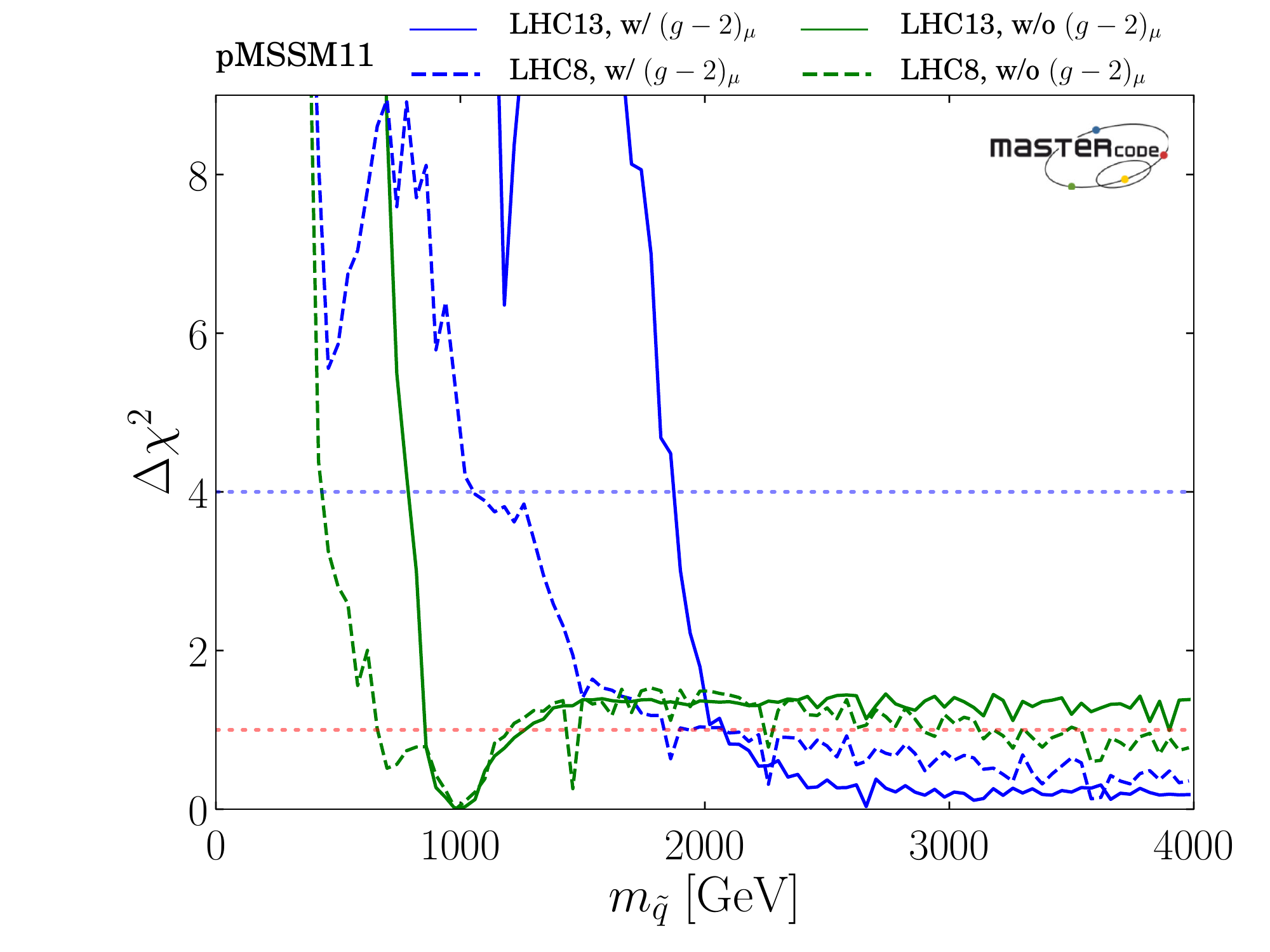}
\includegraphics[width=0.49\textwidth]{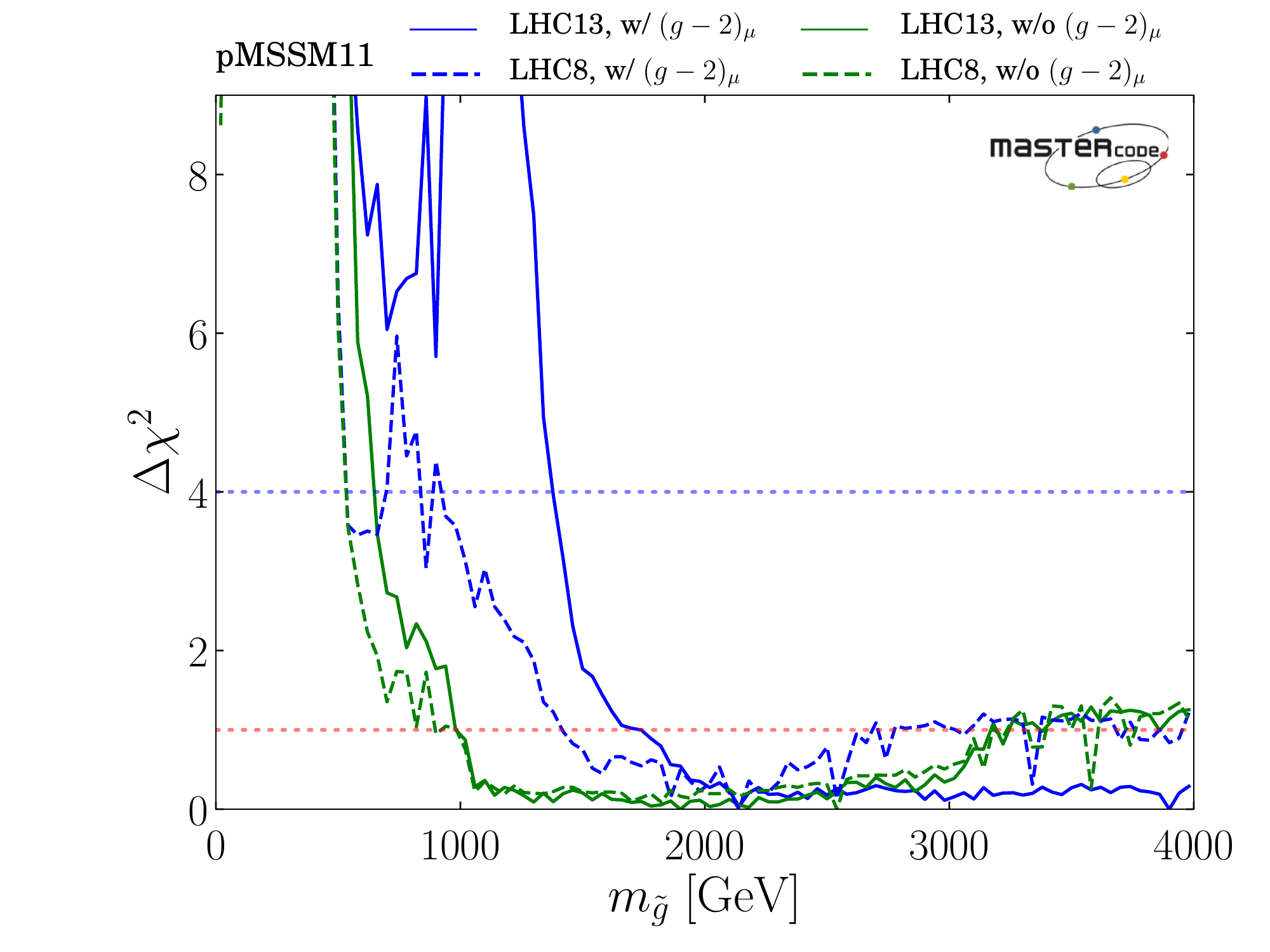} \\
\caption{\it Left panel: one-dimensional profile likelihood functions for the $\sq$ mass in the pMSSM11 with (blue) and
without the \gmt\ constraint (green) and with (solid) and without (dashed) applying the constraints from LHC Run II.
Right panel: similarly for the $\gl$ mass.}
\label{fig:1dglsq}
\end{figure*}

\noindent
{\it Third-generation squarks}\\
An analogous pair of plots showing the profile likelihood functions for the masses of the
$\sto1$ and $\sbot1$ are shown in the left and right panels of \reffi{fig:1stsb}.
{When the LHC 13-TeV data are included we see in the left panel a well-defined local minimum
of the $\chi^2$ function in a compressed-stop region with $\Delta \chi^2 \sim 2.3$ for $\mst1 \sim 400 \gev$.
This is followed by a local maximum that exceeds $\Delta \chi^2 > 9$ for $\mst1 \sim 800 \gev$
when \gmt\ is included (solid blue line) but is lower when \gmt\ is dropped (solid green line).
This is followed in both cases by a monotonic decrease for larger $\mst1$ and
a global minimum of $\chi^2$ for $\mst1 \sim 1800 \gev$.}

{In the case of $\msb1$ (right panel of \reffi{fig:1stsb}).
when the 13-TeV LHC data and \gmt\ are included (solid blue line) there are some irregularities in the $\chi^2$ function
for $\msb1 \sim 1000 \gev$, but no hint of a compressed-sbottom region when \gmt\ is dropped (dashed blue line).
Comparing with the situation when only LHC 8-TeV used, we see that the 13-TeV data
have increased significantly the pressure on scenarios with {$\msb1 \lesssim 1.5 \tev$}.
At larger masses the $\chi^2$ functions $\msb1$ are very similar to those for $\mst1$, whether \gmt\
is included or not.} \\

\begin{figure*}[h]
\centering
\includegraphics[width=0.49\textwidth]{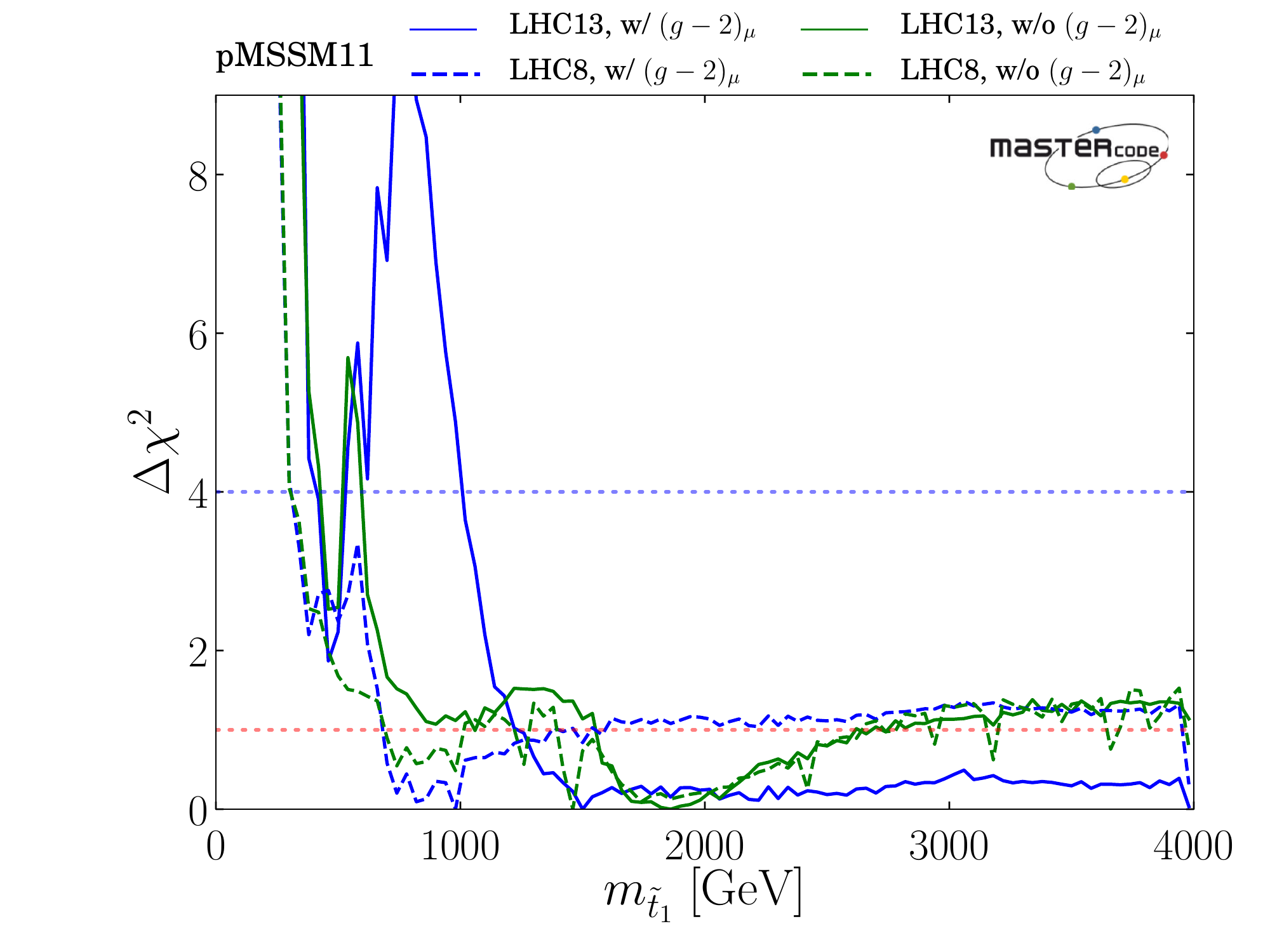}
\includegraphics[width=0.49\textwidth]{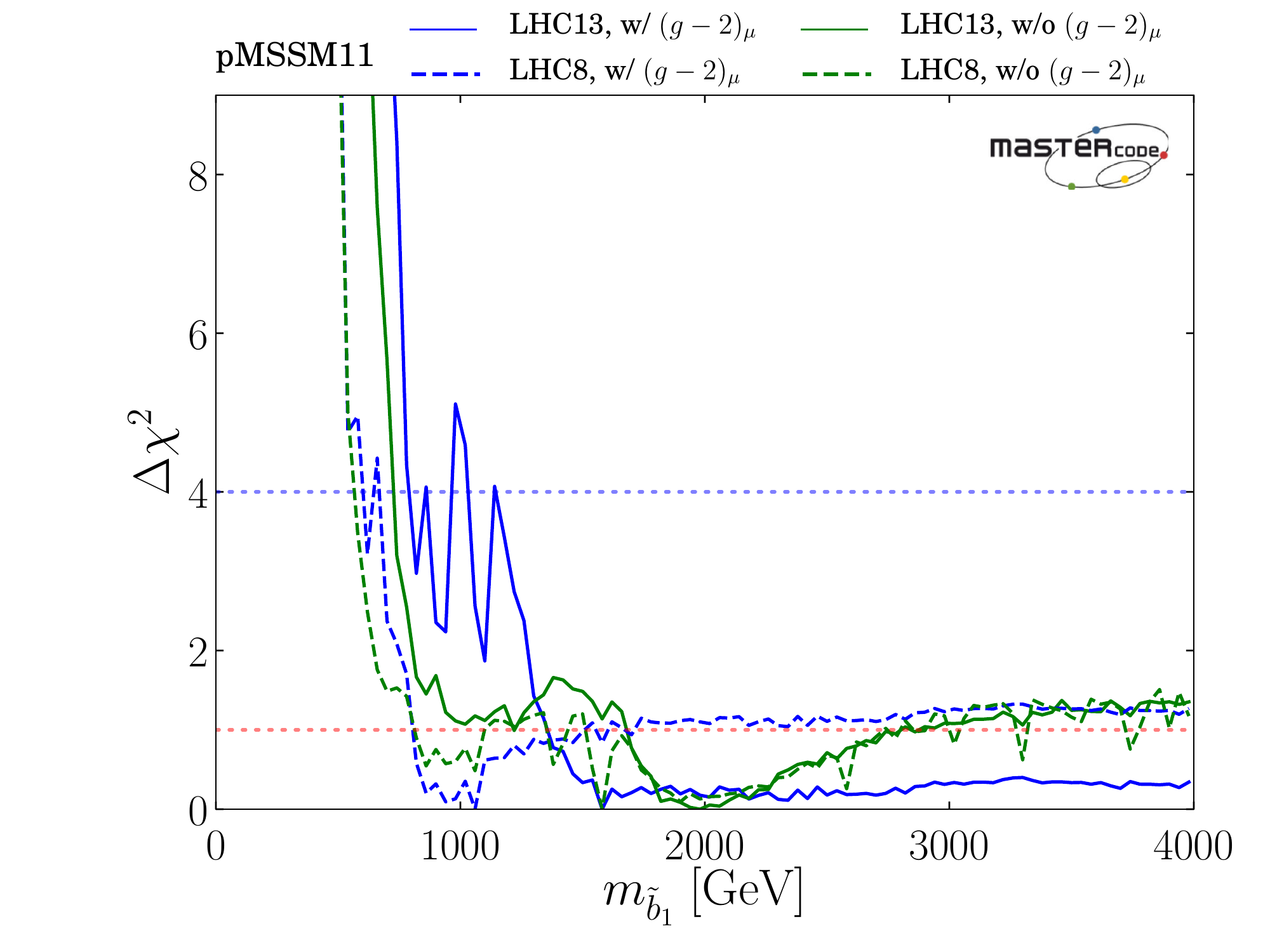} \\
\caption{\it Left panel: one-dimensional profile likelihood functions for the $\tilde t_1$ mass in the pMSSM11 with (blue) and
without the \gmt\ constraint (green) and with (solid) and without (dashed) applying the constraints from LHC Run II.
Right panel: similarly for the $\tilde b_1$ mass.}
\label{fig:1stsb}
\end{figure*}

\noindent
{\it Sleptons}\\
\reffi{fig:1dsmst} displays analogous plots of the profile likelihood functions for $m_{\tilde \mu_R}$
(left panel, those for $m_{\tilde \mu_L}$ and $m_{\tilde e_{L,R}}$ are very similar)
and $\mstaue$ (right panel, {that for $m_{\tilde \tau_2}$ is quite similar}).
When the \gmt\ constraint is implemented (blue lines),
the $\chi^2$ function for $m_{\tilde \mu_R}$ exhibits the expected well-defined minimum
at $m_{\tilde \mu_R} \sim 200$ to $500 \gev$ {when the LHC 13-TeV data are included}.
{In the absence of the \gmt\ constraint (green lines)}, this is replaced by a plateau with $\Delta \chi^2 \sim 2$
that extends to $m_{\tilde \mu_R} \sim 900 \gev$, where the profile likelihood function drops to very small values for
larger $m_{\tilde \mu_R}$. {The drop occurs because this fit prefers $\mneu1 \sim 900$ to $1000 \gev$, and
any heavier ${\tilde \mu_R}$ can decay into a $\neu1$ in this mass range.}

{We see in the right panel of \reffi{fig:1dsmst} that
when \gmt\ is included (blue lines) the profile likelihood function for $\mstaue$ is quite
different from that for $m_{\tilde \mu_R}$, thanks to the decoupling between their soft SUSY-breaking
masses in the pMSSM11. The $\chi^2$ function falls monotonically to a local minimum when $\mstaue \sim 300 \gev$
and remains small for larger $\mstaue$, whether the LHC 13-TeV data are included (solid line), or not
(dashed line). However, when \gmt\ is dropped (green lines),
the profile likelihood function for $\mstaue$ is quite similar to that for $m_{\tilde \mu_R}$,
also exhibiting a plateau with $\Delta \chi^2 \sim 2$ and falling to small values for $\mstaue \gtrsim 900 \gev$
when the LHC 13-TeV data are included.} {This feature appears because,
in order to avoid a charged LSP, a smaller value of $\mstaue$ would require a smaller value of $\mneu1$,
which is disfavoured as seen in the left panel of Fig.~\ref{fig:1dneu} and discussed below.}  \\

\begin{figure*}[h]
\centering
\includegraphics[width=0.49\textwidth]{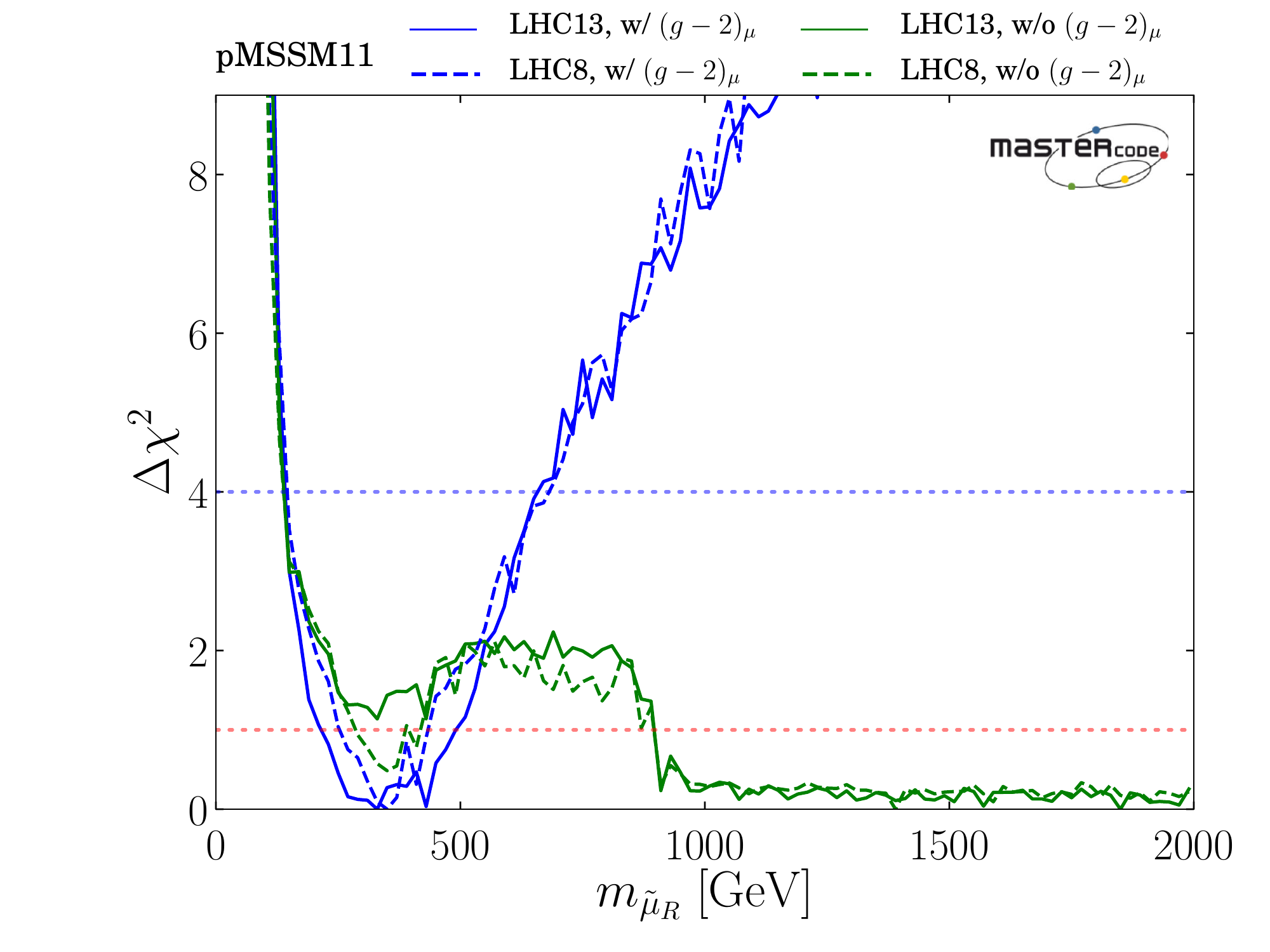}
\includegraphics[width=0.49\textwidth]{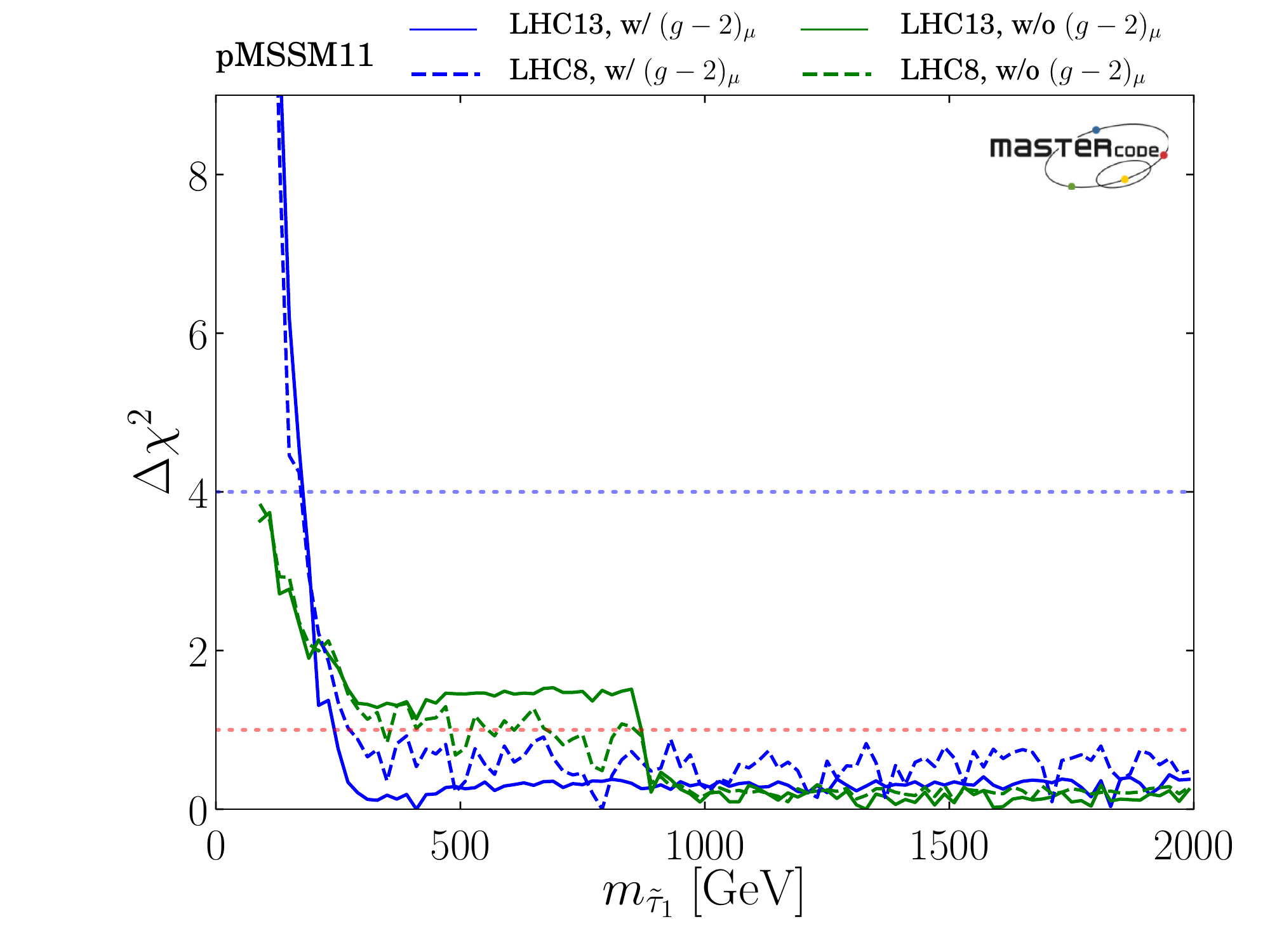} \\
\caption{\it Left panel: one-dimensional profile likelihood functions for the $\tilde \mu_R$ mass in the pMSSM11 with (blue) and
without the \gmt\ constraint (green) and with (solid) and without (dashed) applying the constraints from LHC Run II.
Right panel: similarly for the $\tilde \tau_1$ mass.}
\label{fig:1dsmst}
\end{figure*}

\noindent
{\it Electroweak inos}\\
\reffi{fig:1dneu} shows the profile likelihood functions for the lightest neutralino $\neu1$ (left panel)
and the lighter chargino $\cha1$ (right panel). When the \gmt\ constraint is applied (blue
lines), the $\chi^2$ function for $\mneu1$ {including 13-TeV data} exhibits
a well-defined but broad minimum at $\mneu1 \sim 100$ to $400 \gev$. This preference for small $\mneu1$
was already seen in the upper boundaries of the 68\% and 95\% CL regions in the planes involving
$\mneu1$ shown in the previous Section when the \gmt\ constraint is applied (left panels).

{On the other hand,
  when the \gmt\ constraint is dropped (green lines)} we see a preference
for $\mneu1 \sim 950 \gev$.
{Despite the fact that the LSP is a nearly-pure Higgsino at this best-fit point, this mass of $\sim 950$ GeV is below
the $\sim 1.1 \tev$ mass expected for a Higgsino dark matter candidate.
This arises because, at the best-fit point, several of the squark masses lie close to the
LSP mass, making multiple coannihilation important. Due to the
relatively large number of states with masses close to the Higgsino,
their density actually increases the final LSP relic density~\footnote{This effect was noted previously
in a different context in~\cite{ProPro}.},
thereby pushing the mass of the Higgsino below its nominal $\sim 1.1 \tev$ value.}

Turning now to the profile likelihood functions for the lighter chargino $\cha1$
(right panel of \reffi{fig:1dneu}), we see that when \gmt\ is taken into account (blue lines)
the $\chi^2$ function also features a well-defined minimum for $\mcha1 \sim 200$ to $500 \gev$
{(that for $\neu2$ is very similar)},
reflecting the importance of $\cha1 - \neu1$ coannihilation. This minimum is
followed by a rise to a local maximum at $\mcha1 \sim 600 \gev$, {which is more pronounced
when the 13-TeV data are included (solid blue)},
followed by a slow decrease as  $\mcha1$ increases further. {When the \gmt\ constraint is dropped
and the LHC 13-TeV data are included (solid green line)},
the $\chi^2$ functions for $\mcha1$ {and $\mneu2$ have global minima} at $\mneu1 \sim 1000 \gev$,
accompanied by plateaus with $\Delta \chi^2 \sim 2$ at smaller and larger
values of $\mcha1$. {The dip in the $\chi^2$ function occurs because
the fit to \bsdmm\ is improved for $\mcha1 \simeq \mneu2
\sim \mneu1 \sim 1 \tev$.}
Chargino coannihilation is important around this global minimum of the $\chi^2$ function, and so
are other coannihilation mechanisms, as we discuss later.\\

\begin{figure*}[h]
\centering
\includegraphics[width=0.5\textwidth]{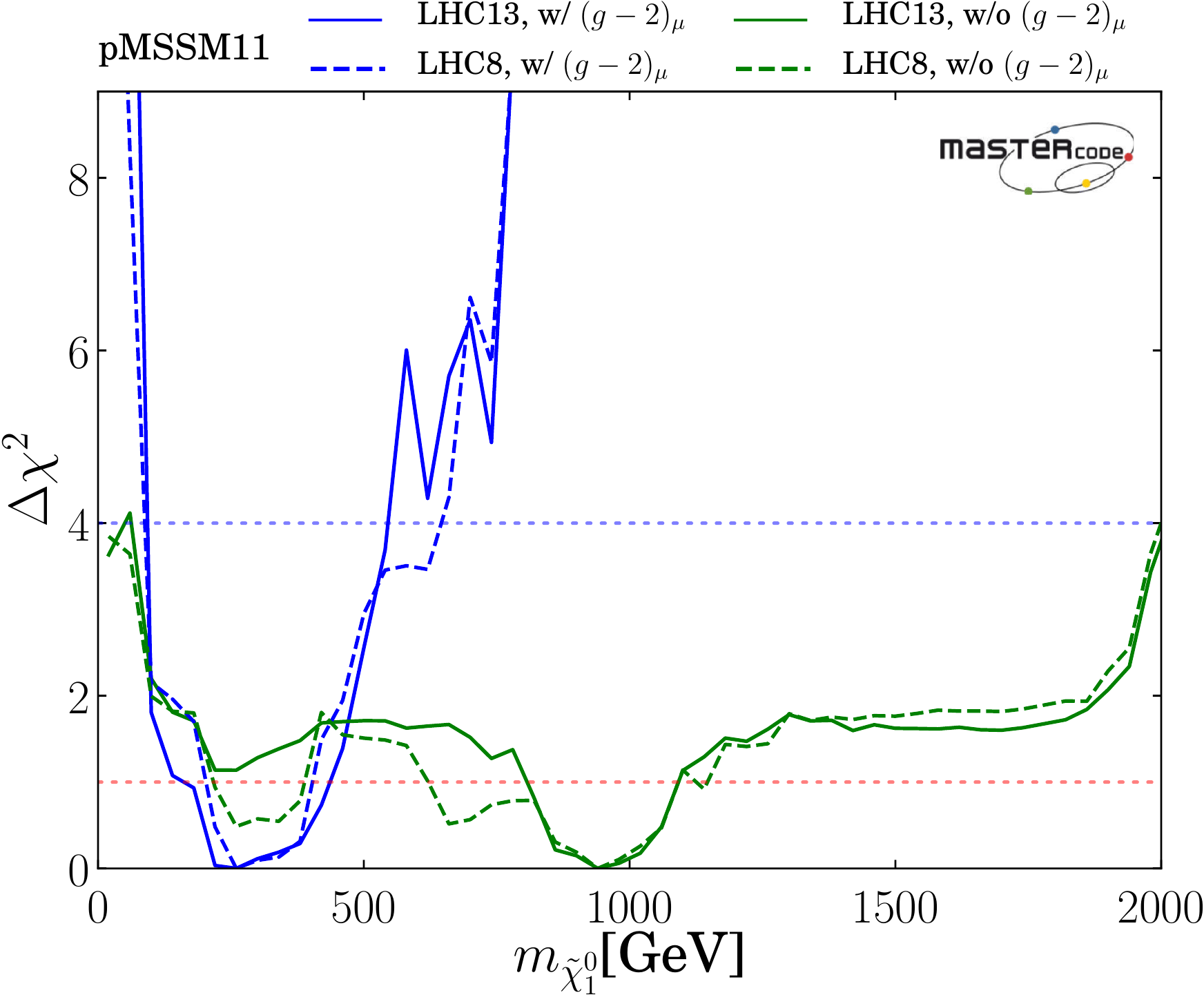}~~~
\includegraphics[width=0.5\textwidth]{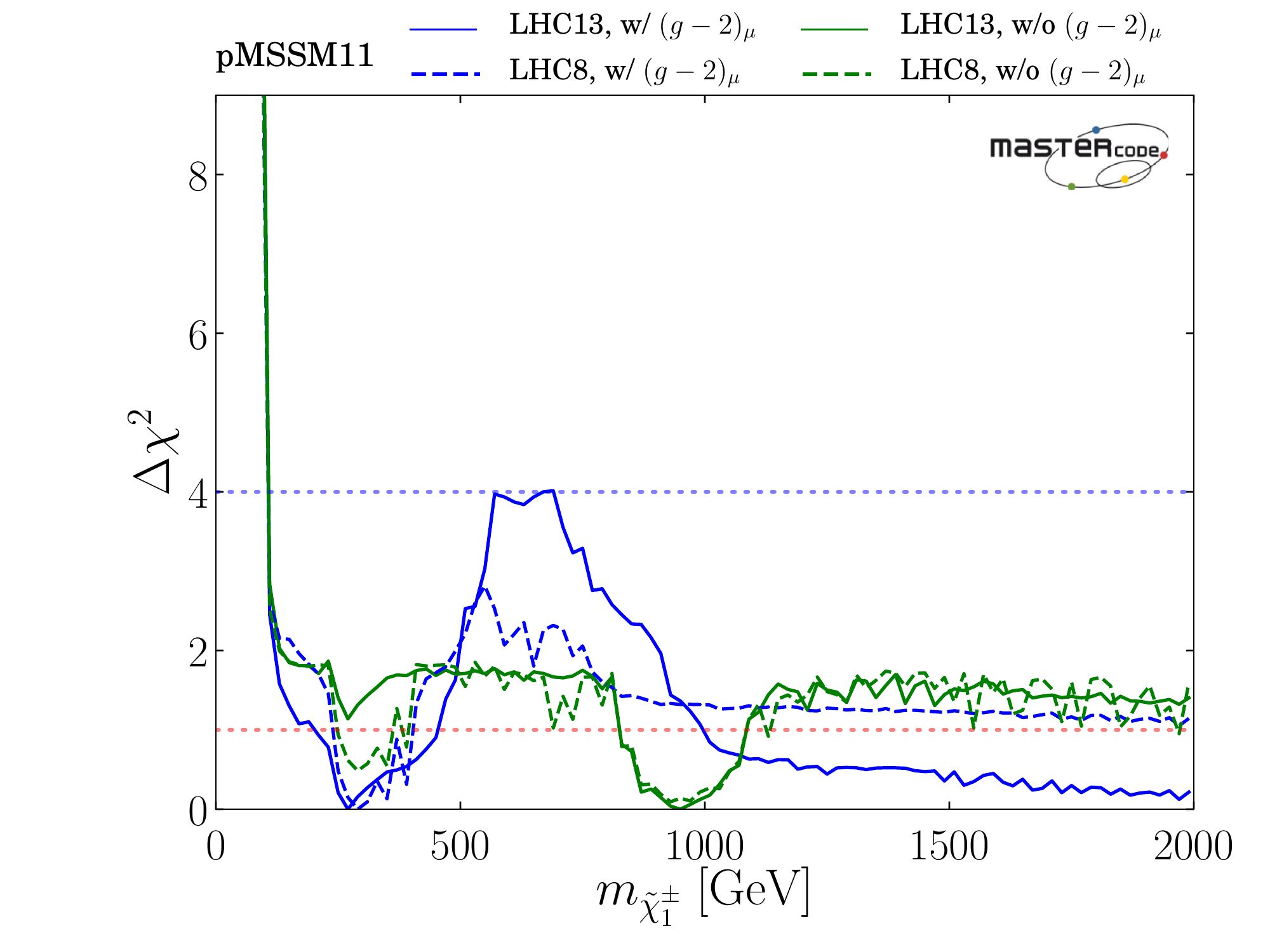} \\
\caption{\it Left panel: one-dimensional profile likelihood functions for the $\neu1$ mass in the pMSSM11 with (blue) and
without the \gmt\ constraint (green) and with (solid) and without (dashed) applying the constraints from LHC Run II.
Right panel: similarly for the $\cha1$ mass.}
\label{fig:1dneu}
\end{figure*}

\subsection{Neutralino Composition}
\label{sec:compo}

It is interesting also to examine the profile likelihood functions for the {amplitudes $N_{1i}$
characterizing the $\neu1$ composition:
\begin{equation}
\neu1 \; = \; N_{11} {\tilde B} + N_{12} {\tilde W^3} + N_{13} {\tilde H_u} + N_{14} {\tilde H_d} \, ,
\label{composition}
\end{equation}
which are shown in \reffi{fig:1hcompo}, {again for the analysis with the 13-TeV data
as solid lines and without them as dashed lines, and with \gmt\ as blue lines and without it
as green lines. The top left panel shows that, when \gmt\ is included, \ an almost pure $\tilde B$
composition of the $\neu1$ is preferred, $N_{11} \to 1$, though the possibility that this component is almost absent
is also allowed at the level $\Delta \chi^2 \sim 4$. On the other hand, when the constraint from \gmt\ is removed, there
is a mild ($\Delta \chi^2 \sim 1$) preference for $N_{11} \to 0$. The reason for this is again the preference for a large $\tilde H_{u,d}$
components in the latter case, where the neutralino mass is allowed to be larger, due to flavor constraints slightly favoring a 1 TeV neutralino as a solution to the observed DM relic density.
The upper right panel shows that a small $\tilde W^3$ component in the $\neu1$ is
preferred in all cases.}~\footnote{{This is because
we only scan $m_{\tilde \ell}$ {and $m_{\tilde \tau} < 2 \tev$,}
hence $\mneu1 < 2 \tev$, so do not probe the expected Wino-like
LSP region where $\mneu1 \sim 3 \tev$.}}.
Finally, the lower panel confirms that small $\tilde H_{u,d}$ components are preferred by $\Delta \chi^2 \gtrsim 4$ when
\gmt\ is included, whereas there would have been a preference for these
components to dominate in the absence of the \gmt\ constraint.}

\begin{figure*}[htbp!]
\centering
\includegraphics[width=0.475\textwidth]{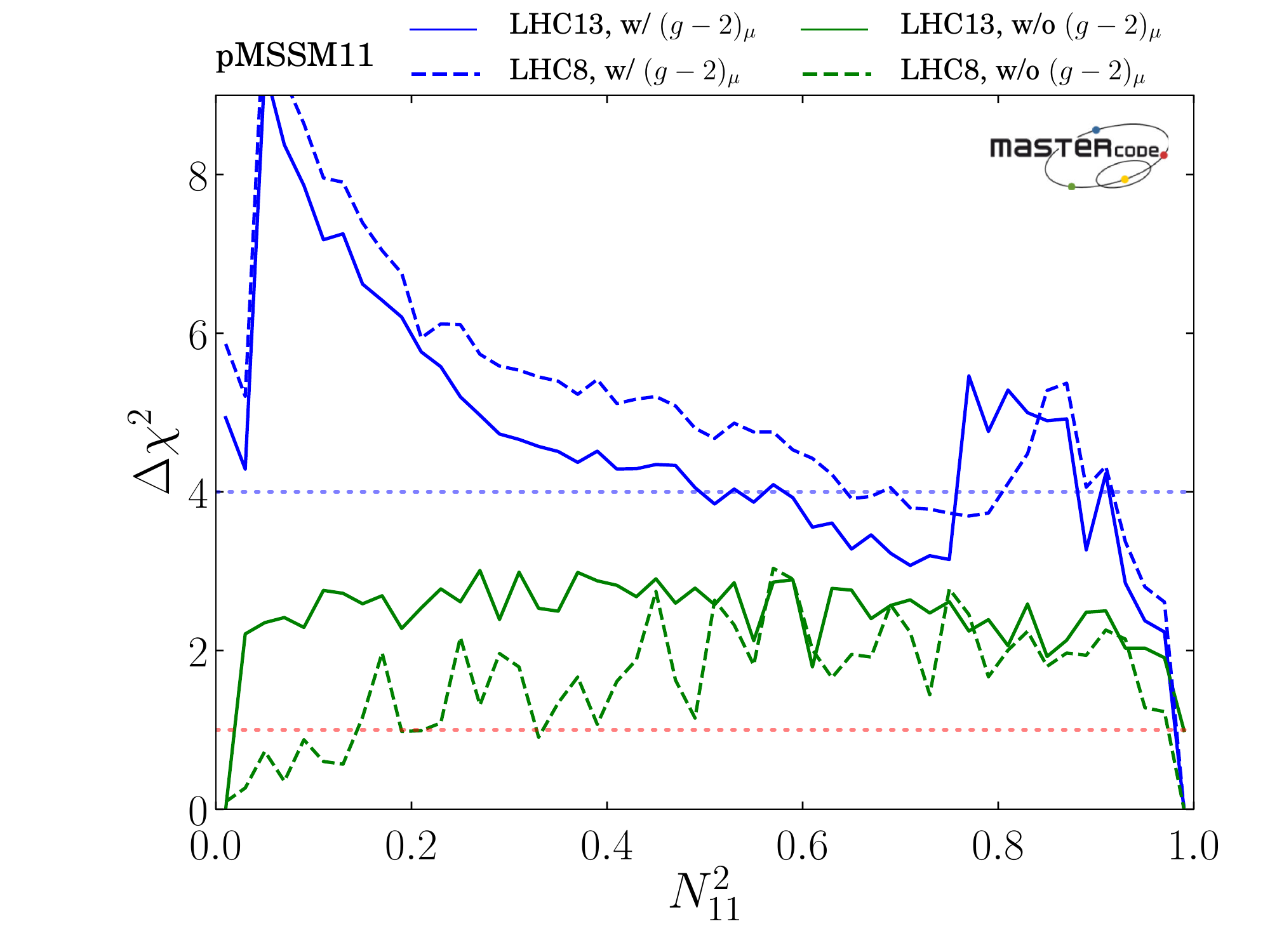}
\includegraphics[width=0.475\textwidth]{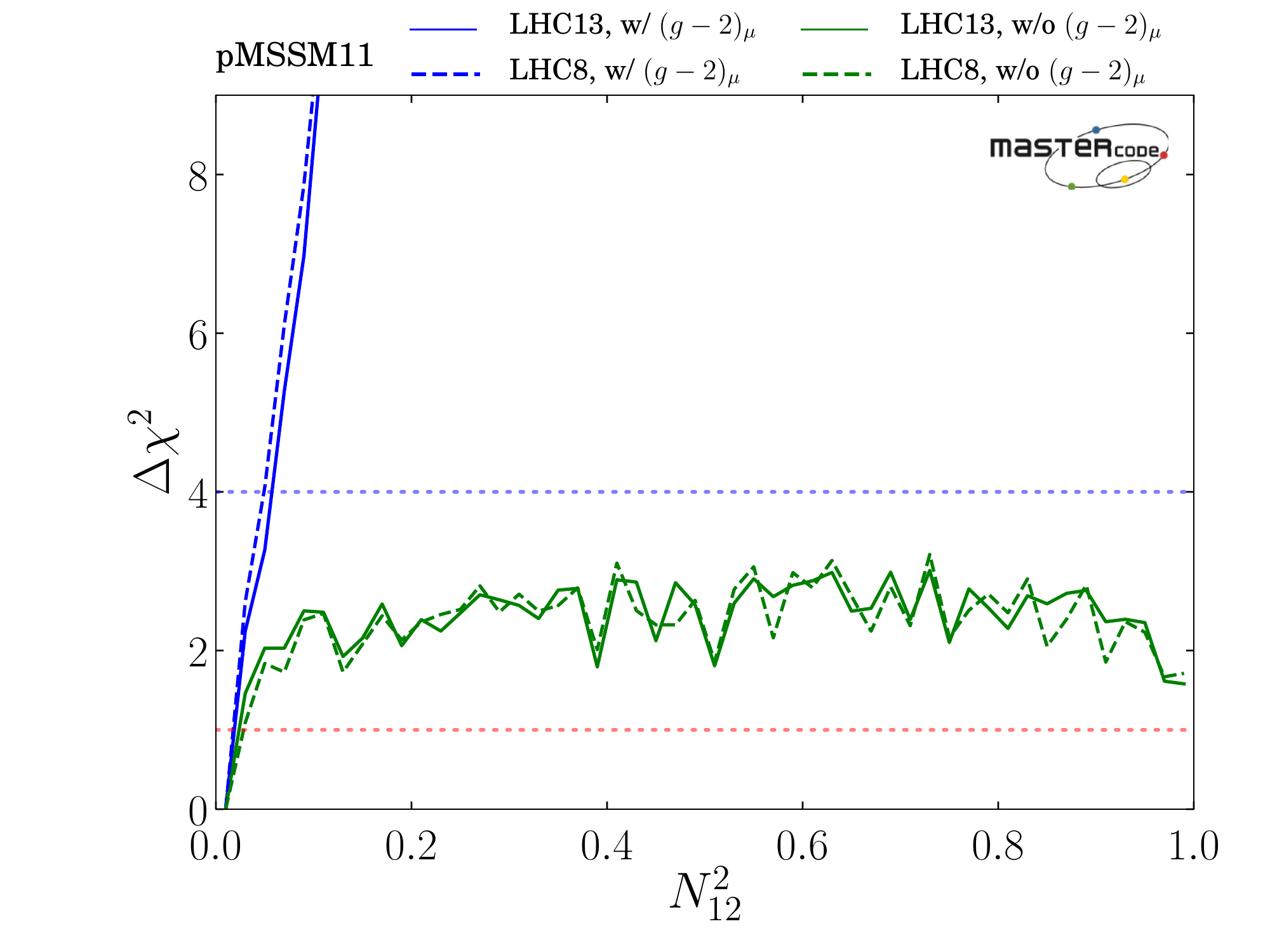} \\
\vspace{2cm}
\centering
\includegraphics[width=0.475\textwidth]{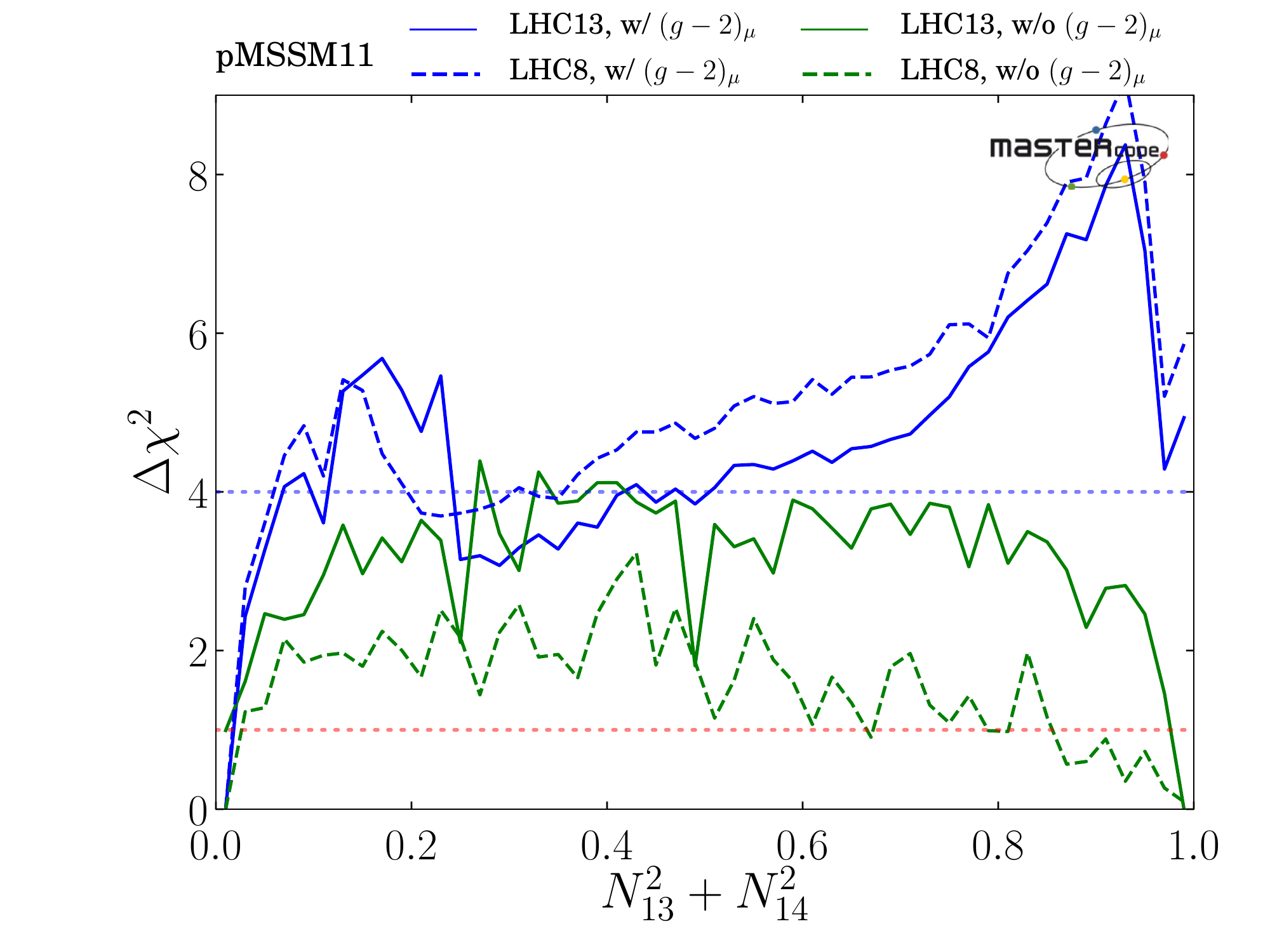}
\caption{\it One-dimensional likelihood plots for the $\tilde B$ fraction in the LSP $\neu1$
composition in the (upper left), for the $\tilde W^3$ fraction (upper right)
and for the $\tilde H_{u,d}$ fraction (lower panel).}
\label{fig:1hcompo}
\end{figure*}

{Fig.~\ref{fig:triangles} displays information about the preferred and
disfavoured $\neu1$ compositions in two triangular panels.
Both are for fits including LHC 13-TeV data (those dropping these data are quite similar), the left panel
includes the \gmt\ constraint and the right panel drops it. The $\Delta \chi^2$ for the best-fit points
at each location in the triangles are colour-coded as indicated. We see in the left panel that in the case with \gmt\ a small Wino
fraction $N_{12}^2 < 0.1$ is strongly favoured, while the relative proportions of the Bino fraction
$N_{11}^2$ and the Higgsino fraction $N_{13}^2 + N_{14}^2$ are relatively unconstrained at the 95\% CL.
On the other hand, the right panel shows that almost all binary combinations of Bino, Wino and Higgsino
(along the edges of the triangle) are allowed at the 95\% CL, but three-way mixtures
(in the interior of the triangle) are strongly disfavoured.}

\begin{figure*}[h]
\centering
\includegraphics[width=0.475\textwidth]{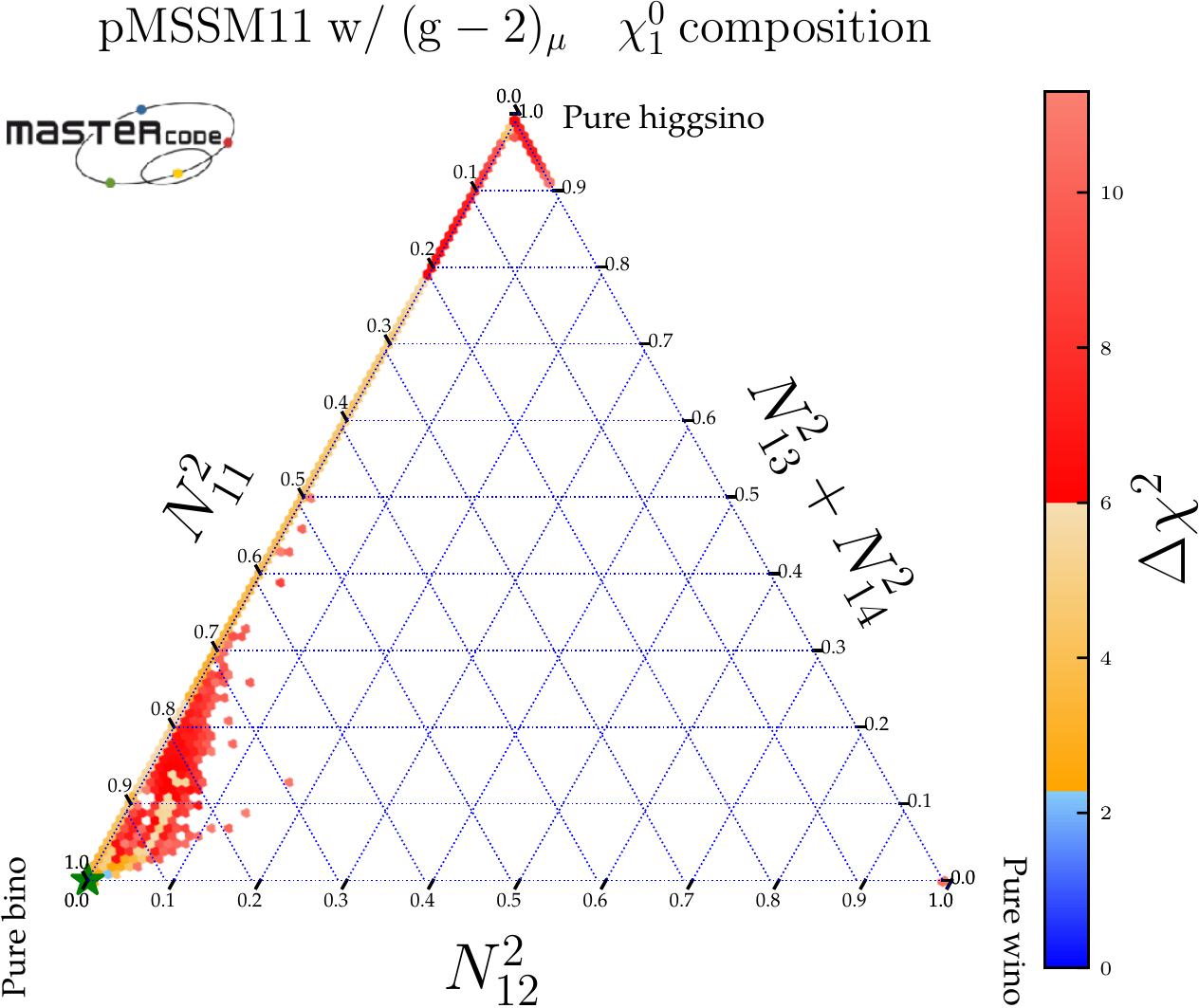}
\includegraphics[width=0.475\textwidth]{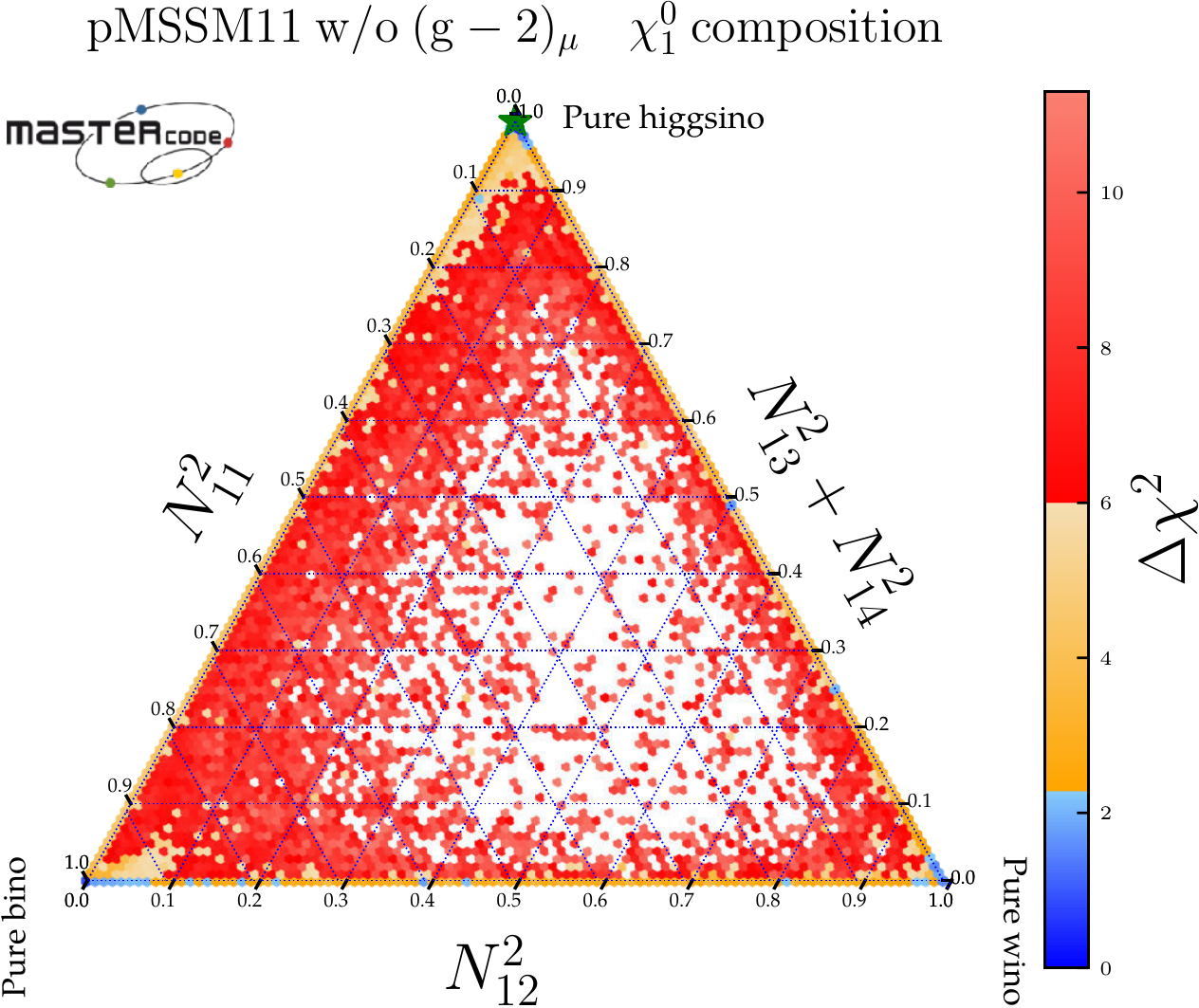} \\
\caption{\it {Triangular presentations of the composition of the $\neu1$ in the fit
with LHC 13-TeV and with (without) the \gmt\ constraint in the left (right) panel.}}
\label{fig:triangles}
\end{figure*}

Table~\ref{tab:lsp} compares the composition of the LSP $\neu1$ found at the best-fit points in our present pMSSM11 analysis
based on LHC 13-TeV data
(with and without the \gmt\ constraint) with the composition at the best-fit point from our previous pMSSM10 analysis
that also applied the \gmt\ constraint~\protect\cite{mc11}.
We see that both the pMSSM11 and pMSSM10 analyses with \gmt\ prefer an almost pure $\tilde B$ composition.
On the other hand, when the \gmt\ constraint is dropped
the pMSSM11 analysis prefers an almost equal mixture of $\tilde H_u$ and $\tilde H_d$ components {with a small
admixture of $\tilde B$ and again a very small admixture of $\tilde W_3$
{because we only scan $m_{\tilde \ell}$ {and $m_{\tilde \tau}$},  hence $\mneu1 < 2 \tev$}.}
{Table~\ref{tab:lsp} also displays the composition of the second-lightest neutralino, \neu2,
and we see that its content is mainly $\tilde W_3$ in the fit to the pMSSM11 with \gmt\
and in the pMSSM10 fit, but is mainly Higgsino in the  fit to the pMSSM11 without \gmt.}

\begin{table*}[!h]
  \centering
  \def\arraystretch{1.5}%
\begin{tabular}{|c|c|cccc|}
\hline
~Model & State & $\tilde B$ & $\tilde W_3$ & $\tilde H_u$ & $\tilde H_d$ \\
\hline
  ~pMSSM11~(with \gmt)& $\tilde{\chi}^0_1$ & 0.99 & -0.03 & 0.04 & -0.01\\
                      & $\tilde{\chi}^0_2$ & 0.03 & 0.99 & -0.06 & -0.01\\
\hline
  ~pMSSM11~(w/o \gmt) & $\tilde{\chi}^0_1$ & 0.01 & 0.04 & 0.71 & 0.70\\
                      & $\tilde{\chi}^0_2$ & 0.09 & 0.02 & -0.70 & -0.70\\
\hline
  ~pMSSM10 & $\tilde{\chi}^0_1$ & 0.99 & -0.11 & 0.09 & -0.04\\
           & $\tilde{\chi}^0_2$ & 0.12 & 0.98 & -0.13 & 0.05\\
\hline
\end{tabular}
\caption{\it The {amplitudes characterizing the decomposition of the LSP $\neu1$ and of the $\neu2$
into interaction eigenstates} at the best-fit points
  in our present pMSSM11 analysis {including LHC 13-TeV data}, with and without the \gmt constraint,
  compared with the composition at the best-fit point
  found in our previous pMSSM10 analysis that also included the \gmt\ constraint,
  {but only LHC 8-TeV data}~\protect\cite{mc11}.}
  \label{tab:lsp}
\end{table*}


\subsection{\boldmath{$B$}-Physics Observables}
\label{sec:flavour}

{\reffi{fig:1dBsmmBsg} displays the one-dimensional profile likelihood functions for $\bsdmm$ in the pMSSM11 (left panel)
and the \bsg\ branching ratio (right panel), with and without the LHC 13-TeV
data and the \gmt\ constraint. {We see in the
left panel that a value of $\bsdmm$ close to the SM value is preferred if both these constraints are applied,
though deviations at the level of $\pm \sim 10$\%
are allowed at the level of $\Delta \chi^2 = 4$ (2\,$\sigma$), corresponding to the 95\% CL.
On the other hand, if \gmt\ is dropped, a larger range of $\bsdmm$ is allowed,
with a larger deviation at the level of $\pm \sim 30$\%
becoming allowed at the level of $\Delta \chi^2 = 4$.
In particular, when the LHC13 data are included but \gmt\ is dropped, the global $\chi^2$ function
is minimized at a value of \bsdmm\ below the SM value, as hinted by the present experimental data,
with the SM value being mildly disfavoured by $\Delta \chi^2 \simeq 1$. It will be interesting to see
how measurements of \bsdmm\ evolve.}

The analogous curves for \bsg\ in the right panel of \reffi{fig:1dBsmmBsg} show preferences for values close the SM
predictions, with 2\,$\sigma$ ranges that are $\pm 20$\%. Discriminating between the SM and the pMSSM11 would
require significant reductions in both the theoretical and experimental uncertainties in \bsg.
}

\begin{figure*}[h]
\centering
\includegraphics[width=0.49\textwidth]{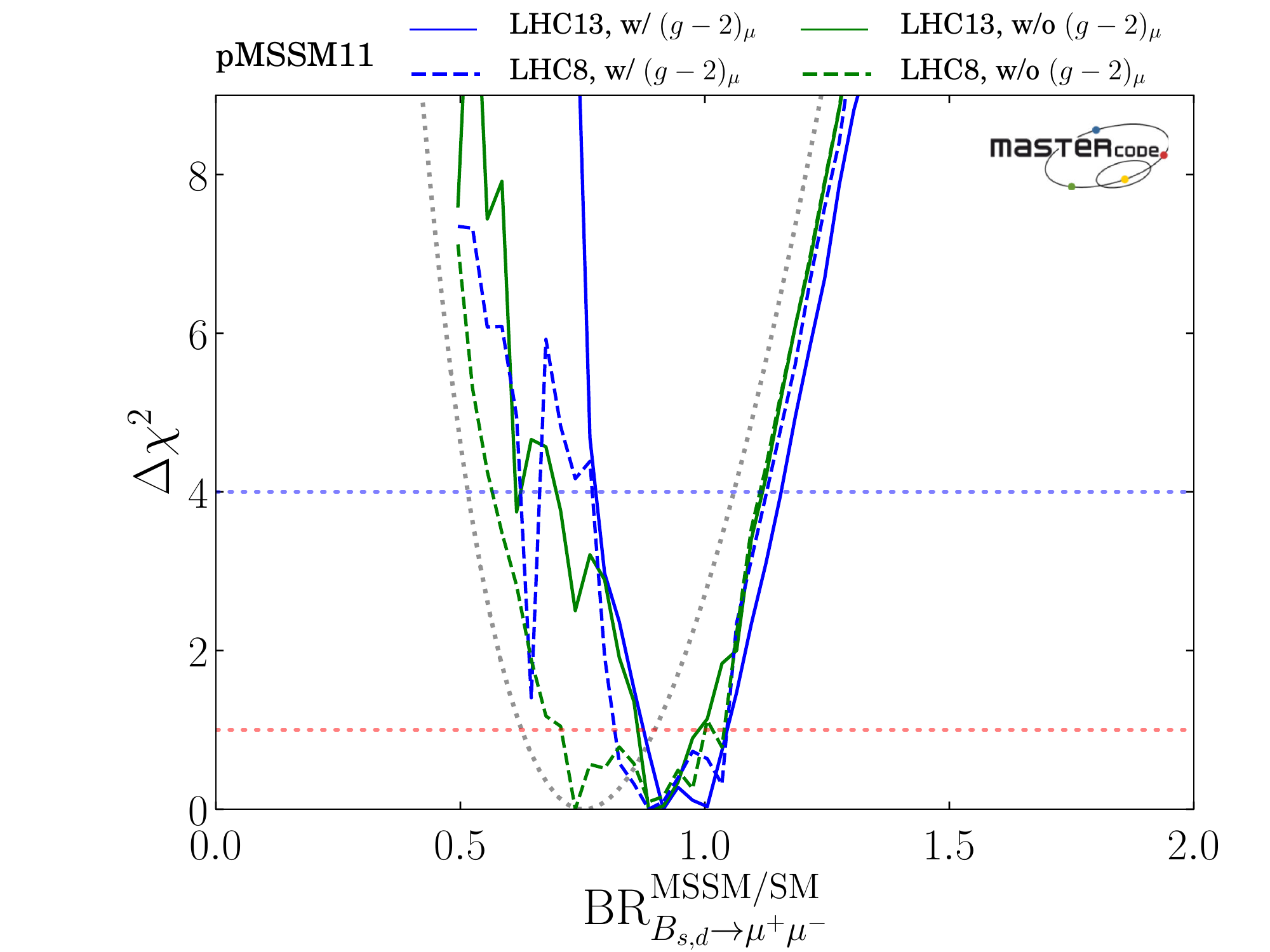}
\includegraphics[width=0.49\textwidth]{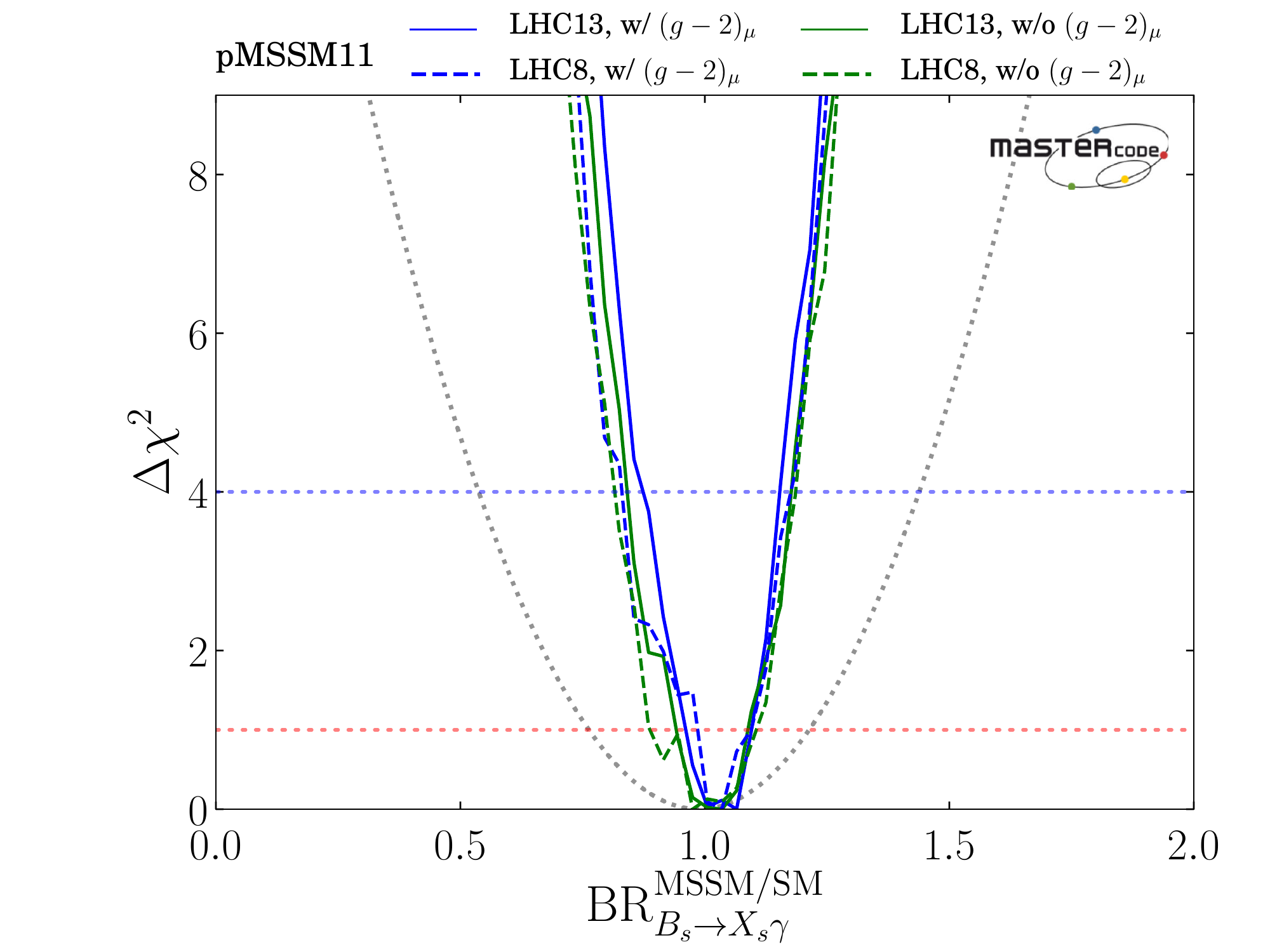} \\
\caption{\it One-dimensional profile likelihood functions for $\bsdmm$ in the pMSSM11 (left panel)
and the \bsg\ branching ratio (right panel), with and without the LHC 13-TeV
data and the \gmt\ constraint. {Also shown as dotted lines are the experimental constraints,
including the corresponding theoretical uncertainties within the Standard Model.}}
\label{fig:1dBsmmBsg}
\end{figure*}

As already mentioned in Section~\ref{sec:others}, the LHCb Collaboration has recently announced the first experimental
measurement of $\tau (B_s \to \mu^+ \mu^-)$, which is related to the quantity $A_{\Delta \Gamma}$
that takes the value $+ 1$ in the SM, but may be different in a SUSY model such as the pMSSM11.
\reffi{fig:ADG} displays the profile likelihood functions {for $A_{\Delta \Gamma}$ (left panel)
and $\tau(B_s \to \mu^+ \mu^-)/\tau_{B_s}$ (right panel), in our pMSSM11 fits with and without the LHC 13-TeV data and \gmt.
{We restrict our attention to positive values of $A_{\Delta \Gamma}$, corresponding to $\tau(B_s \to \mu^+ \mu^-)/\tau_{B_s} > 0.94$.}
We see that all the fits favour values of $A_{\Delta \Gamma}$ close to unity, with that dropping both the LHC 13-TeV data
and \gmt\ allowing the widest range. Values of $\tau(B_s \to \mu^+ \mu^-)/\tau_{B_s}$ close to unity
are also favoured, with $\Delta \chi^2 \gtrsim 9$ for $\tau(B_s \to \mu^+ \mu^-)/\tau_{B_s} = 0.94$.
The new LHCb measurement~\cite{1703.05747} does
not challenge any of these model predictions.}}

\begin{figure*}[h]
\centering
\includegraphics[width=0.495\textwidth]{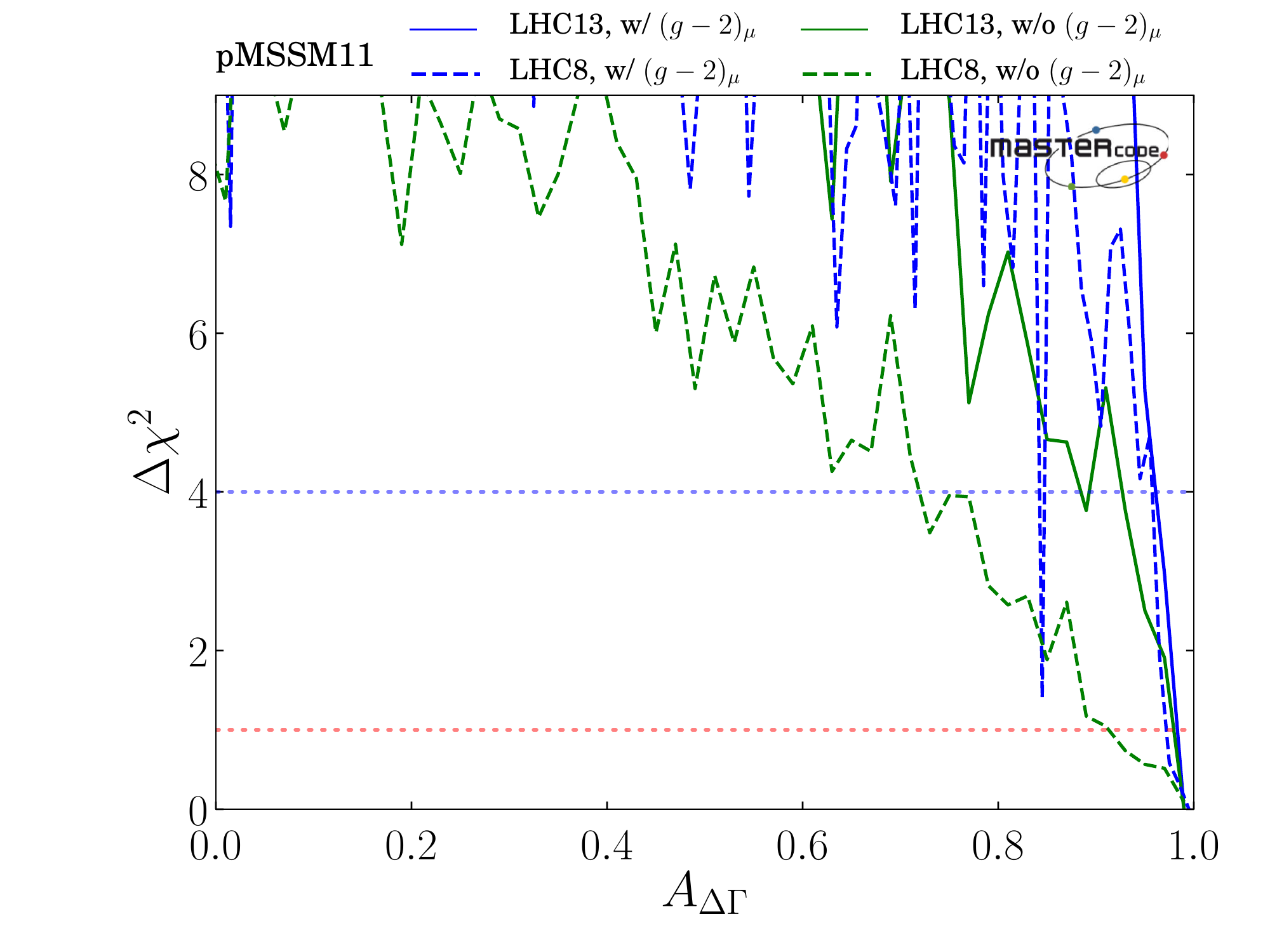}
\includegraphics[width=0.495\textwidth]{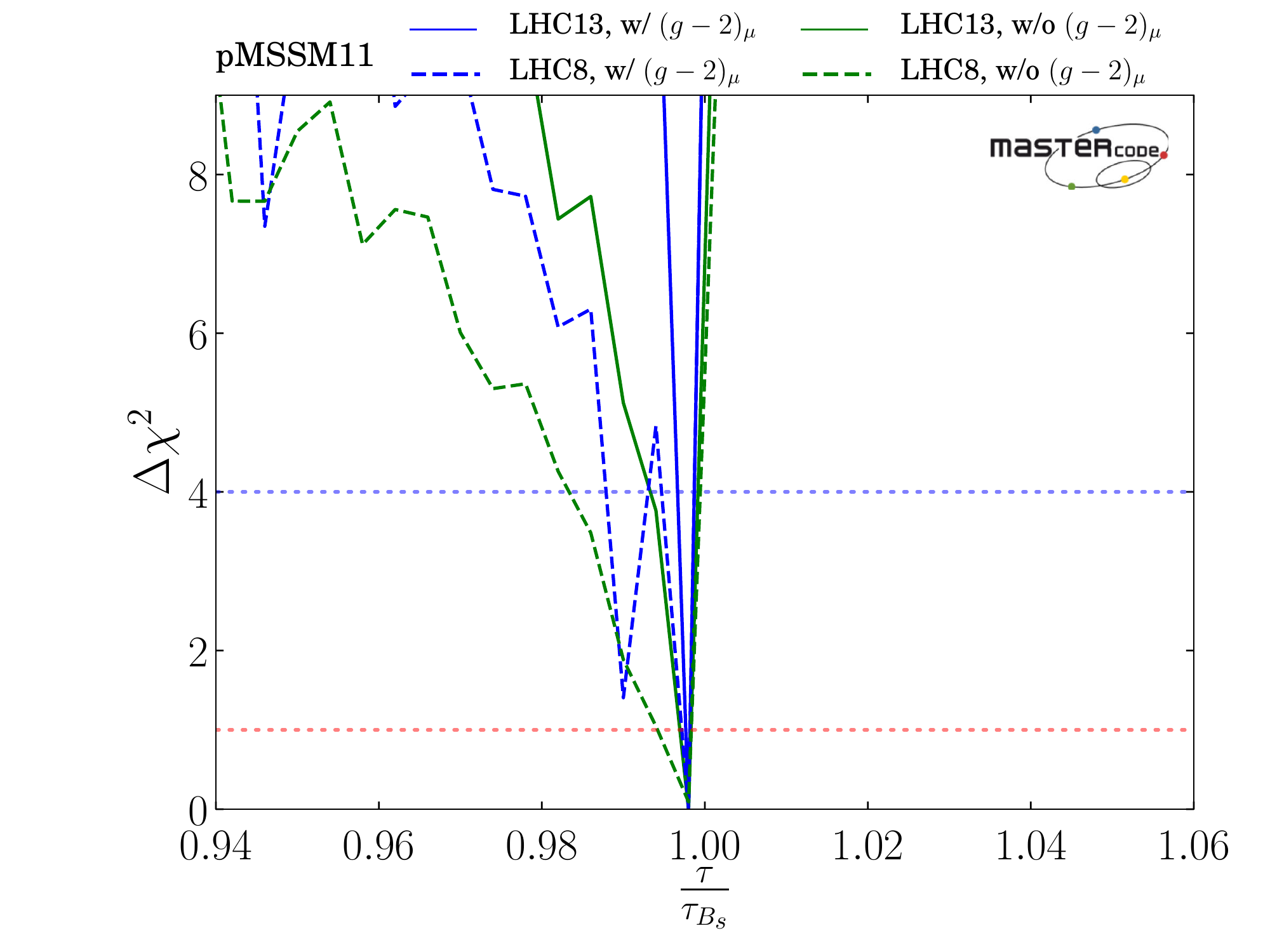} \\
\caption{\it One-dimensional $\chi^2$ profile likelihood functions for $A_{\Delta \Gamma}$ (left panel)
and $\tau(B_s \to \mu^+ \mu^-)/\tau_{B_s}$ (right panel), {in the fits with and without the LHC 13-TeV data and \gmt.}
}
\label{fig:ADG}
\end{figure*}


\subsection{Higgs Observables}
\label{sec:Higgs}

\begin{figure*}[htbp!]
\centering
\includegraphics[width=0.49\textwidth]{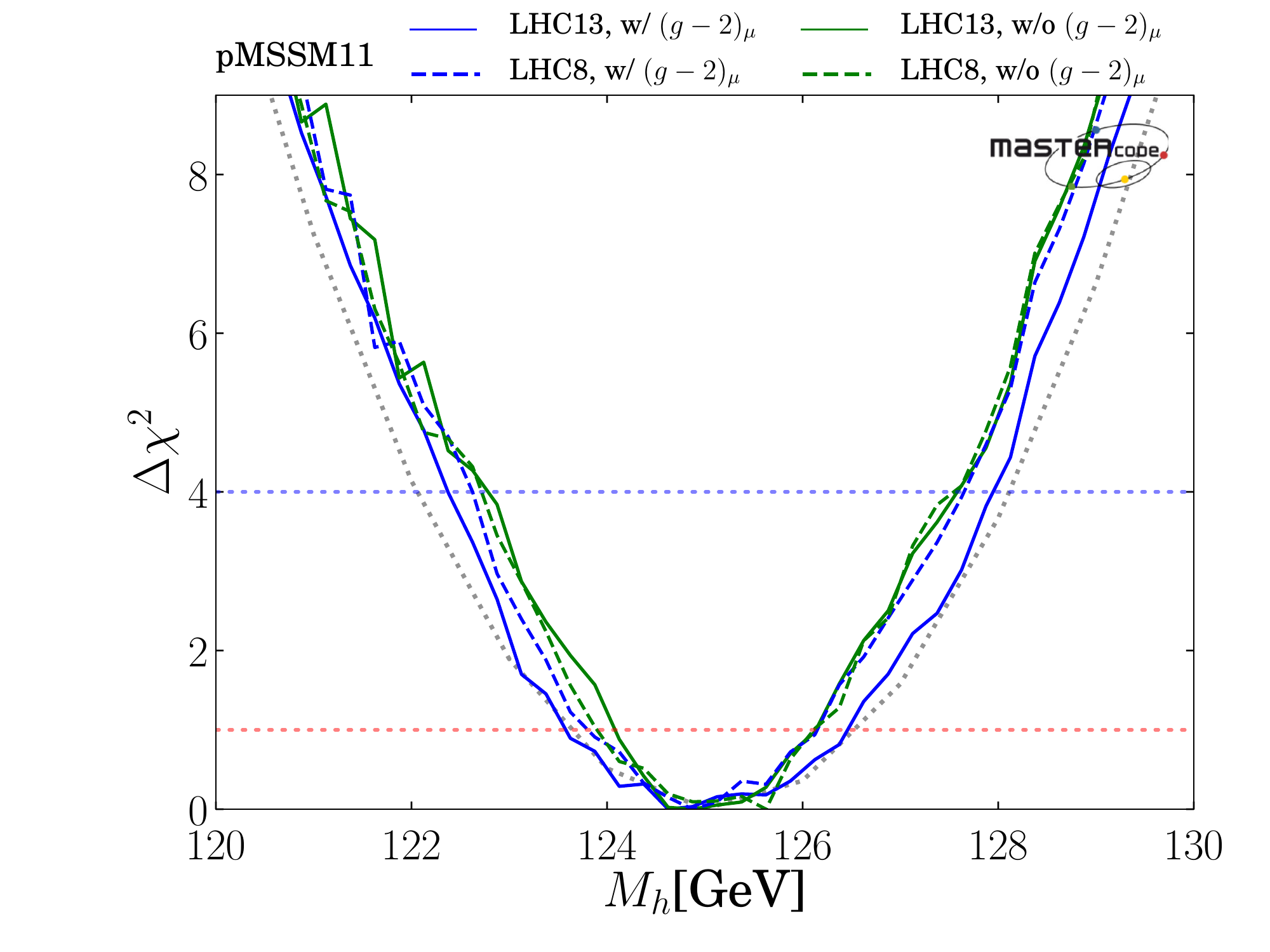}
\includegraphics[width=0.49\textwidth]{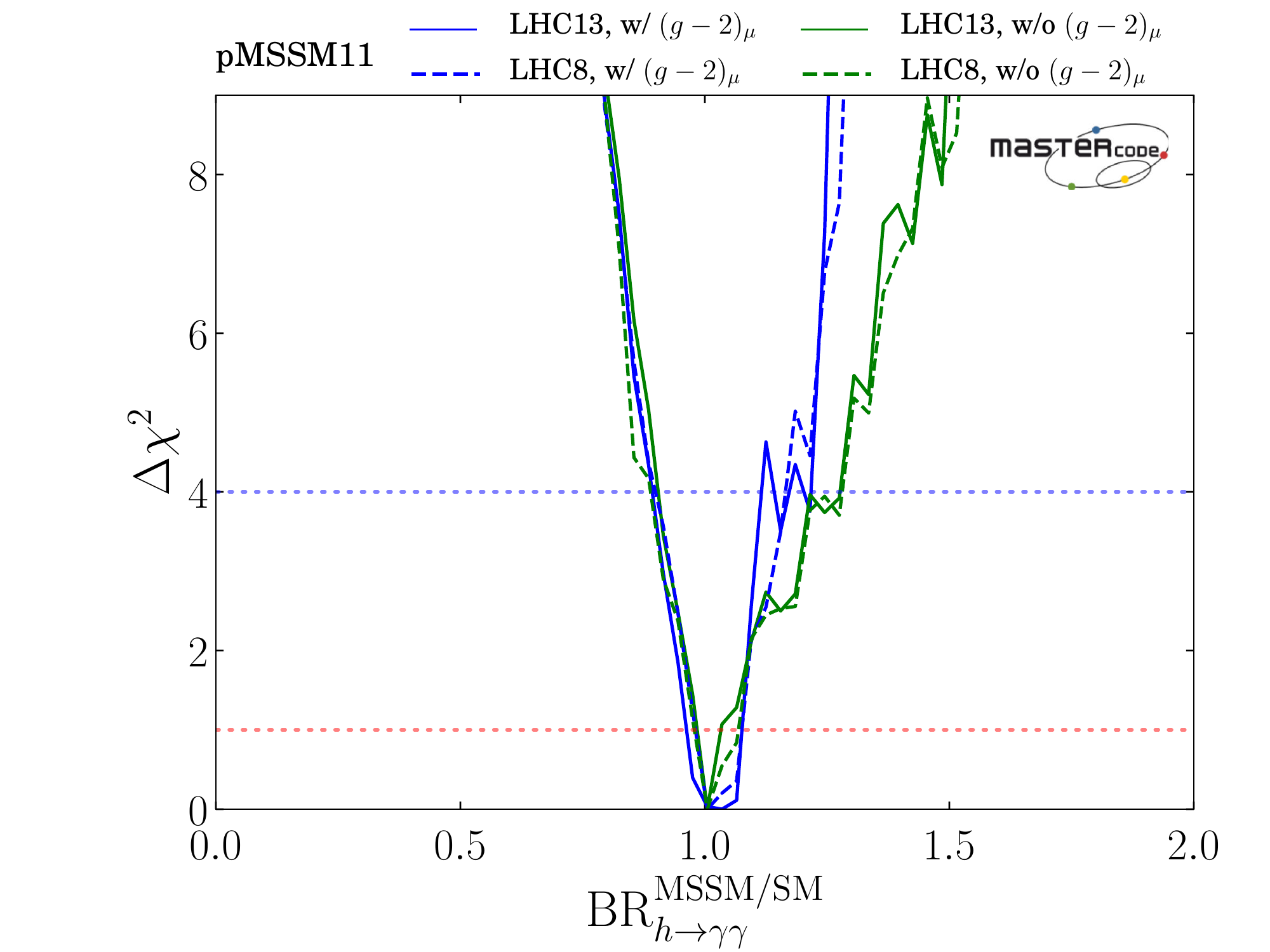} \\
\vspace{2cm}
\centering
\includegraphics[width=0.49\textwidth]{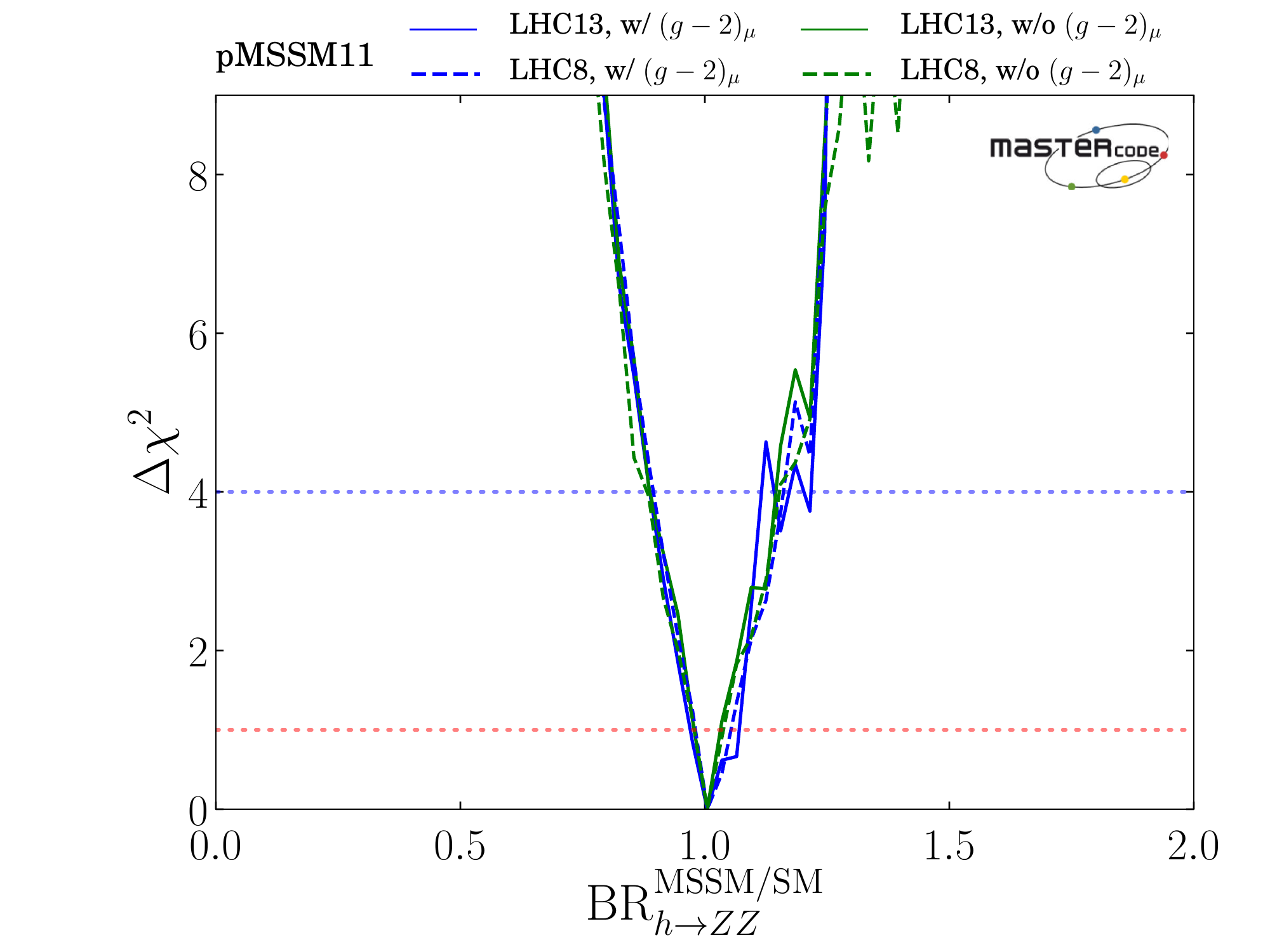}
\includegraphics[width=0.49\textwidth]{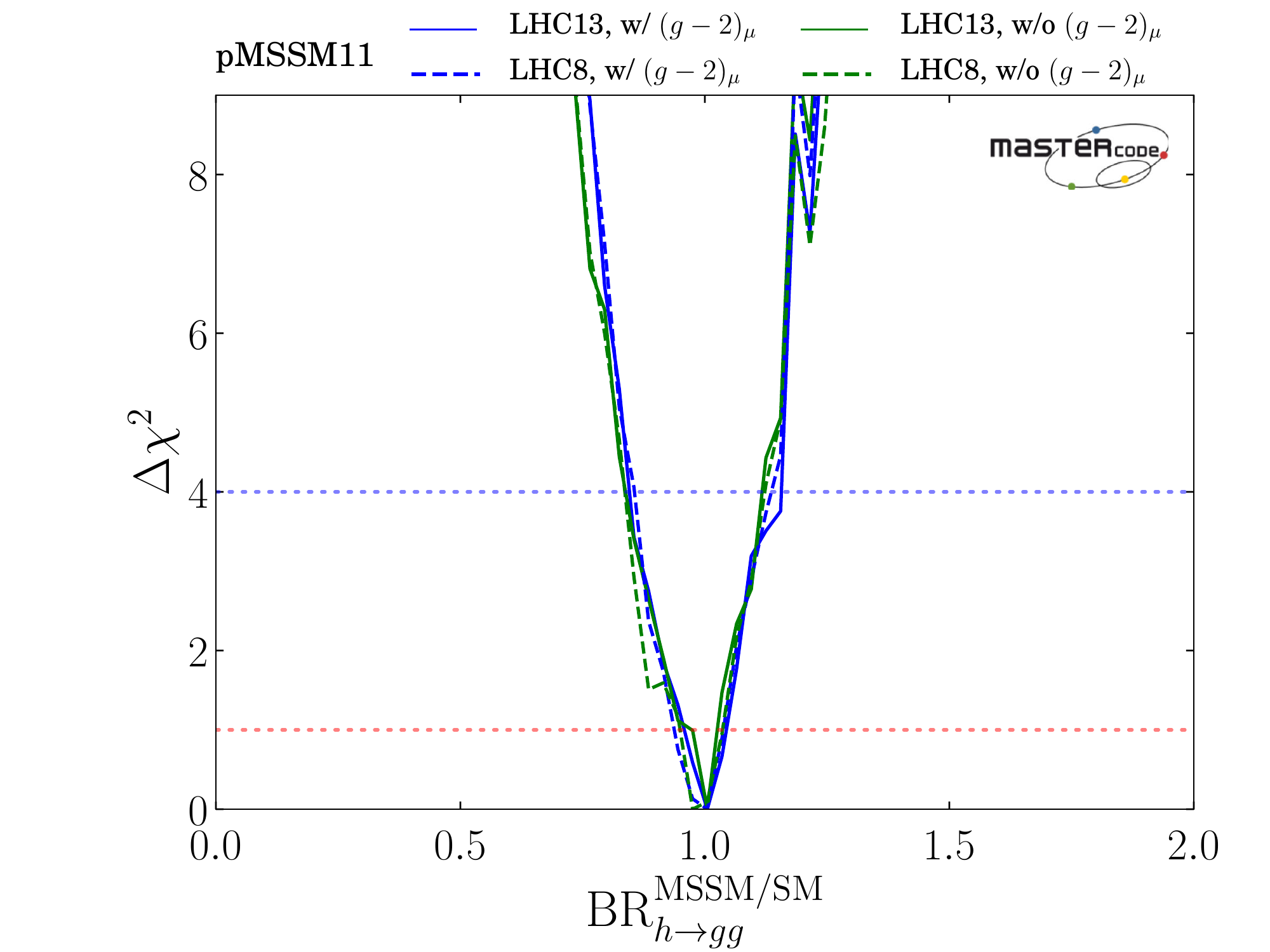}
\caption{\it One-dimensional profile likelihood plots for $\Mh$ (upper left panel),
for the $h \to \gamma \gamma$ branching ratio in the pMSSM11
relative to that in the SM with (upper right panel), for the $h \to Z Z^*$ branching ratio (lower left panel)
and for the $h \to gg$ branching ratio (lower right panel). {In the upper left panel we also show as
a dotted line the experimental constraint combined with
the corresponding theoretical uncertainty within the pMSSM11.}}
\label{fig:1hBR}
\end{figure*}

{\reffi{fig:1hBR} shows similar plots of {$\Mh$ (upper left panel),
and of the ratios of the branching ratios for $h \to \gamma
\gamma, Z Z^*$ and $h \to gg$ {(treated as a proxy for {$\sigma(gg \to h)$})} to their values in the SM
 in the upper right, lower left and lower right panels, respectively. Taking into account the theoretical uncertainties
 in the calculation of $\Mh$ in a supersymmetric model~\cite{FeynHiggs}, which we take to be $\pm 3 \gev$\footnote{We implement the constraint in the fit as a Gaussian likelihood-penalty with $\sigma = 1.5$ GeV, to avoid
   issues which would result from using a flat interval (due to the discontinuity in the p.d.f. at the interval extrema).
   }, there is no tension with
 the global fits. These also favour values of the decay branching ratios that are} similar
to those in the SM {whether \gmt\ is included in the fit, or not}, though with
uncertainties that are typically $\pm \sim 20$\%. As discussed in~\cite{mc-su5},
the global combination of ATLAS and CMS measurements using LHC Run-1 data has significantly larger uncertainties.
}

\subsection{Dark Matter Measures}
\label{sec:measures}

{In Section~2.4 we introduced various possible mechanisms for bringing the
relic $\neu1$ density into the range allowed by Planck and other data, proposing measures of their
prospective importance that we portrayed using different colours in the two-dimensional parameter planes shown in Section~3.
We emphasized there and in the subsequent discussions of one-dimensional profile likelihood functions earlier in Section~4 the
{roles} played by certain of these DM mechanisms. In this Subsection we display profile likelihood functions
for the most interesting of these DM measures, discussing the $\Delta \chi^2$ levels at which they
become relevant. As in the previous Sections, we compare results for
the analysis in which the \gmt\ constraint is applied with those when \gmt\ is discarded.}

\begin{figure*}[htbp!]
\vspace{-1cm}
\includegraphics[width=0.45\textwidth]{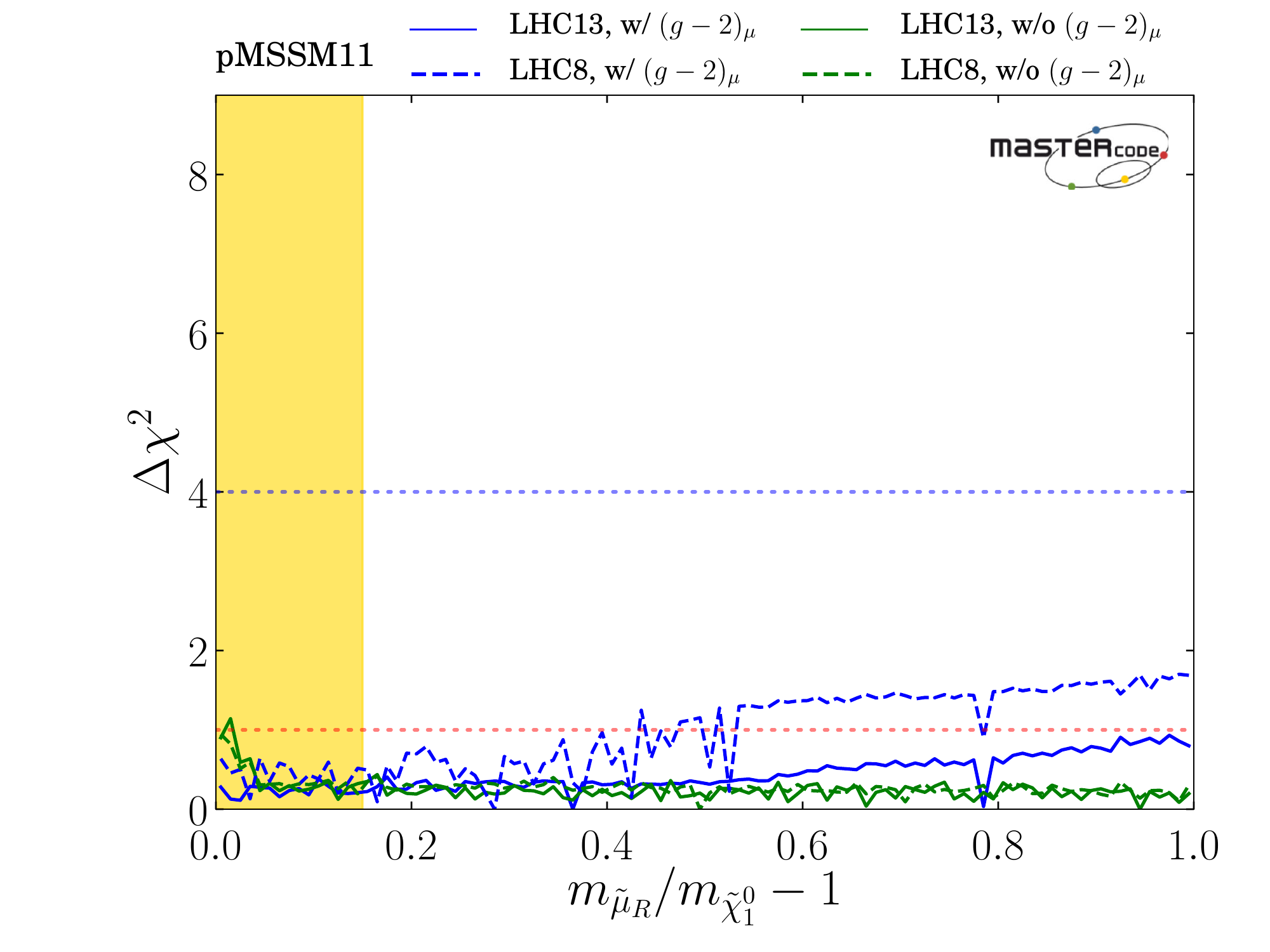}
\includegraphics[width=0.45\textwidth]{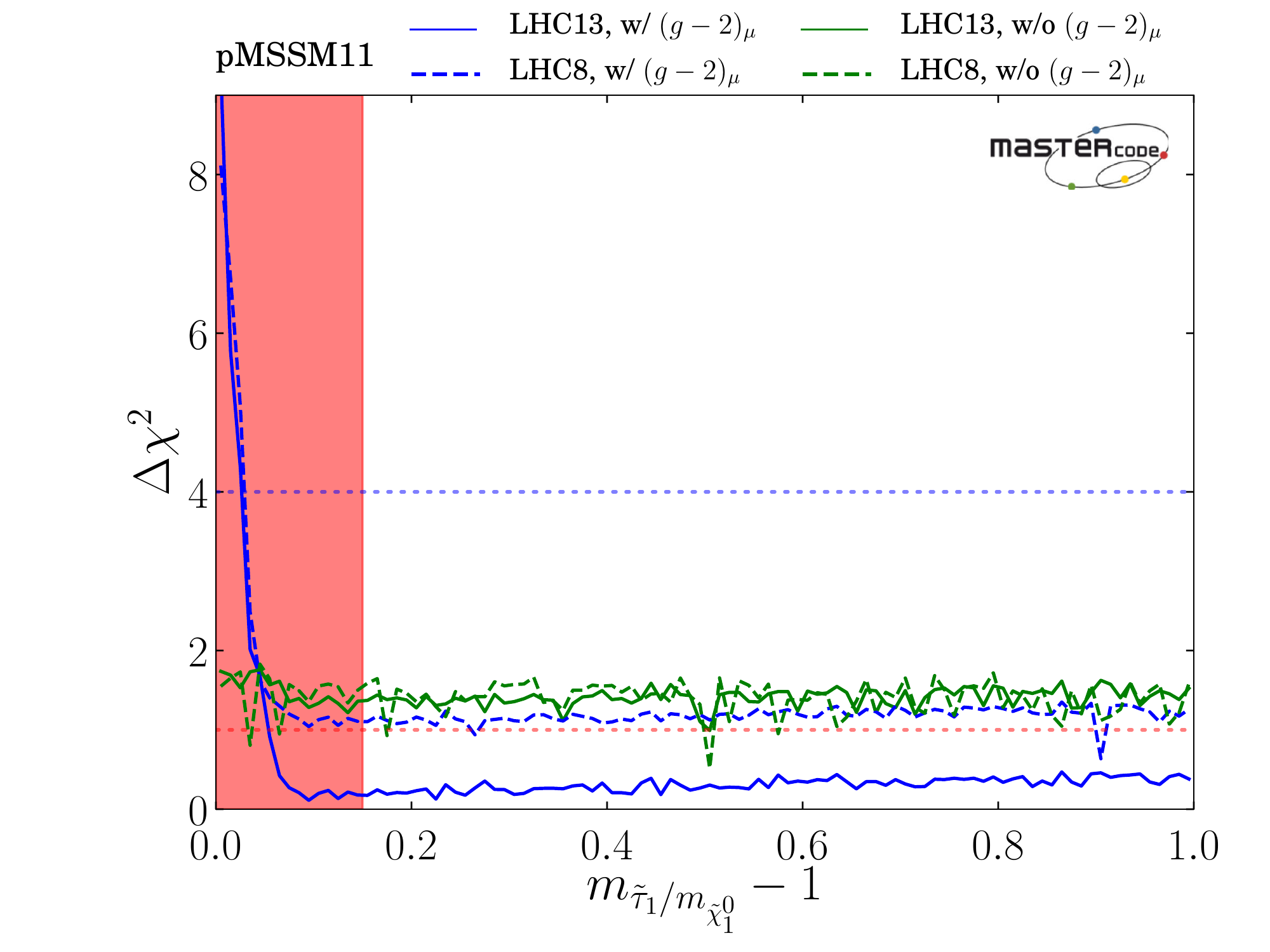}\\
\includegraphics[width=0.45\textwidth]{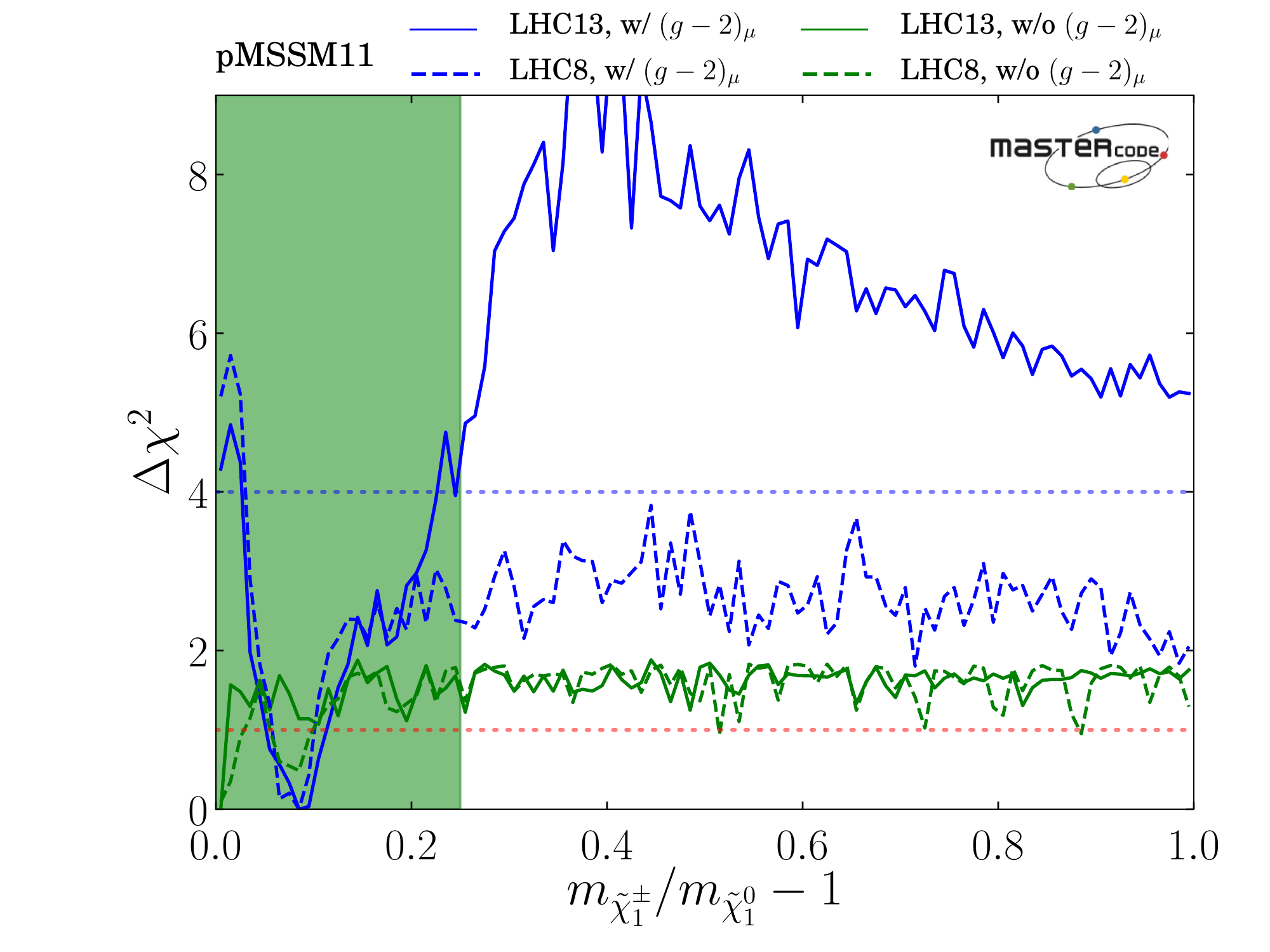}
\includegraphics[width=0.45\textwidth]{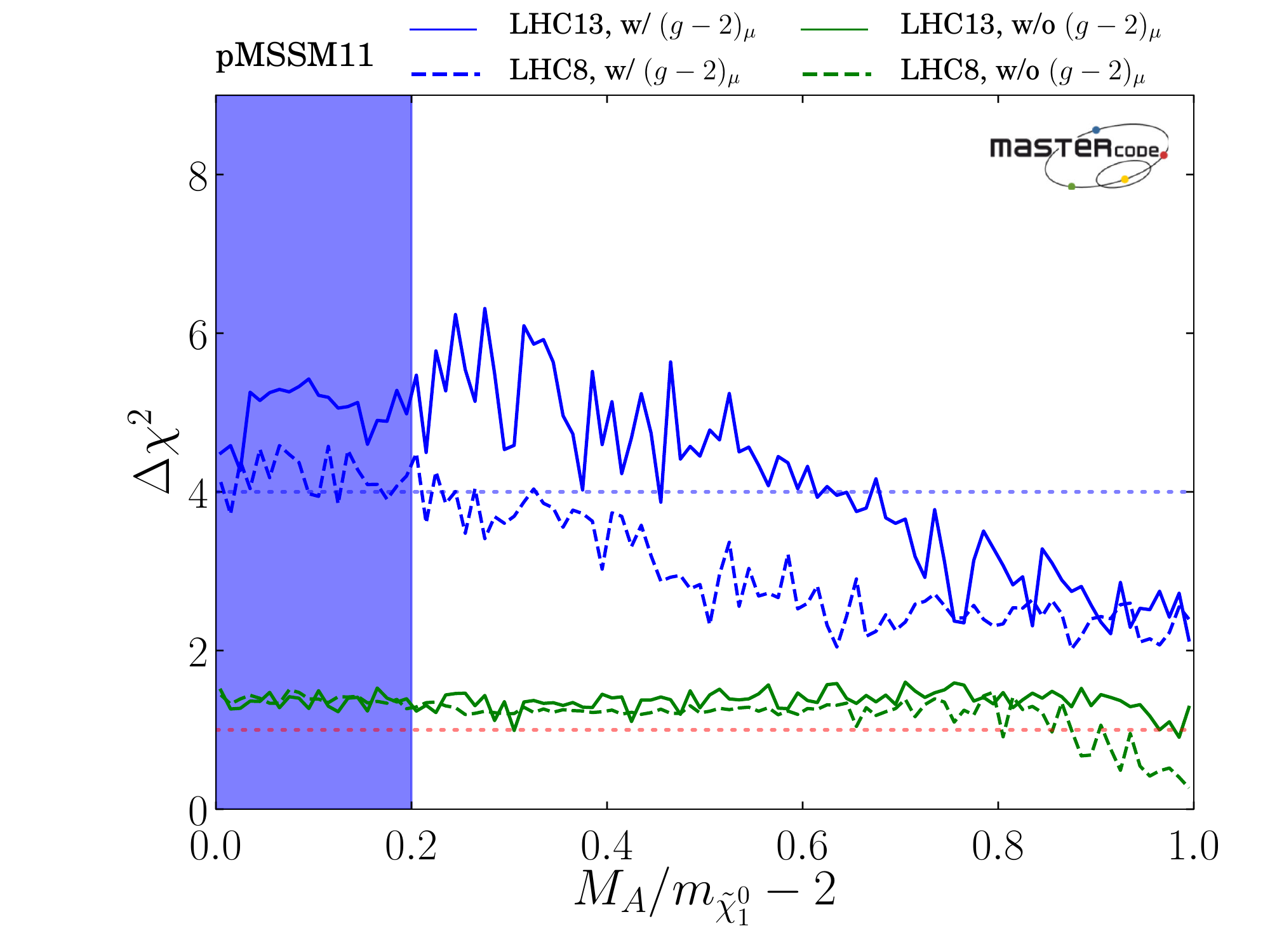} \\
\includegraphics[width=0.45\textwidth]{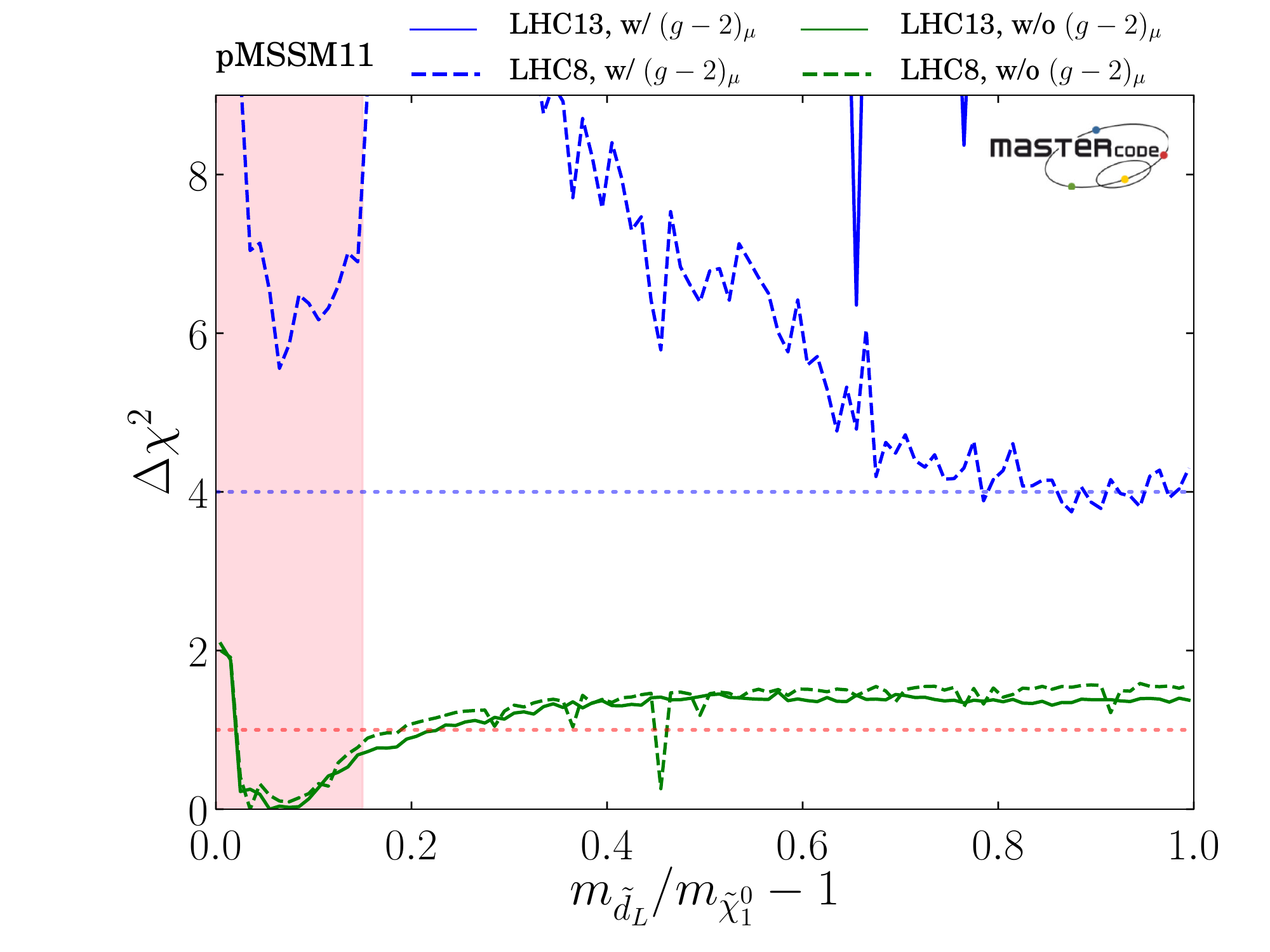}
\includegraphics[width=0.45\textwidth]{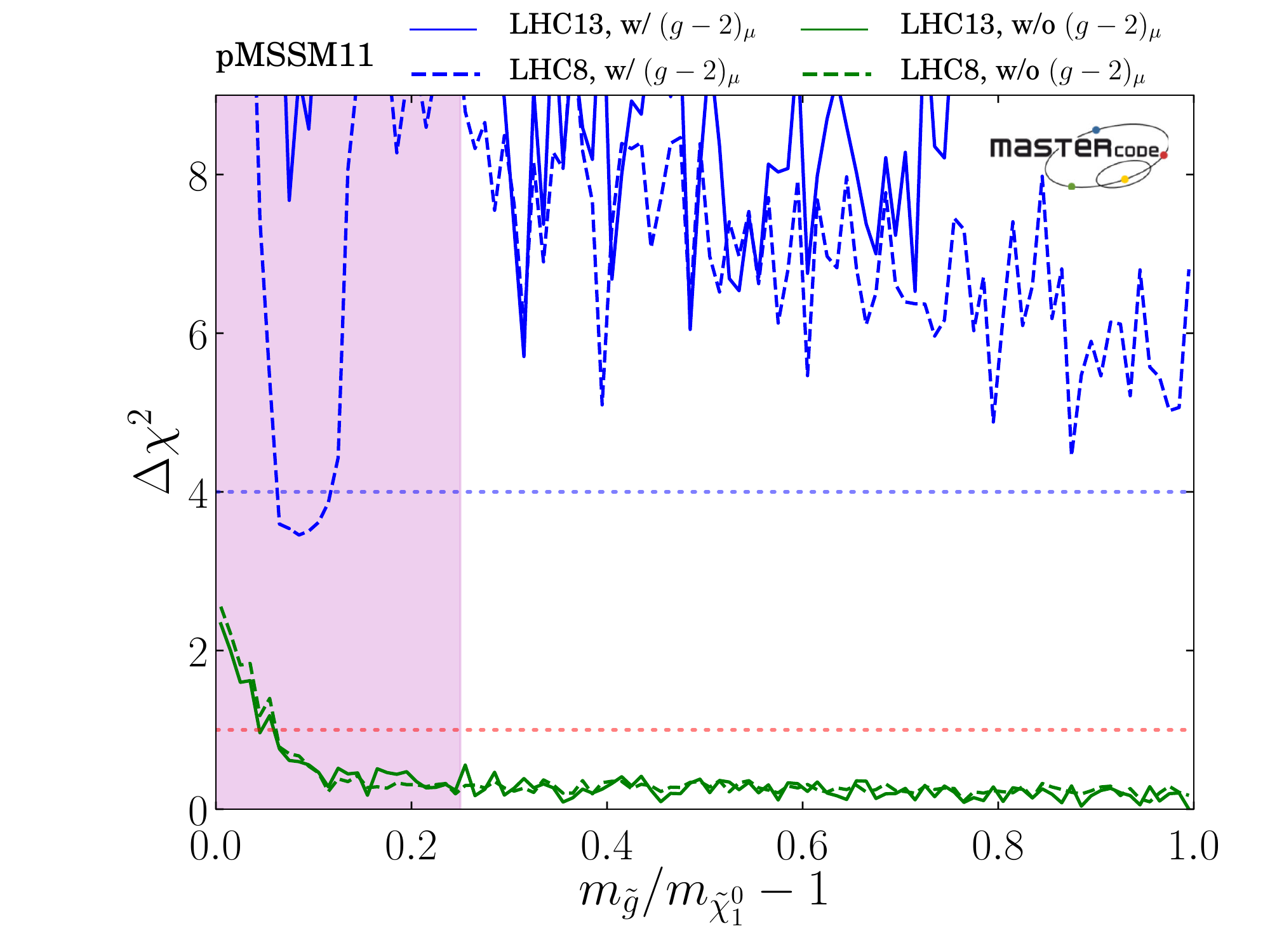} \\
\vspace{-0.5cm}
\caption{\it {One-dimensional profile likelihood plots for the measures of the prospective importance of
$\tilde \mu_R$ coannihilation (top left), $\tilde \tau_1$ coannihilation (top right),
$\cha1$ coannihilation (middle left), rapid annihilation via $A/H$ bosons (middle right),
$\tilde d_L$ coannihilation (bottom left) and gluino coannihilation (bottom right).
The vertical coloured bands correspond to the DM mechanism criteria introduced in Section~2.4.}
 }
\label{fig:EWmeasures}
\end{figure*}

{\reffi{fig:EWmeasures} displays the profile likelihood functions for the selected DM measures.
The top left panel shows the first- and second-generation slepton measure, and we see that
{$\Delta \chi^2$ is generally small throughout this region}.
The $\tilde \tau_1$  measure is shown in the top right panel, and we see that with \gmt\ included,
whether or not the LHC 13-TeV results are included,  the $\chi^2$ function
has a shallow minimum within the region where this mechanism may dominate (shown as the
vertical pink band), but very small values of the $\tilde \tau_1$ coannihilation measure are disfavoured,
and larger values of this measure also appear with a negligible likelihood price.
On the other hand, when \gmt\ is dropped we find that $\Delta \chi^2 \sim 2$ is almost independent
of $\mstaue/\mneu1$.}

{The $\chi^2$ function rises as $\mstaue/\mneu1 \to 1$
when \gmt\ is included, because this constraint prefers small values of $\mneu1$, for
which the relic density constraint cannot be satisfied when  $\mstaue/\mneu1 \to 1$.
However, since the first- and second-generation slepton masses are independent of $\mstaue$
in the pMSSM11 there is no such obstacle disfavouring $m_{\tilde \mu_R}/\mneu1 \to 1$. Therefore the
profile $\chi^2$ function for the first- and second-generation DM measure does not rise in this
limit, as seen in the top left panel of \reffi{fig:EWmeasures}.}

In the case of the $\cha1$ coannihilation measure shown in the middle left panel of \reffi{fig:EWmeasures},
we see that the best-fit pMSSM11 points lie within this shaded band, whether the LHC 13-TeV data
and/or \gmt\ are included or not.
In the case with \gmt, the best-fit point has $\mcha1/\mneu1 \sim 1.1$ whether the LHC 13-TeV data
are included or not, whereas when \gmt\ is dropped there is a
strong preference for $\mcha1/\mneu1$ close to unity, {which is possible in the case because the LSP
is Higgsino-like.} {As in the case of the $\staue$ DM measure,
the relic density constraint disfavours $\mcha1/\mneu1 \to 1$ when \gmt\ is included.
We find some parameter sets with $\mcha1-\mneu1 \lesssim 10 \mev$ that have $\Delta \chi^2 \gtrsim 4$,
which occur when $M_1$ is negative, near the border of a region where the LSP would be the $\cha1$.}

In the case of the $A/H$ measure shown in the middle right panel, we see that {$\Delta \chi^2 > 3$ in this region
when the \gmt\ and LHC 13-TeV constraints are both used. However, the $\chi^2$ price of rapid annihilation
through the $A/H$ poles is reduced if either of these constraints is dropped}.
Indeed, including the \gmt\ constraint forces the neutralino mass to be at most $\simeq 500$ GeV, in which case
the funnel condition implies an upper bound on $M_A \lesssim 1~\mathrm{TeV}$, well within the reach of LHC 13-TeV searches for $\tan\beta \gtrsim 15$.

The bottom left panel of \reffi{fig:EWmeasures} displays
the profile likelihood function for the squark coannihilation measure $m_{\tilde q_L} /\mneu1 - 1$.
{We see that before the LHC-13 data the best-fit point with \gmt\ included was in the
squark coannihilation region with $m_{\tilde q_L} /\mneu1 < 1.1$, though this feature was absent when
\gmt\ was dropped. Including the LHC 13-TeV data, the best-fit points with and without \gmt\ have
$m_{\tilde q_L} \gg \mneu1$, but there is still a vestige of the squark coannihilation region with
$\Delta \chi^2 < 4$ when \gmt\ is dropped. The reason for this is that lifting the \gmt\ constraint
  allows for a heavier neutralino, which in turn implies heavier squark masses still allowed by LHC-13 TeV data}
Finally, the bottom right panel of \reffi{fig:EWmeasures}
shows the gluino coannihilation measure, and we see that this may also play a role when $\Delta \chi^2 < 4$,
{unless both the LHC 13-TeV data and \gmt\ are included}.\\

\subsection{NLSP Lifetimes}
\label{sec:lifetimes}

\begin{figure*}[htb!]
\centering
\includegraphics[width=0.7\textwidth]{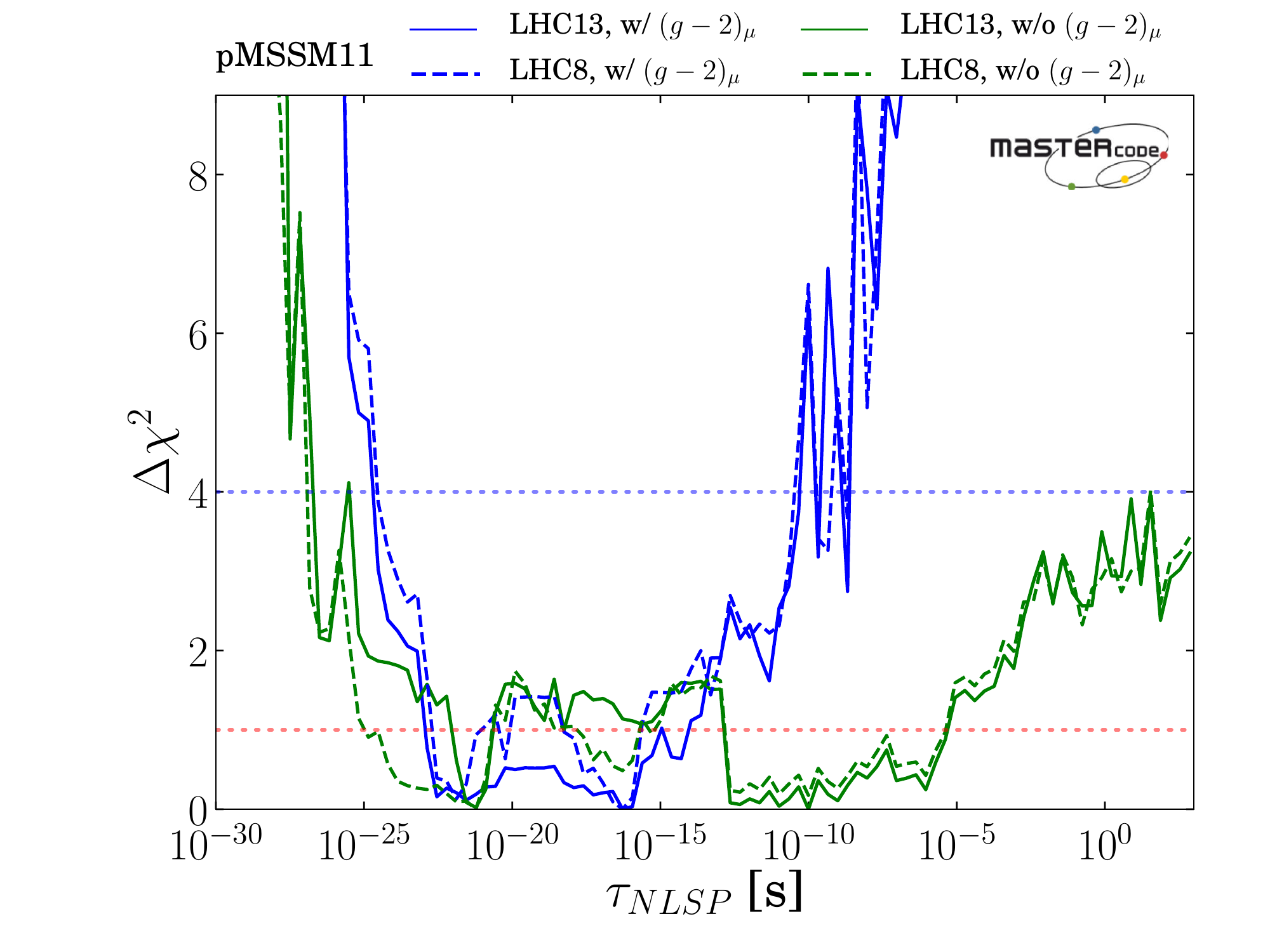}
\caption{\it {One-dimensional profile likelihood plot for the NLSP lifetime, $\tau_{\rm NLSP}$, including all possible NLSP species.}}
\label{fig:life}
\end{figure*}

{We display in Fig.~\ref{fig:life} the one-dimensional profile likelihood
for the NLSP lifetime, $\tau_{\rm NLSP}$, including all possible NLSP species. There is little
difference between the $\Delta \chi^2$ functions with \gmt, whether or not the LHC 13-TeV
data are included (blue curves). In both cases, we find that $\Delta \chi^2 \gtrsim 4$ for
$\tau_{\rm NLSP} \gtrsim 10^{-10}$~s. On the other hand, when the \gmt\ constraint is dropped
(green curves), we see that values of $\tau_{\rm NLSP} \lesssim 10^3$~s are allowed at the
$\Delta \chi^2 \lesssim 4$ level, again whether or not the LHC 13-TeV data are included (green curves).
As already mentioned, we exclude from our scan parameter sets with NLSP lifetimes exceeding $10^3$~s, as they could alter the
successful predictions of standard Big Bang nucleosynthesis~\cite{BBN}.}

\begin{figure*}[htbp!]
  \centering
  \includegraphics[width=0.5\textwidth]{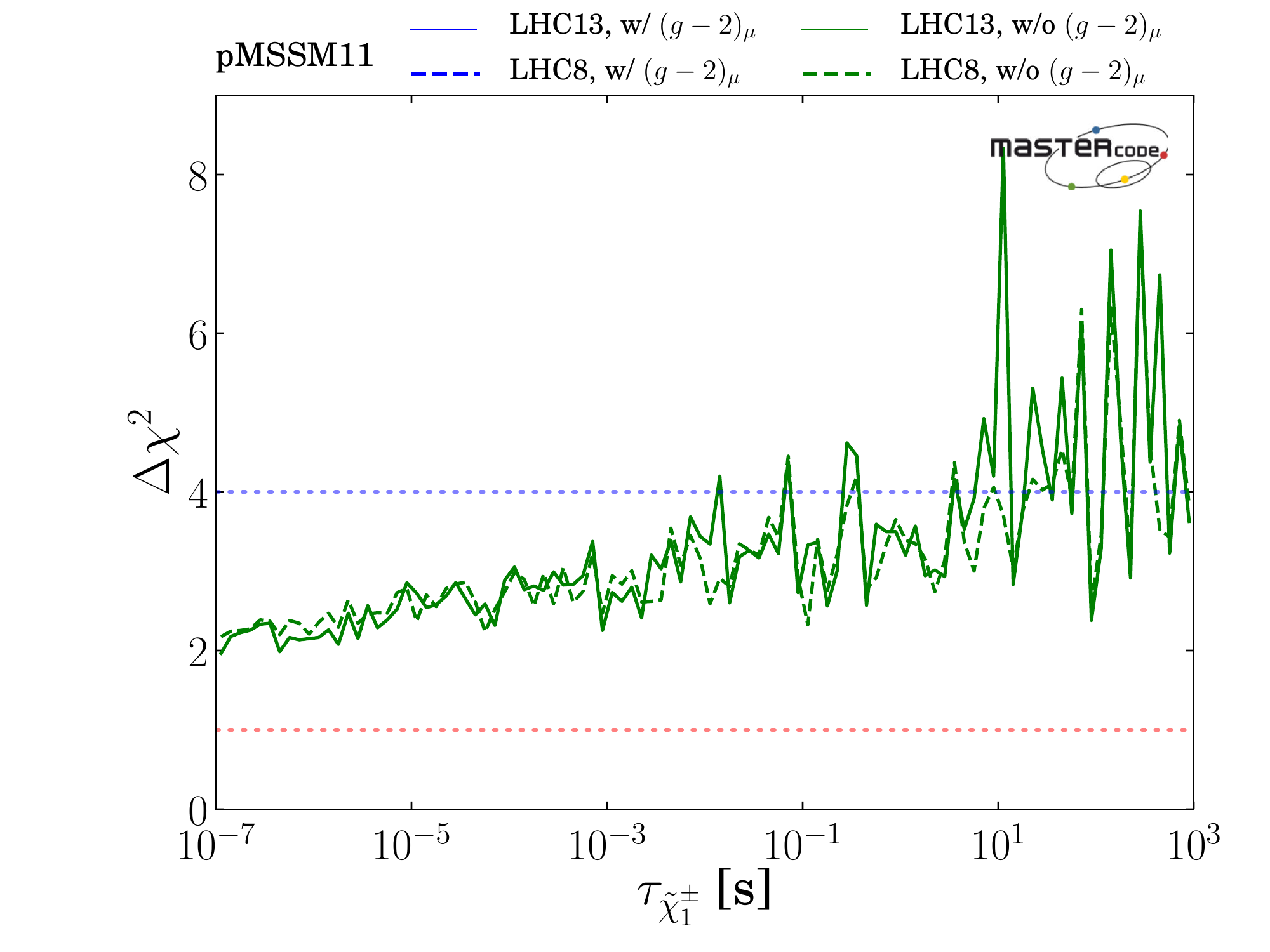}~\includegraphics[width=0.5\textwidth]{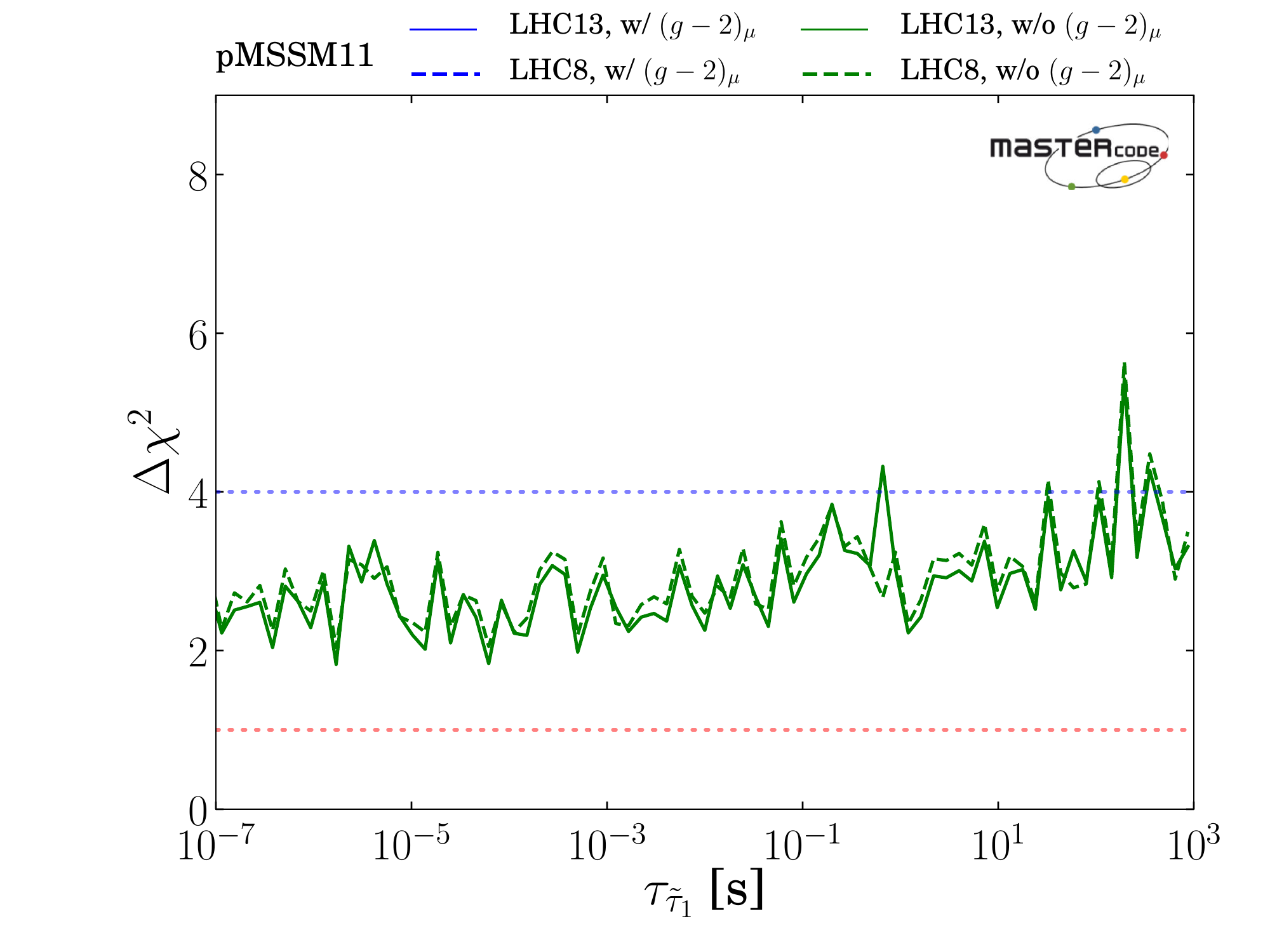}\\
  \vspace{2cm}
  \centering
 \includegraphics[width=0.5\textwidth]{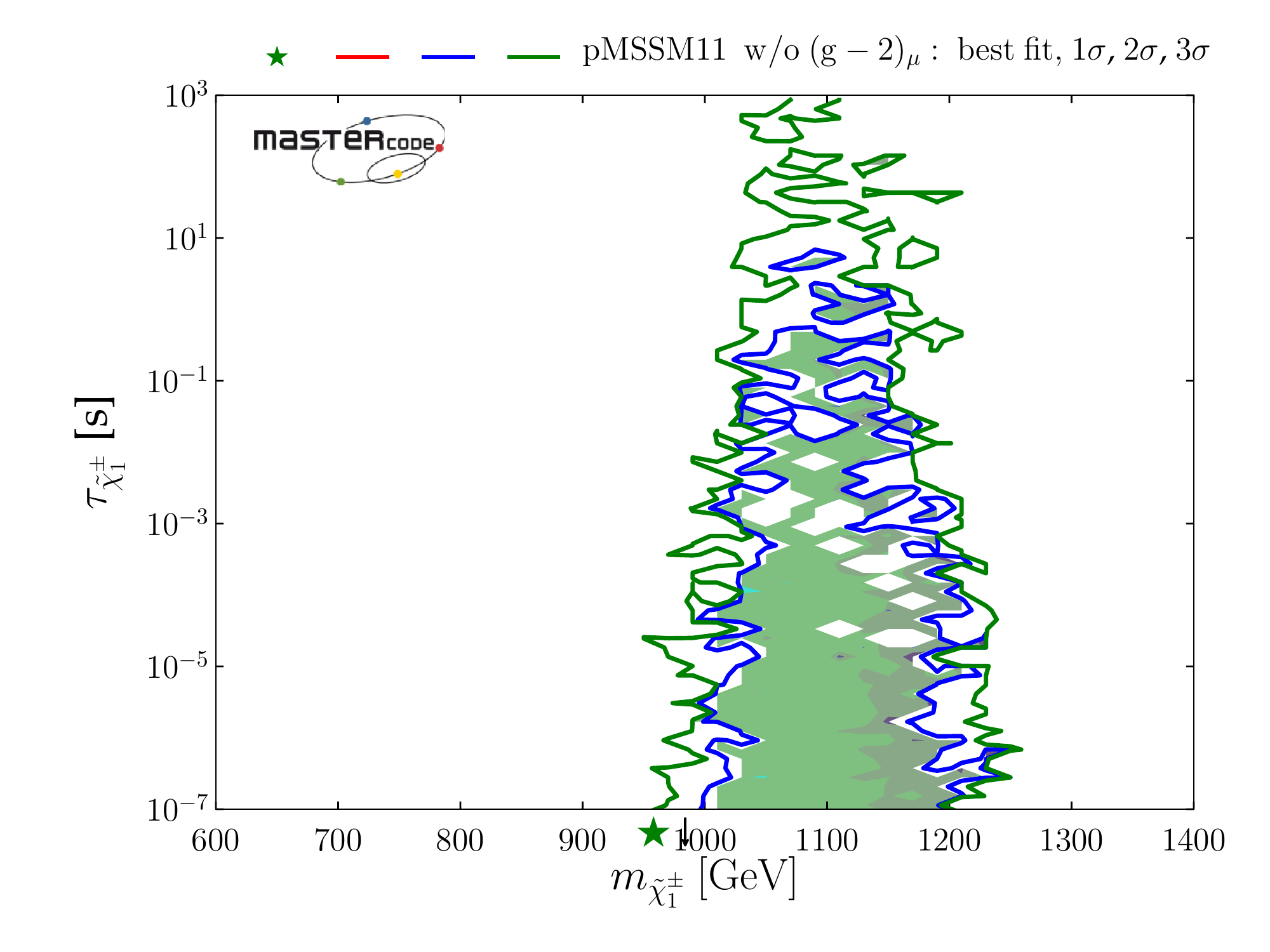}~\includegraphics[width=0.5\textwidth]{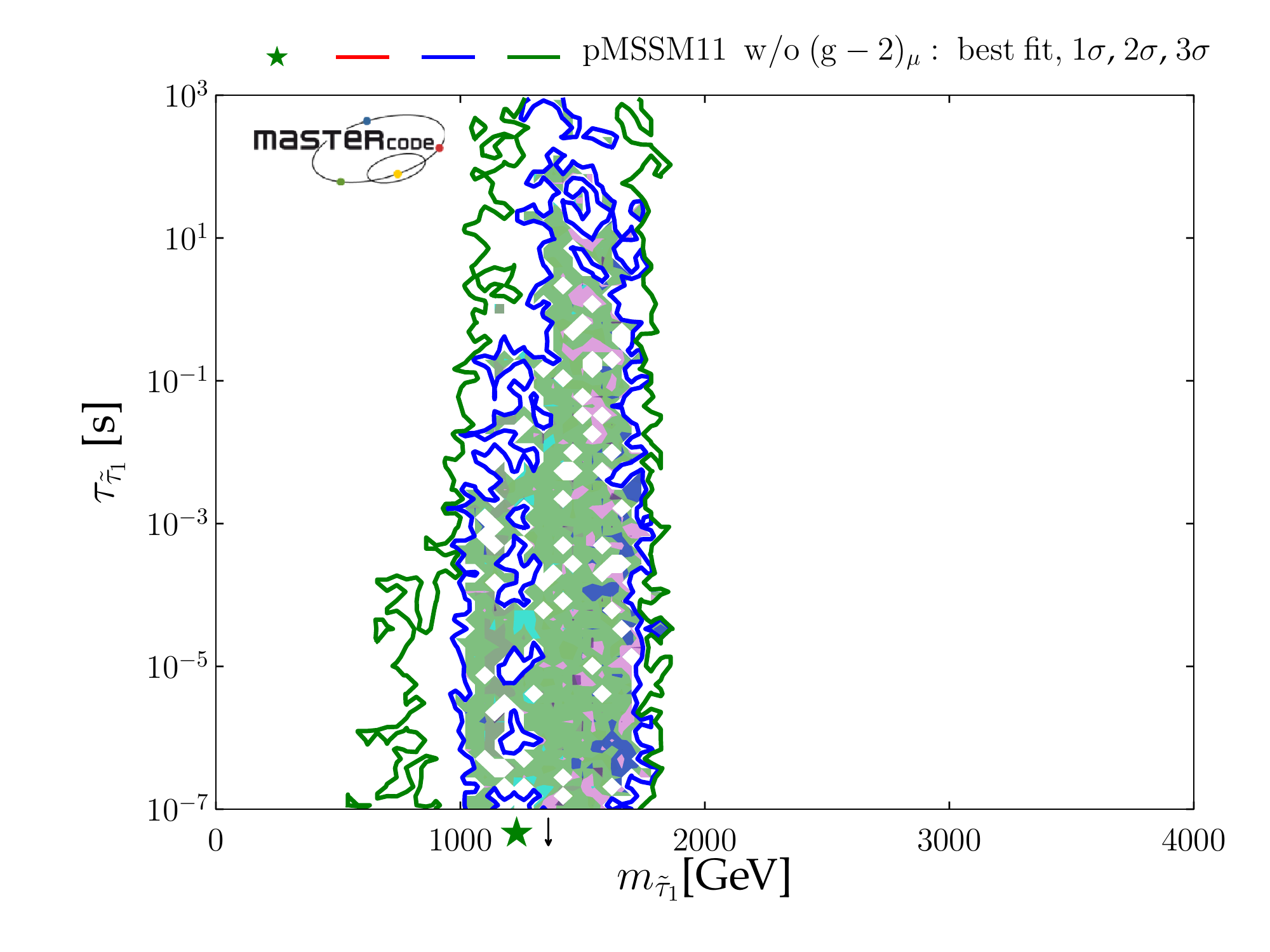}
  \caption{\it Upper panels: one-dimensional profile likelihood plots for the lifetime of the $\cha1$ (left) and the $\staue$ (right).
  Lower panels: the corresponding mass-lifetime planes for the $\cha1$ and $\staue$, {with the 95\% CL regions shaded according
  to the dominant DM mechanisms.}}
\label{fig:morelife}
\end{figure*}

{The upper panels of Fig.~\ref{fig:morelife} display the $\Delta \chi^2$ distributions for chargino
(left) and stau lifetimes (right) between $10^{-7}$~s and $10^3$~s, for the fits omitting \gmt\ (fits including
\gmt\ give $\Delta \chi^2$ outside the displayed range). We see that, whereas shorter lifetimes
are favoured, lifetimes as long as $10^3$~s are allowed at the 95\% CL for both sparticle species when \gmt\ is
dropped, whether or not the LHC 13-TeV data are included. The lower panels of Fig.~\ref{fig:morelife} display the
corresponding mass-lifetime planes for the chargino and stau. We see that a long-lived chargino would
have a mass $\mcha1 \sim 1.1 \tev$, and a long-lived stau would have a mass $\mstaue \sim 1.5 \tev$,
both beyond the reaches of current LHC searches for long-lived charged particles. We have also checked the
possible lifetimes of other NLSP candidates, finding that squarks and gluinos generally have lifetimes
$\lesssim 10^{-17} (10^{-10})$~s at the 95\% CL in fits including LHC 13-TeV with (without) the \gmt\ constraint,
with just a few points having longer lifetimes.
Hence they also do not offer good prospects for LHC searches for long-lived particles.}

\subsection{Spin-Independent Scattering Cross Section}
\label{sec:ssi}

{We now discuss the prospects for direct detection of $\neu1$ DM via spin-independent elastic scattering.
\reffi{fig:2dSIplanes} shows $(\mneu1, \ssi)$ planes, including the LHC 13-TeV data,
with (left panel) and without (right panel) the \gmt\ constraint.
The values of \ssi\ displayed are the nominal values calculated using the central values
of the matrix elements in the {\tt SSARD} code.
The pale green shaded region is that excluded by the combined LUX~\cite{lux16},
XENON1T~\cite{XENON1T} and PandaX-II~\cite{pandax} limit,
which is shown as a solid black line~\footnote{{For completeness, we also show the constraints on \ssi\
from the CRESST-II~\cite{CRESST-II}, CDMSlite~\cite{CDMSlite} and CDEX~\cite{CDEX}
experiments, which are most important at low values of \mneu1
that are excluded by our analysis.}}. The yellow shaded region lies below
the neutrino `floor', which is shown as an orange dashed line. We see that $\mneu1 \gtrsim 100 \gev$ in both the
cases with and without the \gmt\ constraint, with upper limit $\mneu1 \lesssim 550$ {at the 95\% CL when \gmt\
is included. When this constraint is dropped, the 95\% CL range extends up to $2 \tev$, the upper limit for
which our analysis is applicable, because we have limited our scan to slepton masses $\le 2 \tev$.}

{We see that the nominal prediction
for \ssi\ at the best-fit point is at the level of the sensitivities projected for the planned
LUX-Zeplin (LZ) and XENON1T/nT experiments (solid purple line) when the \gmt\ constraint is dropped, and
somewhat higher if \gmt\ is included. However, we emphasize that there are considerable uncertainties in
the estimate of \ssi, which are reflected in the fact that the range of nominal
{\tt SSARD} predictions extends above the current combined limit
from the LUX~\cite{lux16}, XENON1T~\cite{XENON1T} and PandaX-II~\cite{pandax} experiments. There is no
incompatibility when the uncertainties in the \ssi\ estimate are taken into account.
The 68 and 95\% CL ranges of the nominal values of \ssi\ extend slightly
below the neutrino `floor' in the case with \gmt\ included, and much lower in the case where \gmt\ is dropped.
In both cases, large values of \ssi\ occur in the chargino coannihilation region {(green shaded area), with
other DM mechanisms including squark coannihilation yielding large values of \ssi\ for $\mneu1 \gtrsim 1 \tev$.
However,} this and the other DM mechanisms indicated also allow much smaller
values of \ssi.} {As in the case of the pMSSM10 studied in~\cite{mc11}, {we expect that
points with very small values of \ssi\ would, in general, have similarly small values for the spin-independent
scattering cross section on neutrons.}}

\begin{figure*}[htbp!]
\vspace{-1cm}
\centering
\includegraphics[width=0.45\textwidth]{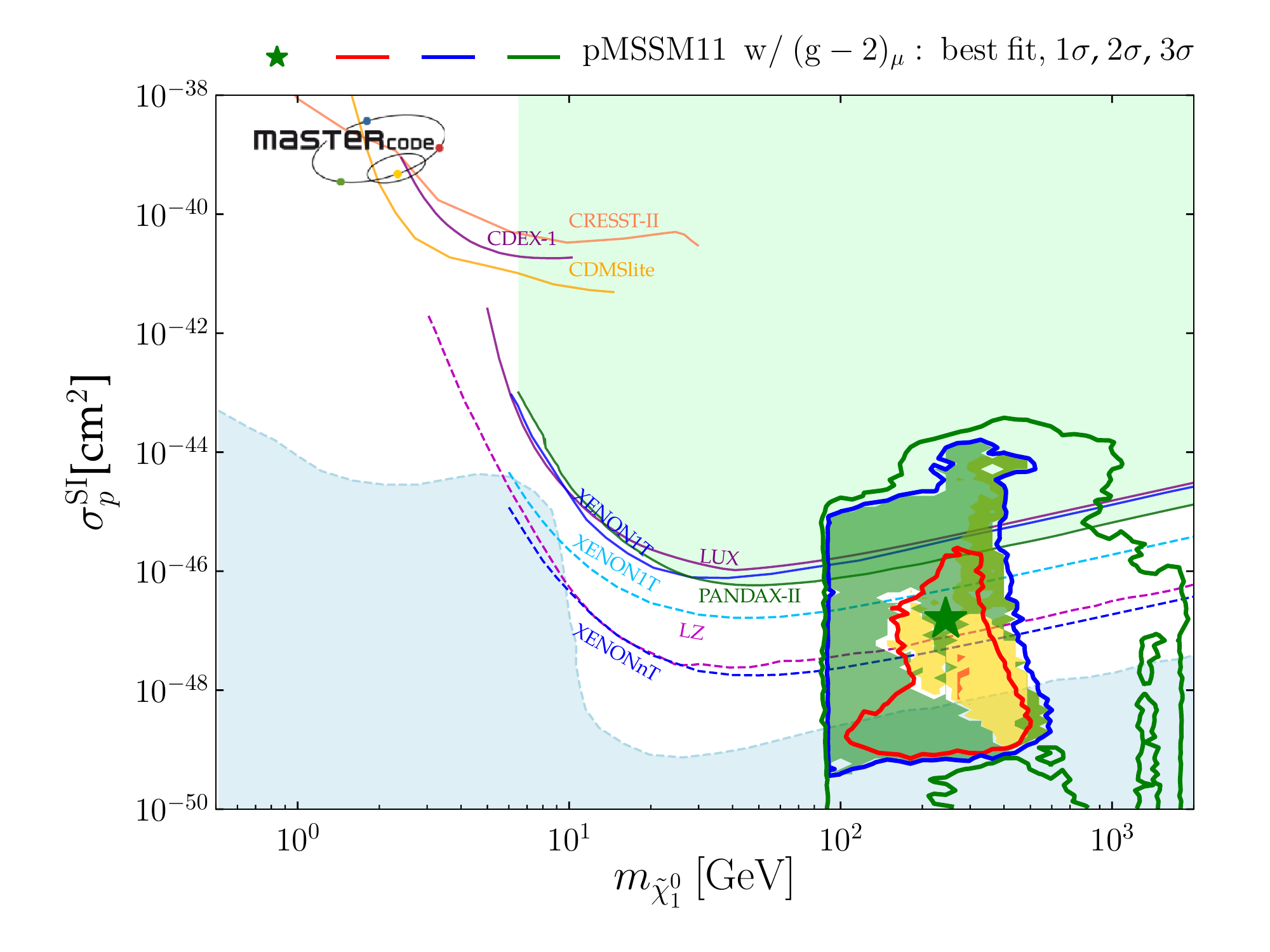}
\includegraphics[width=0.45\textwidth]{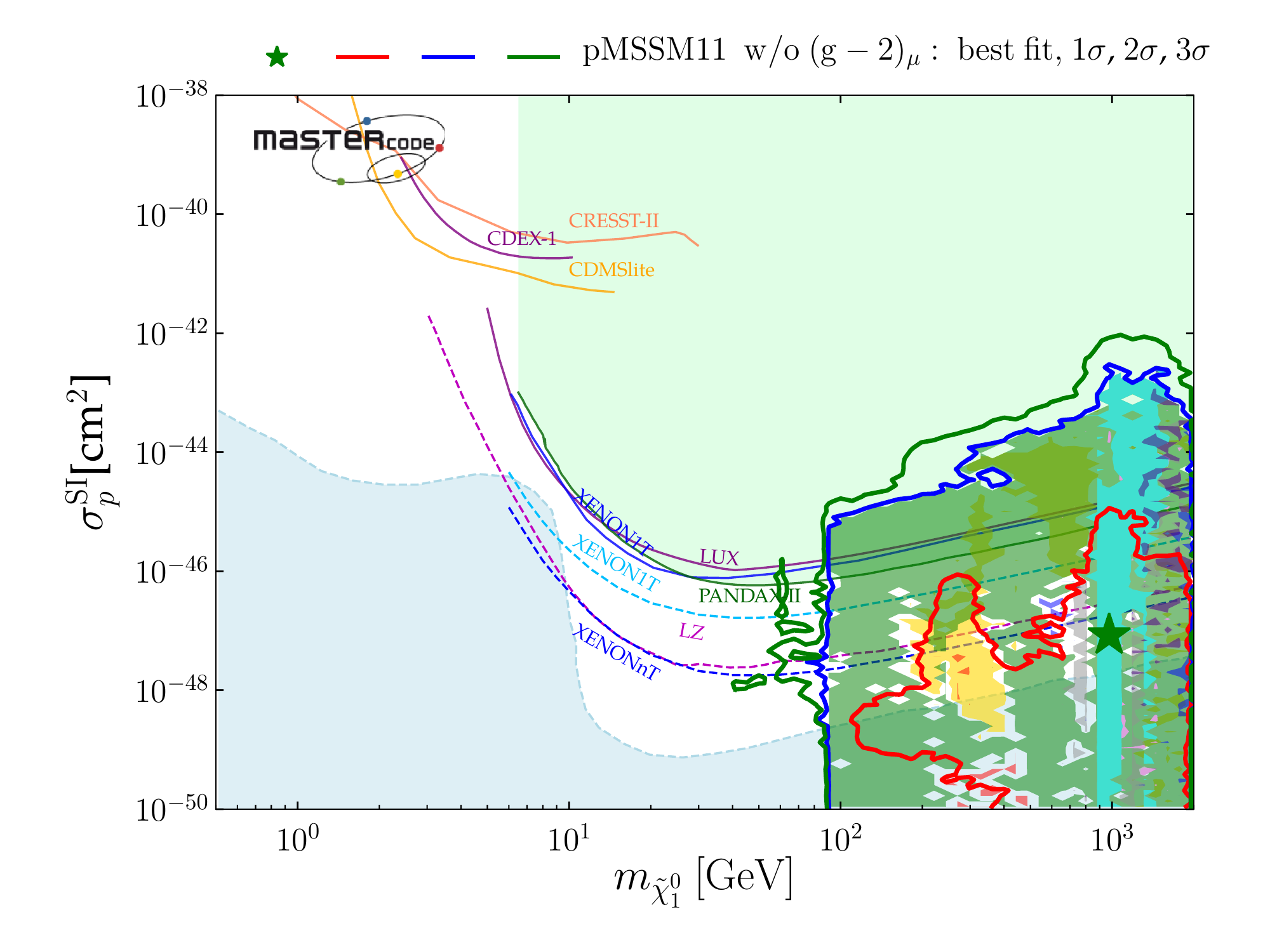} \\
\includegraphics[width=0.8\textwidth]{figs/pMSSM11_dm_legend.pdf}
\caption{\it Planes of $(\mneu1, \ssi)$ with (left panel) and without (right panel) the \gmt\ constraint applied,
where the values of \ssi\ displayed are the nominal values calculated using the {\tt SSARD} code.
The upper limits established by the {LUX~\protect\cite{lux16}, XENON1T~\protect\cite{XENON1T}
and PandaX-II~\protect\cite{pandax} Collaborations are shown as green, magenta and blue contours, respectively,
and the combined limit is indicated by a black line with green shading above.}
The projected {future} 90\% CL exclusion sensitivities of the
LUX-Zeplin (LZ)~\cite{Mount:2017qzi} and XENON1T/nT~\cite{Aprile:2015uzo} experiments are shown as solid purple and dashed blue lines, respectively,
and the neutrino background `floor' is shown as a dashed light-blue line with a shading of the same colour below. }
\label{fig:2dSIplanes}
\end{figure*}

\subsection{Spin-Dependent Scattering Cross Section}
\label{sec:ssd}

{Fig.~\ref{fig:2dSDplanes} displays the corresponding planes of $(\mneu1, \ssd)$ with (left panel) and without (right panel)
the \gmt\ constraint applied. {Here the neutrino `floor' is taken from~\cite{Ng}.}
As in the \ssi\ case, we see that the allowed ranges of $\mneu1$ extend from
$\sim 100 \gev$ to $\sim 550 \gev$ when \gmt\ is included and up to the sampling limit of $2 \tev$ when \gmt\ is dropped.
The uncertainties in the calculation of \ssd\ are significantly smaller than those for \ssi, and we see
that the ranges of the 68 and 95\% regions in the nominal \ssd\ calculations lie below the upper limit
from the PICO experiment~\cite{PICO} (solid purple line). In both the left and right panels,
the nominal predictions for the best-fit points lie some $\sim 3$ orders of magnitude below the current PICO limit.}
{For completeness, we also show the upper limits from SuperKamiokande~\cite{SK} and IceCube~\cite{IceCube} searches for
energetic solar neutrinos, assuming that the LSPs annihilate predominantly into $\tau^+ \tau^-$ (which is not
always the case in the pMSSM11) and neglecting the uncertainties in interpretation mentioned earlier:
see the discussion in the following Section.}

\begin{figure*}[htb!]
\centering
\includegraphics[width=0.45\textwidth]{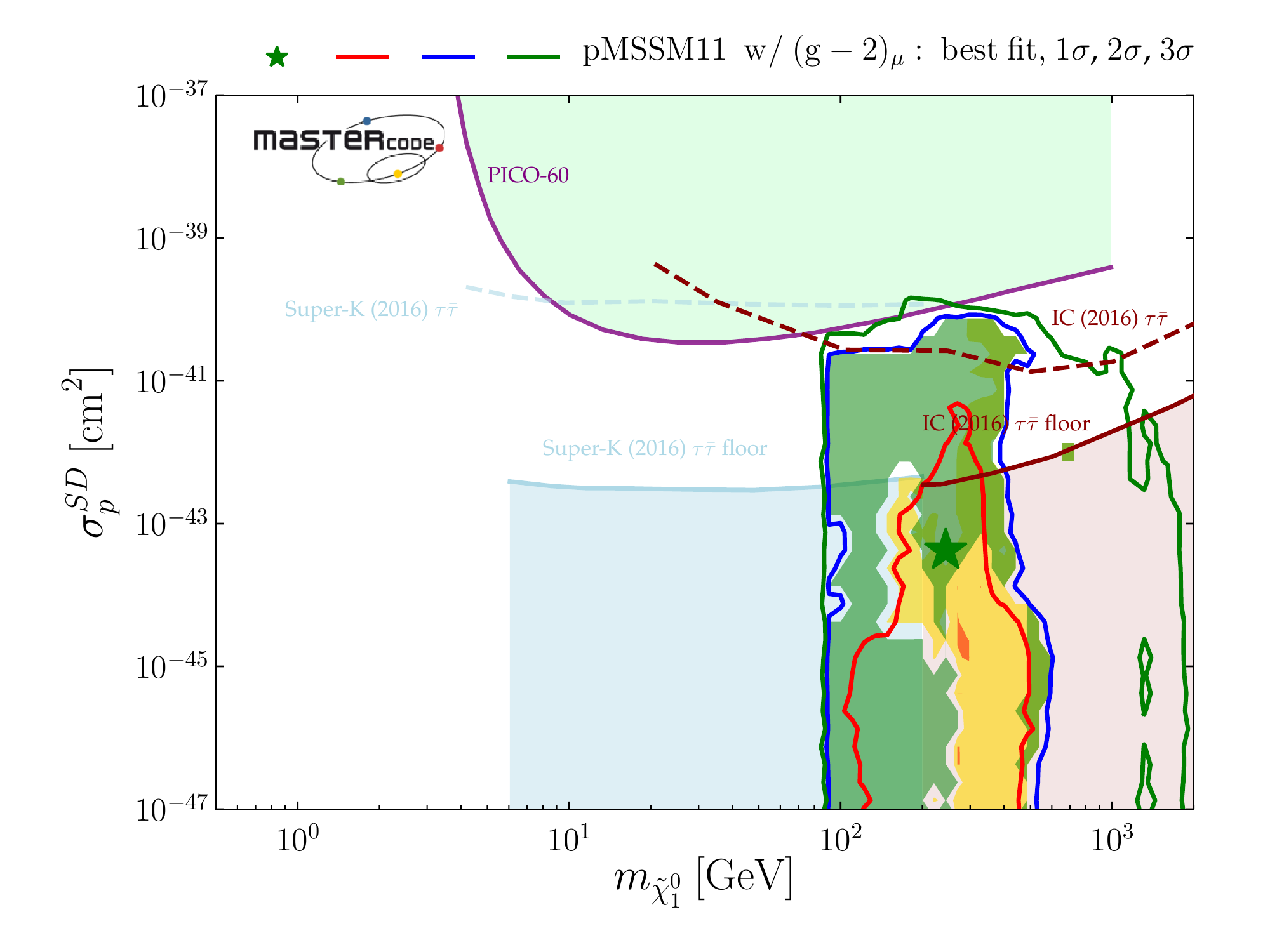}
\includegraphics[width=0.45\textwidth]{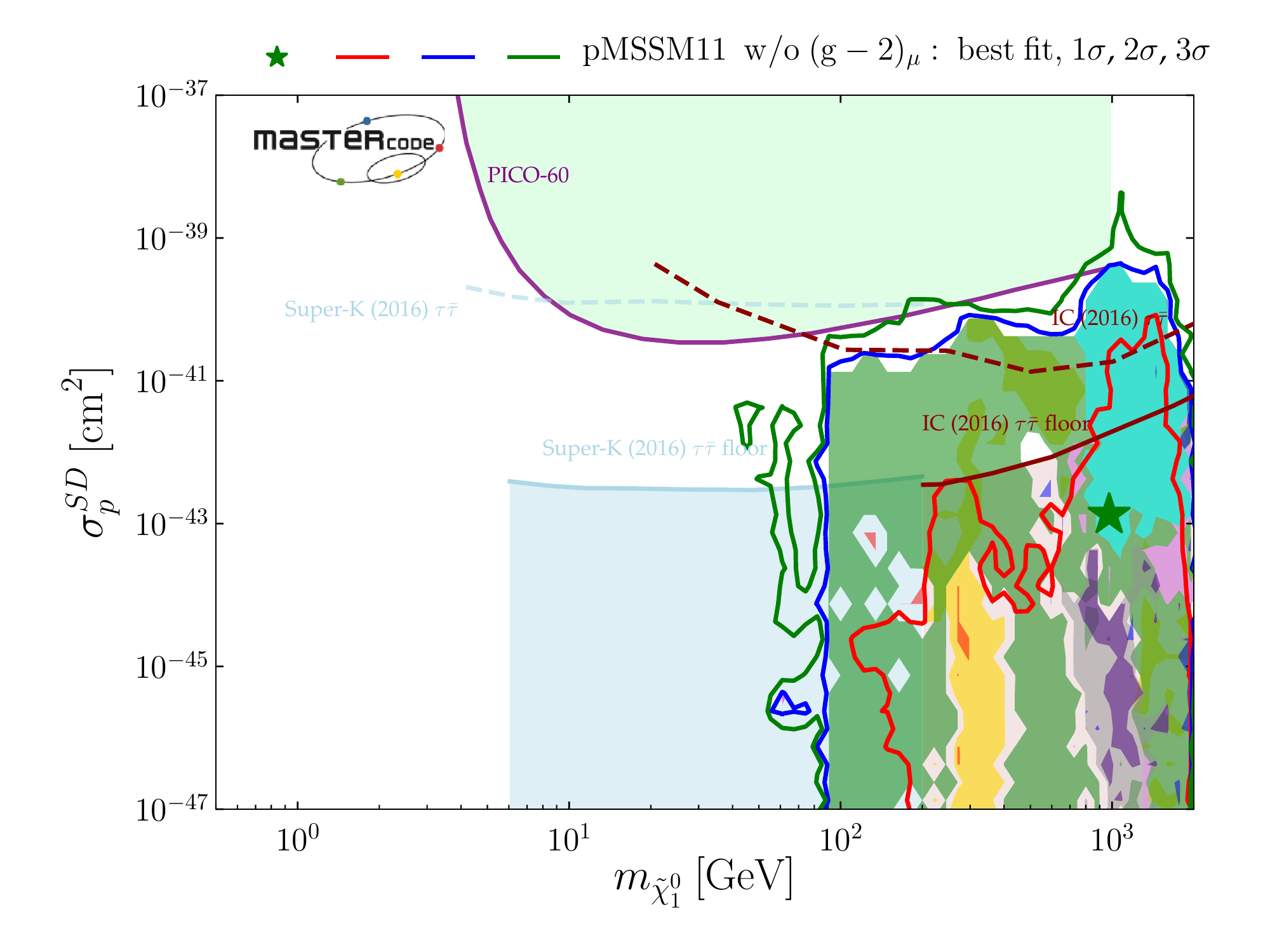} \\
\includegraphics[width=0.8\textwidth]{figs/pMSSM11_dm_legend.pdf}
\caption{\it Planes of $(\mneu1, \ssd)$ with (left panel) and without (right panel) the \gmt\ constraint applied,
where the values of \ssd\ displayed are the nominal values calculated using the {\tt SSARD} code~\protect\cite{SSARD}.
The upper limit established by the PICO Collaboration~\protect\cite{PICO} is shown as a purple contour,
with green shading above. {The neutrino `floor' for \ssd\ is taken from~\protect\cite{Ng}.
We also show the indicative upper limits from
SuperKamiokande~\protect\cite{SK} and IceCube~\protect\cite{IceCube} searches for
energetic solar neutrinos obtained assuming that the LSPs annihilate predominantly into $\tau^+ \tau^-$,
which are subject to the caveats discussed in the text.} }
\label{fig:2dSDplanes}
\end{figure*}

{We see in the left panel of Fig.~\ref{fig:2dSDplanes}
(when \gmt\ is included) that points with chargino coannihilation as the dominant DM mechanism yield
nominal predictions for \ssd\ that extend over many orders of magnitude below the current PICO limit
{and well below the $\tau^+\tau-$ floor}.
Points for which slepton coannihilation is the dominant DM mechanism do not reach so close to the
PICO limit, but may also lie many orders of magnitude below it. We see in the right panel (when \gmt\ is dropped)
similar ranges of nominal \ssd\ values. We also see that when $\mneu1 \gtrsim 1 \tev$ many competing DM
mechanisms come into play, and may give small values of \ssd. However, in the case of squark coannihilation
\ssd\ may lie within $\sim 3$ orders of magnitude of the PICO upper limit.}

\subsection{Indirect Astrophysical Searches for Dark Matter}
\label{sec:IDM}

{We have explored the possible impact of indirect searches for DM via annihilations into
neutrinos inside the Sun. If the DM inside the Sun
is in equilibrium between capture and annihilation, the annihilation is quadratically sensitive to the local
Galactic DM density. However, as discussed earlier, equilibrium is not always a good approximation. We note also
that the capture rate is not determined solely by spin-dependent scattering on protons in the Sun,
but also depends on the amount of spin-independent scattering on Helium and heavy nuclei. As we
have seen, the \ssi\ matrix element is more uncertain than that for \ssd, {and this uncertainty should be
propagated into the constraint on \ssd.} Finally, we note that the
greatest sensitivity of the IceCube search for energetic neutrinos from the Sun~\cite{IceCube} is for
annihilations into $\tau^+ \tau^-$ and $W^+ W^-$, which are not always the dominant final states
in the pMSSM11 models of interest.

{Using the nominal values of the matrix elements from {\tt SSARD}
and neglecting the astrophysical uncertainties, we have calculated the signals in the IceCube detector
for a subset of our pMSSM11 points that are consistent with the PICO constraint~\cite{PICO}.
We find that the IceCube $W^+ W^-$ constraint~\cite{IceCube} has negligible impact on these parameter sets,
and that only a fraction are affected by the IceCube $\tau^+ \tau^-$
constraint. In view of this and the uncertainties in the interpretation of the IceCube searches, we
have not included them in our fits.}

{
\section{Impacts of the LHC 13-TeV and New Direct Detection Constraints}
\label{sec:impacts}


\begin{figure*}[htbp!]
\centering
\includegraphics[width=0.45\textwidth]{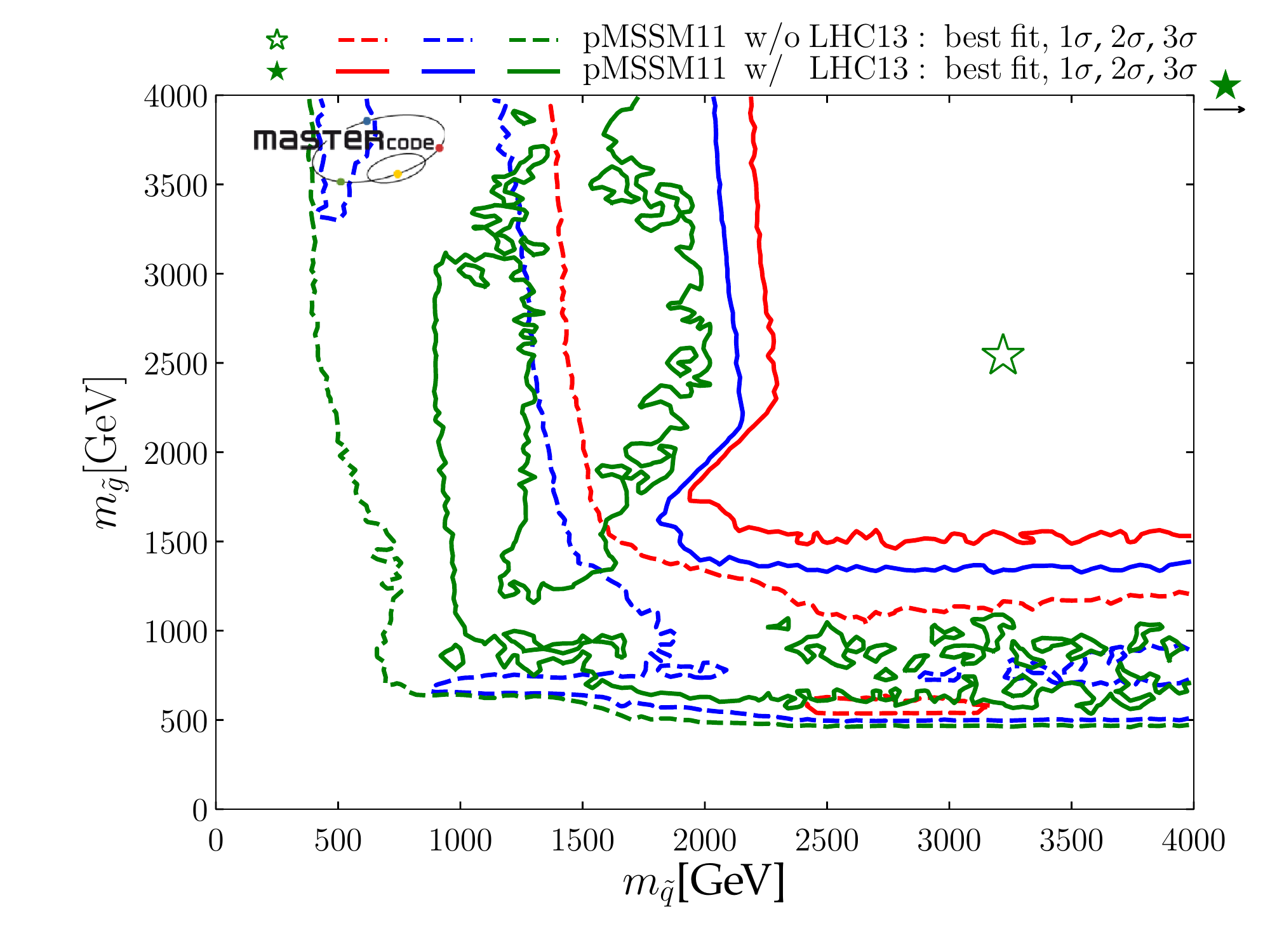}
\includegraphics[width=0.45\textwidth]{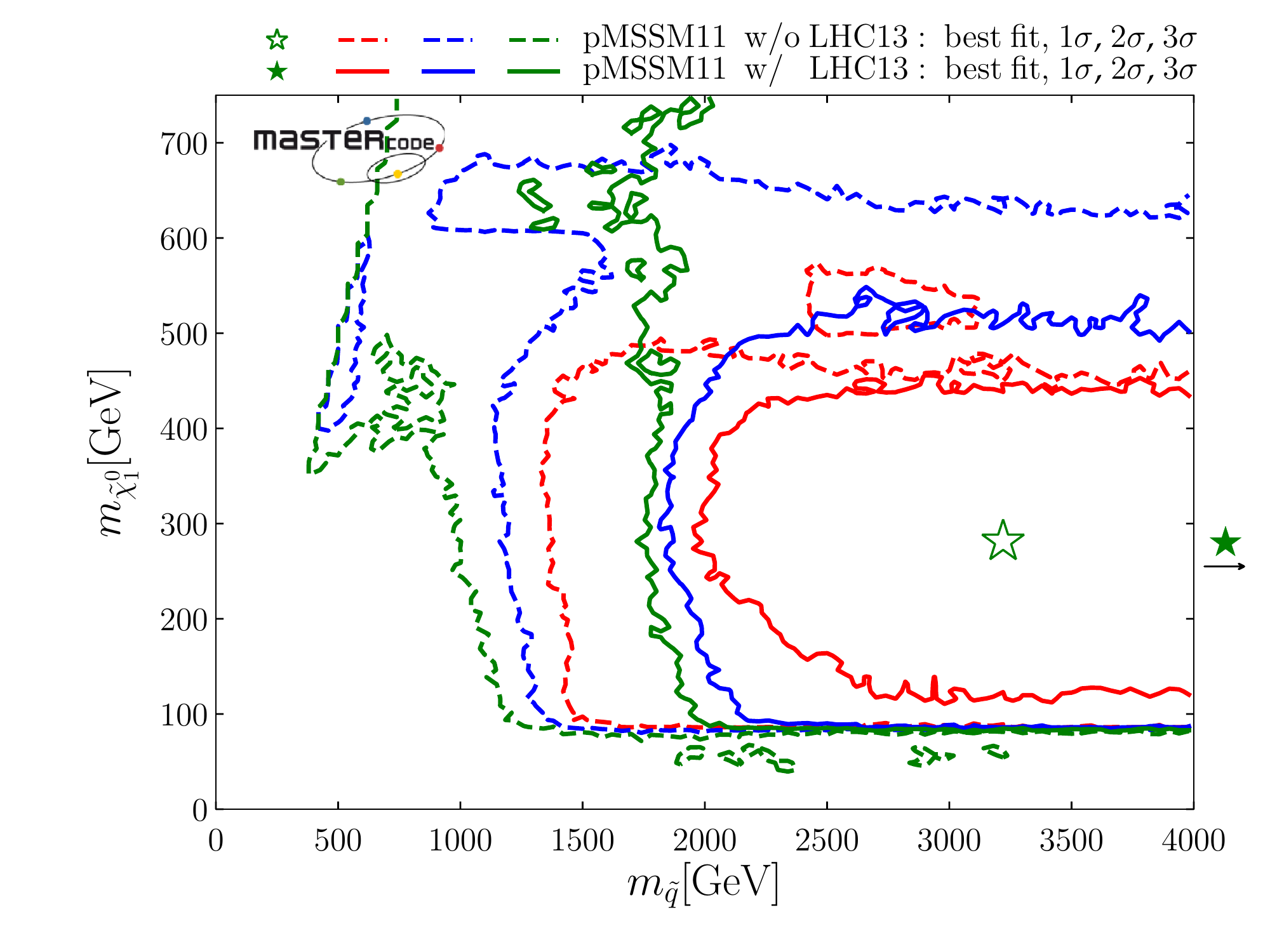} \\
\vspace{2cm}
\centering
\includegraphics[width=0.45\textwidth]{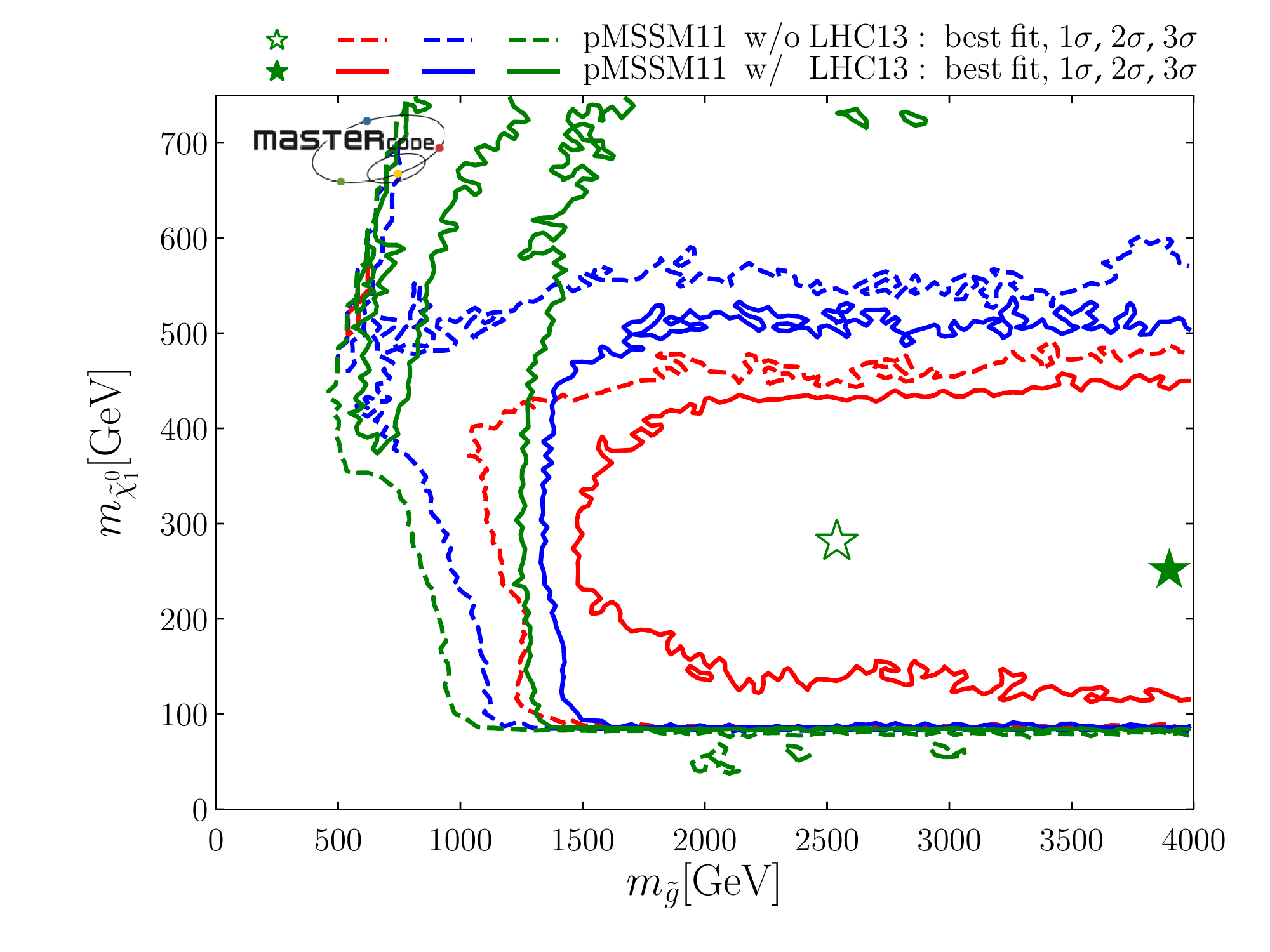}
\includegraphics[width=0.45\textwidth]{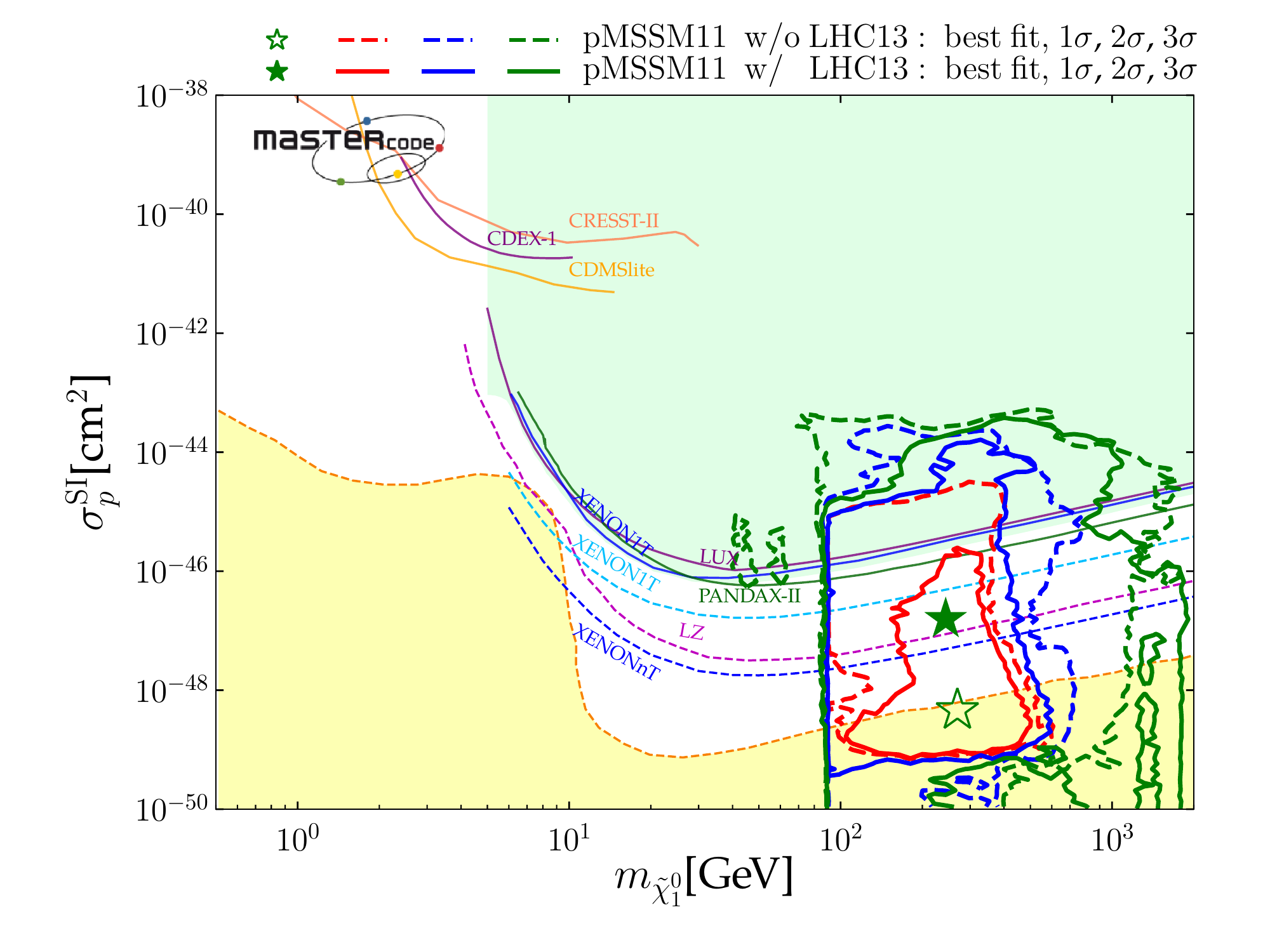} \\
\centering
\caption{\it {{Two-dimensional projections of the global likelihood function for the pMSSM11 in the $(\msq, \mgl)$
and $(\msq, \mneu1)$ planes (upper panels) and the $(\mgl, \mneu1)$ and $(\mneu1, \ssi)$ planes (lower panels).}
The plots compare the regions of the pMSSM11 parameter space
favoured at the 68\% (red lines), 95\% (blue lines) and 99.7\% CL (green lines) in a global fit including
the  LHC 13-TeV data and recent results from the Xenon-based direct detection experiments
LUX, XENON1T, and PandaX-II~\protect\cite{lux16,XENON1T,pandax} (solid lines), and omitting them (dashed lines).}
}
\label{fig:impact}
\end{figure*}
}


{In this Section we illustrate the impact of the LHC 13-TeV data and the recent updates from the
Xenon-based direct detection experiments LUX, XENON1T, and PandaX-II~\cite{lux16,XENON1T,pandax} on relevant pMSSM11 parameter planes.
In the left panel of \reffi{fig:impact} we display the impact of the new results on the $(\msq, \mgl)$ plane:
the solid red, blue and green lines are the current 68\%, 95\% and 99.7\% CL contours, and the dashed lines
are those for the corresponding 68, 95\% and 99.7\% CL contours in a global fit omitting the LHC 13-TeV constraints
and those from the Xenon-based direct detection experiments.
The right panel of \reffi{fig:impact} makes a similar
comparison of the 68, 95 and 99.7\% CL regions in the $(\mneu1, \ssi)$ plane found in global fits including LHC
13-TeV and Xenon-based detector data (solid lines) and omitting these data (dashed lines).

We see in the {upper} left panel of \reffi{fig:impact} that the LHC 13-TeV constraints exclude bands of parameter space
at low $\msq$ and $\mgl$, disallowing in particular a squark coannihilation region at $\msq \sim 500 \gev$
and large $\mgl$ and a gluino coannihilation strip at $\mgl \sim 500 \gev$ that were allowed by the LHC 8-TeV
data. {The impact on the gluino and squark coannihilation strips can also be appreciated from the upper right and lower left panels,
where they appear as dashed-blue islands along the diagonal where the mass is
degenerate with the neutralino that disappear completely after the inclusion of the
LHC 13-TeV constraints.}
The bottom right panel of \reffi{fig:impact} shows that low values of \ssi\ that would have been allowed in a fit without the
LHC 13-TeV data are now disallowed. This effect is in addition to the downwards pressure on \ssi\ exerted
by the new generation of Xenon-based direct detection experiments.}

\section{Best-Fit Points, Spectra and Decays}
\label{sec:spectra}

{Following our previous discussions of some two-dimensional projections of the pMSSM11 parameter space and
various one-dimension profile likelihood functions, we now discuss in more detail the best-fit points in the pMSSM11 fits
{incorporating the LHC 13-TeV data, both with and
without the \gmt\ constraint, whose input pMSSM11 parameter values were given in the first and third columns
of Table~\ref{tab:points}. We note, however,} that the likelihood functions are very flat for larger masses, so these
best-fit points should not be taken as definite predictions.
}

{\reffi{fig:spectra} displays the spectra of Higgs bosons and sparticles at the best-fit points for the
pMSSM11 including (upper panel) and excluding (lower panel) the \gmt\ constraint~\footnote{{This
figure was prepared using {\tt PySLHA}~\cite{PySLHA}.}}. In each case we also show
the decay paths with branching ratios  $> 5\%$, the widths of the lines being proportional to the branching ratios.
The heavier Higgs bosons $H, A, H^\pm$, are lighter in the case without \gmt, whereas the sleptons and the electroweak inos
are heavier. {The branching ratio patterns differ in the two cases, with the Higgs bosons mainly decaying to
SM particles when \gmt\ is not imposed.} We note that the first- and second-generation sleptons are much
lighter than the third-generation sleptons in the case with \gmt.
The third-generation squarks are also heavier when \gmt\ is dropped,
whereas the gluino and the first- and second-generation squarks are lighter in this case.
In both cases, the third-generation squarks may lie within reach of future LHC runs,
whereas the first- and second generation squarks would be accessible only if \gmt\ is dropped.
The gluino would also be accessible in this case, and possibly also if \gmt\ is included.}

{We re-emphasize that the remarks in the previous paragraph apply to the best-fit points, and that the spectra might
differ significantly, as the likelihood functions are quite flat for large masses. The 68 and 95\% CL ranges are
displayed in \reffi{fig:spectrum} as orange and yellow bands, respectively, with the best-fit values indicated by
blue lines. We see that for most sparticles the 95 and even 68\% CL ranges extend into the ranges accessible
to future LHC runs. As was to be expected, the best prospects for measuring sparticles at a linear
$e^+ e^-$ collider such as ILC~{\cite{ILC,ILC2}} or CLIC~\cite{CLIC}
are offered by first- and second-generation sleptons and the lighter electroweak inos
$\neu1, \neu2$ and $\cha1$ in the case with the \gmt\ constraint applied.}

\begin{figure*}[htbp!]
  \includegraphics[width=\textwidth]{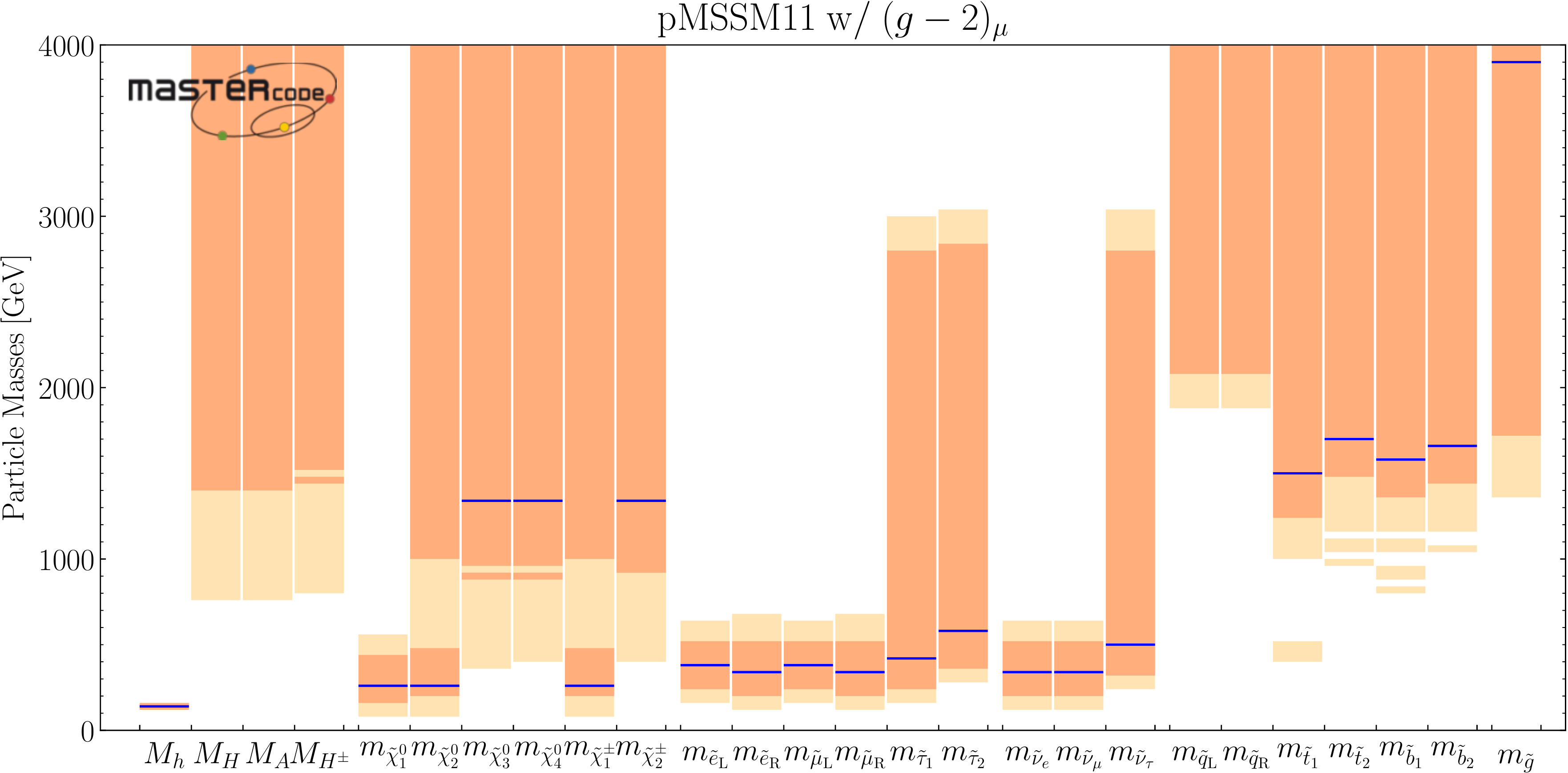}\\
  \vspace{0.3cm}\\
  \includegraphics[width=\textwidth]{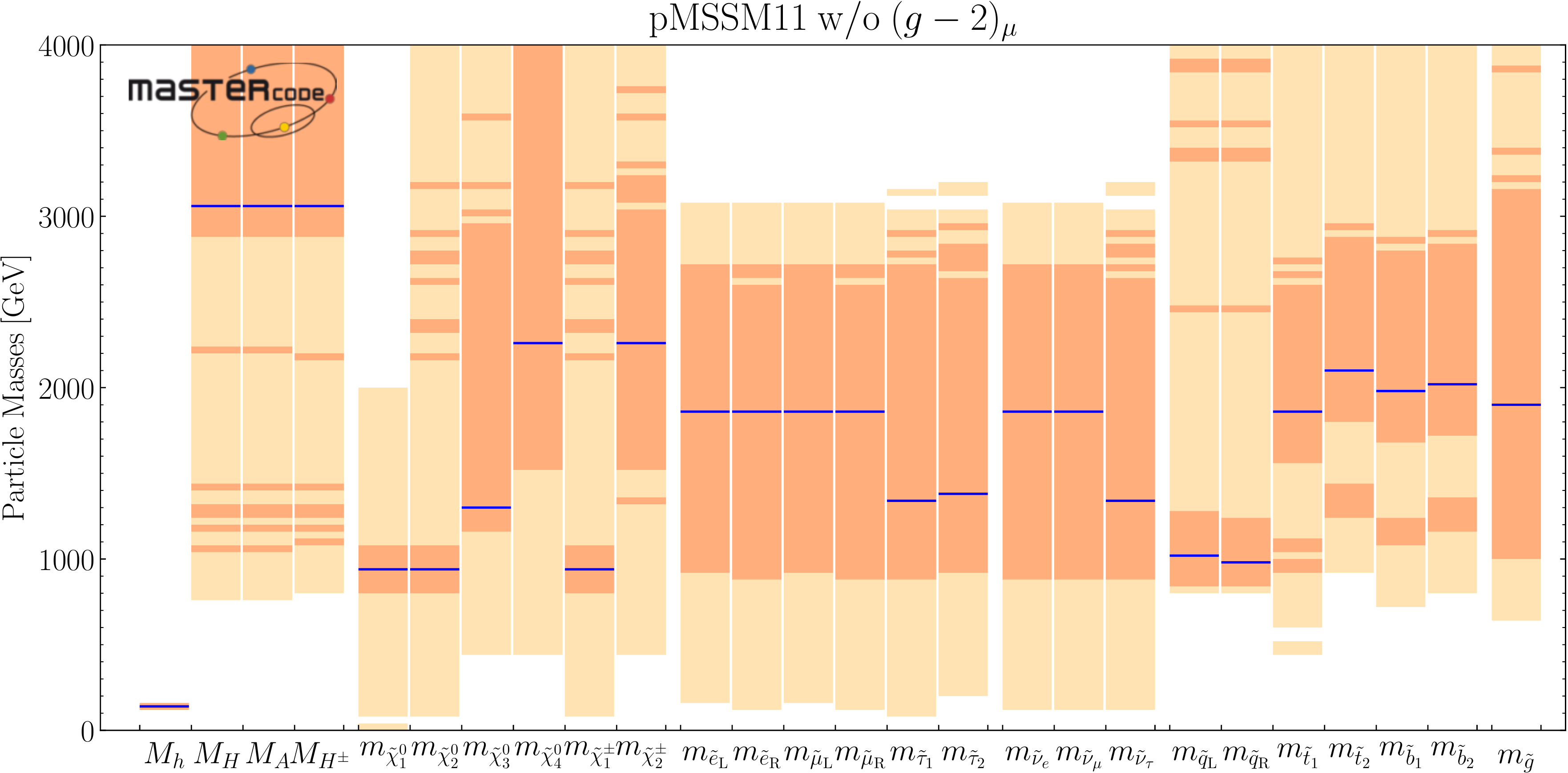}\\
  \caption{\it Higgs and sparticle spectrum for the pMSSM11 with and without the $(g-2)_{\mu}$ constraint applied (upper
  and lower panels, respectively). {The values at the best-fit points are indicated by blue lines,
  the 68\% CL ranges by orange bands, and the 95\% CL ranges by yellow bands.}}
  \label{fig:spectrum}
\end{figure*}

{\reffi{fig:chi2break} displays the breakdowns of the global $\chi^2$ functions
in the cases with (left panel) and without (right panel) the \gmt\ constraint~\footnote{{The corresponding horizontal
bar has diagonal hatching, to recall that it is not included in the fit.}}.
{The different classes of observables are grouped together and colour-coded.
We see that $\MW$ makes only a small contribution,
and that the total contribution to the global $\chi^2$ function of the precision electroweak observables
are quite similar in the two cases. The total contribution of the flavour sector is slightly
reduced when \gmt\ is dropped: $\Delta \chi^2 \sim - 1.2$, largely because of a better
fit to \bsmm, but this improvement is not very significant. The contributions of the
Higgs, LEP, LHC and DM sectors are again very similar in the fits with and without \gmt.} \\

\begin{figure*}[htbp!]
\centering
\includegraphics[width=0.48\textwidth]{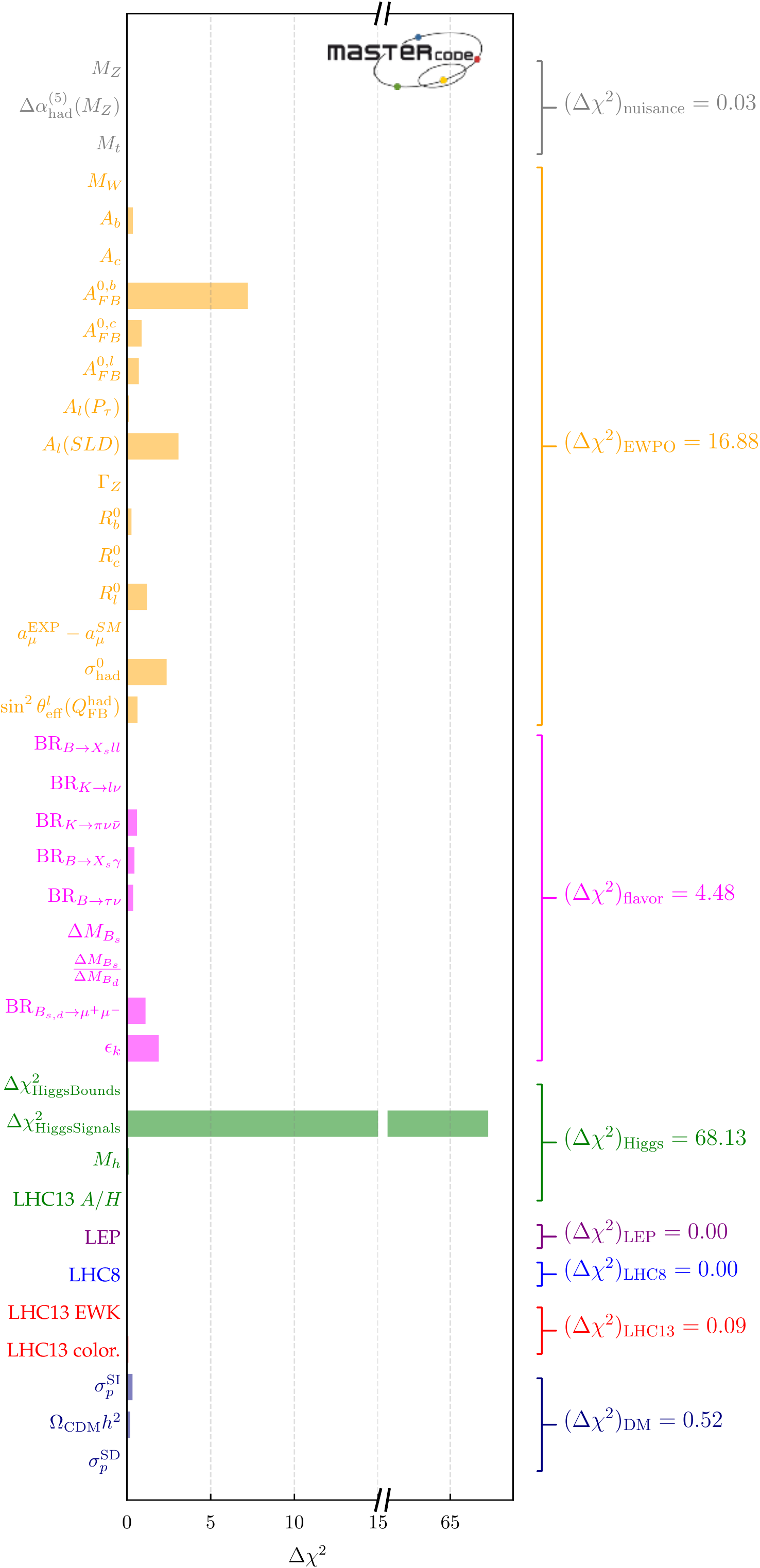}
\includegraphics[width=0.48\textwidth]{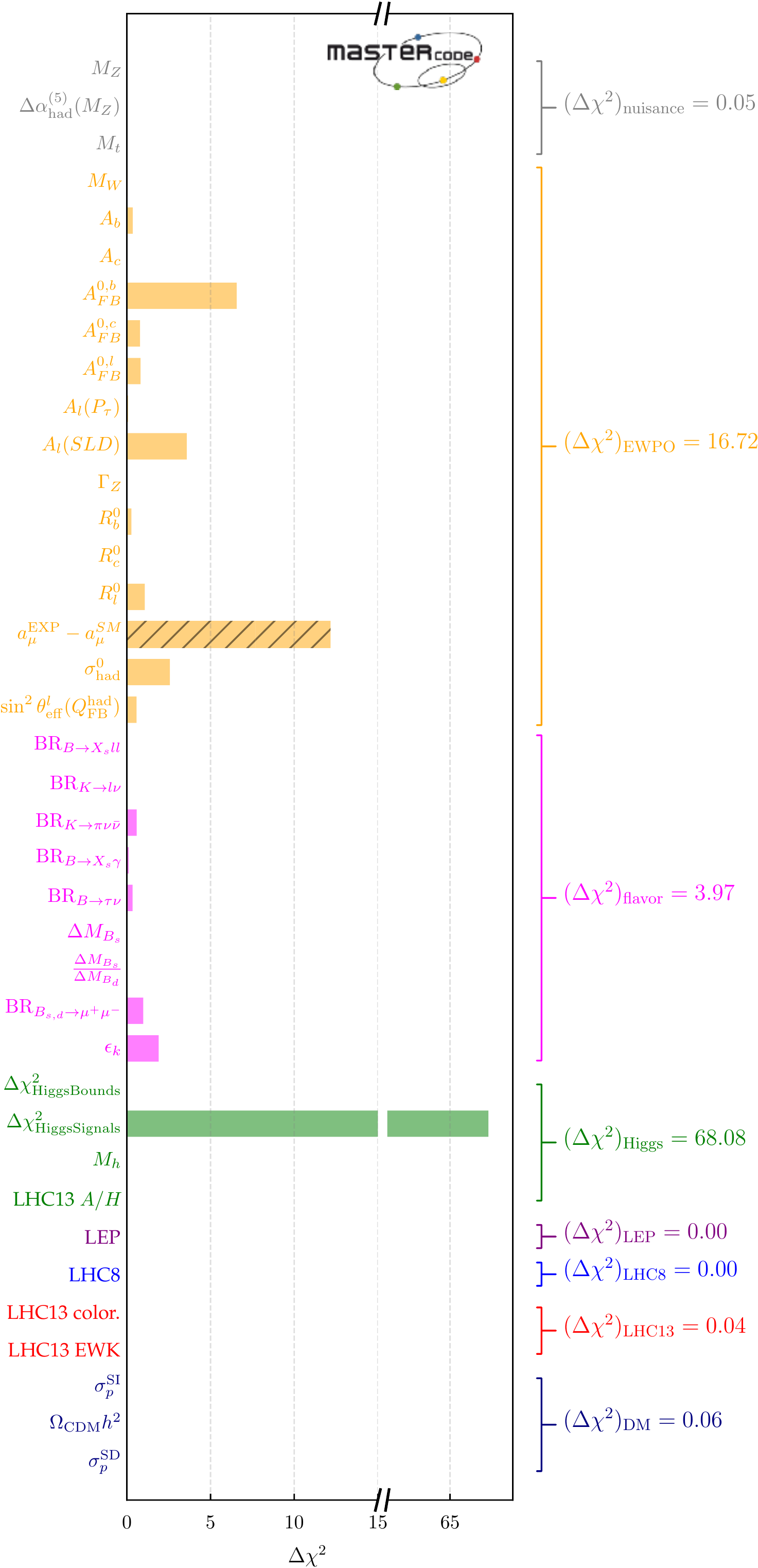} \\
\caption{\it The $\chi^2$ pulls at the best-fit points in the pMSSM11 including (left) and without the \gmt\ constraint (right).
  {In the rightmost plot, the $\chi^2$ pull from \gmt\ is shown (hatched orange bar), but its penalty is not included in the fit.}
  }

  \label{fig:chi2break}
\end{figure*}


\section{Conclusions}
\label{sec:conx}

{In this paper we have used the {\tt MasterCode} tool to analyze the constraints
on the parameter space of the pMSSM11 model, in which the soft SUSY-breaking
contributions to the masses of the first- and second-generation sleptons are allowed
to vary independently from the third-generation slepton mass. We have taken into account the available
constraints on strongly- and electroweakly-interacting sparticles
from $\sim 36$/fb of LHC data at 13 TeV~\cite{cms_0lep-mt2,cms_1lep-MJ,sus-16-039} and the most recent limits from the
LUX, PICO, XENON1T and PandaX-II experiments~\cite{lux16,PICO,XENON1T,pandax} searching directly for DM
scattering. In addition, we have updated the constraint from the measurement of $\MW$
and some constraints from flavour observables, as described in Table~2. We have
presented the results from two global fits, one including the \gmt\ constraint and
without it. We have also made various comparisons with fits without the LHC 13-TeV
data. {Comparing with our earlier fit to the pMSSM10~\cite{mc11}, we note that
the freedom for $\mslep \ne m_{\tilde \tau}$ plays an important role in best
fits. {Furthermore, }
there is a big difference between $M_1$ and $M_2$ at the best-fit point without \gmt.}}

{The most visible impact of the LHC 13-TeV constraints has been on the masses of the
strongly-interacting sparticles: see the left panels of Figs.~\ref{fig:1dglsq} and \ref{fig:1stsb} and compare
the solid and dashed curves. On the other hand, the impact of
the LHC constraints on electroweak inos has been less marked: see Fig.~\ref{fig:1dneu}.
As was to be expected, the importance of the \gmt\ constraint is seen in the likelihood functions
for charged slepton masses and electroweak inos: compare the blue and green curves
in Figs.~\ref{fig:1dsmst} and \ref{fig:1dneu}.
The composition of the LSP $\neu1$ is also different in the cases with and without \gmt:
as seen in Fig.~\ref{fig:1hcompo} and Table~\ref{tab:lsp}, a $\tilde B$ LSP is preferred
when \gmt\ is included, whereas a $\tilde H$ LSP is preferred when \gmt\ is dropped.
Moreover, the inclusion of the \gmt\ constraint also has significant indirect implications for the
squark masses, as also seen in Figs.~\ref{fig:1dglsq} and \ref{fig:1stsb}. This analysis
reinforces the importance of clarifying the interpretation of the difference between the
experimental measurement and the SM calculation of \gmt. We therefore welcome the
advent of the Fermilab \gmt\ experiment~\cite{FNALg-2} and continued efforts to refine the SM calculation.}

{We have also analyzed in this paper the importances of different mechanisms for
bringing the relic LSP density into the range favoured by Planck 2015 and other data:
see the shadings in Figs.~2, 4, 5, 6, 19, 20 and 21, and the profile $\chi^2$ functions for the DM
measures in Fig.~\ref{fig:EWmeasures}. As we see there, important roles are played by
chargino coannihilation, slepton coannihilation and rapid annihilation via direct-channel
$H/A$ boson exchange, though other mechanisms such as stau and squark coannihilation
may be important in limited regions of parameter space~\footnote{Compared
to the pMSSM7 analysis in~\cite{GAMBIT7}, we find that
stop coannihilation is less prominent, and that rapid annihilation through the $Z$ and
the light Higgs boson is of very limited importance. {In these respects the more general
      realization of the MSSM with four additional free parameters yields
      substantially different results.}}.
In the case where the \gmt\ constraint
is dropped, there is a preference for a region where $\mneu1 \sim \mcha1 \sim \msq \sim \mgl$
where multiple coannihilation processes play a role, and the compressed spectrum reduces
the sensitivity of the LHC sparticle searches.}

{In general, our analysis favours quite small deviations from the SM predictions for
electroweak, flavour and Higgs observables: see Figs.~12 and 14, in particular. We have
also analyzed the pMSSM11 predictions for the $A_{\Delta \Gamma}$ and
$\tau (B_s \to \mu^+ \mu^-)$ observables recently measured for the first time by the
LHCb Collaboration~\cite{1703.05747}. As seen in Fig.~13, the pMSSM11 predictions for these
observables are very similar to those in the SM, deviating by much less than the
current experimental uncertainties. Accordingly, we do not include $A_{\Delta \Gamma}$
and $\tau (B_s \to \mu^+ \mu^-)$ in our global fits.}

{We find that current LHC searches for long-lived particles do not impact our
scan of the pMSSM11 parameter space. However, the pMSSM11 still offers
significant prospects for the discovery of long-lived particles.
When the \gmt\ constraint is imposed, we find that $\Delta \chi^2 \gtrsim 4$ for
$\tau_{\rm NLSP} \gtrsim 10^{-10}$~s. However, when the \gmt\ constraint is dropped,
values of $\tau_{\rm NLSP}$ as long as $10^3$~s (the limit we impose in order to maintain
successful Big Bang nucleosynthesis) are allowed at the $\Delta \chi^2 \lesssim 4$ level,}

{As seen in Figs.~\ref{fig:2dSIplanes} and \ref{fig:2dSDplanes}, the pMSSM11 offers interesting prospects
for the detection of supersymmetric DM. In both the spin-independent and -dependent cases,
cross sections close to the present experimental upper limits are favoured at the 68\% CL, whether or not
\gmt\ is included in the set of constraints. Interestingly, in the case of \ssi\ with \gmt\ included,
there is a lower limit that is not far below the neutrino `floor'~\footnote{{However, we repeat
that the uncertainties in the calculation of \ssi\ are large, and these remarks apply within the framework
of a calculation of \ssi\ using {\tt SSARD}.}}, whereas \ssi\ may be much lower when
\gmt\ is dropped, and low values of \ssd\ are allowed in both cases.}

{We turn finally to the prospects for discovering sparticles in future runs of the LHC,
or with a future linear $e^+ e^-$ collider. As seen in Fig.~21, whether or not \gmt\ is included in the global fit,
the third-generation squarks may well be within reach of future LHC runs, and the first-
and second-generation squarks and the gluino may also be accessible if the \gmt\
constraint is dropped. {If it is included, on the other hand,} there are
also good prospects for discovering
electroweakly-interacting sparticles at an $e^+ e^-$ collider, in particular the
${\tilde e}, {\tilde \mu}, \neu1, \neu2$ and $\cha1$.

It is often said that the night is
darkest just before dawn, and the same may be true for supersymmetry.}


\section*{Acknowledgements}

{We thank Gino Isidori for useful discussions.}
The work of E.B. and G.W. is supported in part by the Collaborative Research Center
SFB676 of the DFG, ``Particles, Strings and the early Universe''.
The work of K.S. is partially supported by the National Science Centre, Poland,
under research grants DEC-2014/15/B/ST2/02157, DEC-2015/18/M/ST2/00054 and DEC-2015/19/D/ST2/03136.
K.S. thanks the TU Munich for hospitality during the final stages of this work and
has been partially supported by the DFG cluster of excellence EXC 153 ``Origin and Structure of the Universe'',
by the Collaborative Research Center SFB1258.
The work of M.B., I.S.F. and D.M.S. is supported by the European Research Council
via Grant BSMFLEET 639068.
{The work of J.C.C. is supported by CNPq (Brazil).}
The work of M.J.D. is supported in part by the Australia Research Council.
The work of J.E. is supported in part by STFC (UK) via the research grant ST/L000326/1
and in part via the Estonian Research Council via a Mobilitas Pluss grant,
and the work of H.F. is also supported in part by STFC (UK). The work of S.H. is
supported in
part by the MEINCOP Spain under contract FPA2016-78022-P, in part by the
Spanish Agencia Estatal de Investigaci{\' o}n (AEI) and the EU Fondo Europeo de
Desarrollo Regional (FEDER) through the project FPA2016-78645-P, in part by
the AEI through the grant IFT Centro de Excelencia Severo Ochoa SEV-2016-0597,
and by the Spanish MICINN Consolider-Ingenio 2010 Program under Grant
MultiDark CSD2009-00064. The work of M.L. and I.S.F. is supported by XuntaGal.
The work of K.A.O. is supported in part by DOE grant de-sc0011842 at the
University of Minnesota. The work of G.W. is also supported in part by the European Commission
through the ``HiggsTools'' Initial Training Network PITN-GA-2012-316704.
{During part of this work we used the middleware suite {\tt udocker}~\cite{udocker} to deploy {\tt MasterCode} on
clusters, developed  by the EC H2020 project INDIGO-Datacloud (RIA 653549).
We are particularly grateful to Jorge Gomes for his kind support.
We also thank DESY and especially the DESY IT department for making us available the
computational resources of the BIRD/NAF2 cluster, which have been used intensively to carry out this work.}



\end{document}